\documentclass[universe,conference report,accept,oneauthor,pdlaftex,10pt,a4paper]{Definitions/mdpi}

\usepackage{bm}
\usepackage{latexsym}
\usepackage{amsfonts,amssymb,amsmath}
\usepackage{graphicx,epsfig}
\usepackage{psfrag}
\usepackage{subfigure}
\usepackage{ulem}
\usepackage{hyperref}
\usepackage{color}
\usepackage{appendix}

\newcommand{\ie}{{i.e.~}}

\newcommand{\dd}{\mathrm{d}}

\newcommand{\sss}[1]{{\scriptscriptstyle{#1}}}

\newcommand{\uPl}{\mathrm{Pl}}

\newcommand{\uini}{\mathrm{ini}}

\newcommand{\usssPl}{\sss{\uPl}}

\newcommand{\setR}{\mathbb{R}}

\newcommand{\Mp}{M_\usssPl}

\newcommand{\beq}{\begin{equation}}
\newcommand{\eeq}{\end{equation}}
\newcommand{\bea}{\begin{eqnarray}}
\newcommand{\eea}{\end{eqnarray}}

\newlength{\wsingfig}
\newlength{\wdblefig}
\newlength{\wquadfig}
\newlength{\wtriplefig}
\newcommand{\Eqs}[1]{Eqs.~(\ref{#1})}

\newcommand{\Ref}[1]{Ref.~{\cite{#1}}}

\makeatletter
\gdef\@fpheader{}
\g@addto@macro\bfseries{\boldmath}
\makeatother

\hypersetup{
    bookmarks=true,         
    unicode=false,          
    pdftoolbar=true,        
    pdfmenubar=true,        
    pdffitwindow=false,     
    pdfstartview={FitH},    
    pdfnewwindow=true,      
    colorlinks=true,        
    linkcolor=red,          
    citecolor=cyan,         
    filecolor=magenta,      
    urlcolor=blue,         
    linktocpage=true
}

\firstpage{1} 
\pubvolume{xx}
\issuenum{1}
\articlenumber{1}
\pubyear{2018}
\copyrightyear{2018}
\history{Received: date; Accepted: date; Published: date}
\updates{yes} 
\usepackage[utf8]{inputenc}
\usepackage{bm}

\Title{Cosmic Inflation, Quantum Information and the Pioneering Role
  of John S Bell in Cosmology}

\Author{J\'er\^ome Martin$^{1}$}

\AuthorNames{J'er\^ome Martin}

\address{%
$^{1}$\, Institut d'Astrophysique de Paris, UMR 7095 CNRS, Universit\'e Pierre et Marie Curie, 98bis boulevard Arago, 75014 Paris, France.}

\simplesumm{I discuss whether a signature of the quantum origin of
  large scale structures observed in our Universe can be found.}

\abstract{According to the theory of cosmic inflation, the large scale
  structures observed in our Universe (galaxies, clusters of galaxies,
  Cosmic Background Microwave - CMB - anisotropy \dots ) are of
  quantum mechanical origin. They are nothing but vacuum fluctuations,
  stretched to cosmological scales by the cosmic expansion and
  amplified by gravitational instability. At the end of inflation,
  these perturbations are placed in a two-mode squeezed state with the
  strongest squeezing ever produced in Nature (much larger than
  anything that can be made in the laboratory on Earth). This article
  studies whether astrophysical observations could unambiguously
  reveal this quantum origin by borrowing ideas from quantum
  information theory. It is argued that some of the tools needed to
  carry out this task have been discussed long ago by J.~Bell in a, so
  far, largely unrecognized contribution. A detailed study of his
  paper and of the criticisms that have been put forward against his work
  is presented. Although J.~Bell could not have realized it when he
  wrote his letter since the quantum state of cosmological
  perturbations was not yet fully characterized at that time, it is
  also shown that Cosmology and cosmic inflation represent the most
  interesting frameworks to apply the concepts he investigated. This
  confirms that cosmic inflation is not only a successful paradigm to
  understand the early Universe. It is also the only situation in
  Physics where one crucially needs General Relativity and Quantum
  Mechanics to derive the predictions of a theory and, where, at the
  same time, we have high-accuracy data to test these predictions,
  making inflation a playground of utmost importance to discuss
  foundational issues in Quantum Mechanics.}

\keyword{cosmology; inflation; quantum mechanics}

\begin{document}

\section{Introduction}
\label{sec:intro}

The theory of cosmic
inflation~\cite{Starobinsky:1980te,Starobinsky:1982ee,
  Guth:1980zm,Linde:1981mu,Albrecht:1982wi,Linde:1983gd} is considered
as the leading paradigm for describing the physical conditions that
prevailed in the early Universe. It is a very successful theory
because it solves the puzzles of the standard model of Cosmology but,
also, because it has made predictions that have been observationally
verified (for a recent assessment of the scientific status of
inflation, see Refs.~\cite{Chowdhury:2019otk,Martin:2019zia}). For
instance, it predicts the presence of Doppler peaks in the Cosmic
Microwave Background (CMB) multipoles moments, a vanishing spatial
curvature or a power spectrum of cosmological perturbations close to
scale invariance but not exactly scale
invariant~\cite{Starobinsky:1979ty,Mukhanov:1981xt,Mukhanov:1982nu,Guth:1982ec,
  Hawking:1982cz,Bardeen:1983qw} (for reviews, see {\textit e.g.}
Refs.~\cite{Martin:2004um,Martin:2007bw}). This last prediction has
been verified, at a statistical significant level, only recently,
thanks to the release of the European Space Agency (ESA) satellite
Planck
data~\cite{Ade:2015xua,Ade:2015lrj,Ade:2015ava,Aghanim:2018eyx,Akrami:2018odb}.

But inflation is also interesting because it combines Quantum
Mechanics and General Relativity. Indeed, according to inflation, all
structures in our Universe are of quantum-mechanical origin. This
claim, although very strong, seems to be empirically correct in the
sense that all conclusions that can be derived from this assumption
fit well the data at our disposal. However, it would clearly be
interesting to go beyond an indirect proof and to be able to find an
explicit and unambiguous signature of the quantum origin of the
structures present in our Universe.

This article is devoted to this question and discusses the tools,
often borrowed to Quantum Information Theory, that can be used in
order to address these
problems~\cite{Martin:2012pea,Martin:2015qta,Martin:2016nrr,Martin:2016tbd,Martin:2017zxs}.

We also argue that crucial insights into those issues were anticipated
by John Bell in an, so far, unrecognized contribution ``{\it EPR
  correlations and EPW distributions}''~\cite{1986NYASA.480..263B}
reproduced in his book ``{\it Speakable and unspeakable in quantum
  mechanics}''~\cite{1987suqm.book.....B}. This letter was written
after the invention of inflation but before the quantum state of
cosmological perturbations was fully characterized by Grishchuk and
Sidorov~\cite{Grishchuk:1990bj}. It discusses important ideas related
to the classical limit of Quantum Mechanics. We study Bell's paper but
also the criticisms that have been put forward against
it~\cite{Johansen:1997wz}. There was indeed a long and controversial
discussion about the validity of the results obtained by Bell. What
was not realized before, however, is that the domains most applicable
to the ideas developed by Bell in his article are Cosmology and the
scenario of cosmic inflation.

The article is organized as follows. In the next section,
Sec.~\ref{sec:quantfluctu}, we first review the theory of inflationary
cosmological perturbations, at the classical level in
Sec.~\ref{subsec:classicalpert} and, then, in
Sec.~\ref{subsec:quantumpert} at the quantum level. In
Sec.~\ref{subsec:quantumstate}, we study in more detail the quantum
state in which inflationary perturbations are placed, namely a
two-mode squeezed state. In Sec.~\ref{subsec:interpretation}, we
present some simple considerations that allow us to intuitively
understand what a squeezed state is and, in
Sec.~\ref{subsec:gaussian}, we show that this type of states in fact
belong to a larger class of quantum states known as Gaussian
states. In Sec.~\ref{sec:transition}, we study the
quantum-to-classical transition of cosmological perturbations. In
Sec.~\ref{subsec:stocha}, we investigate if the fluctuations can be
described by a classical stochastic process. In
Sec.~\ref{subsec:discord}, we use tools borrowed from quantum
information theory, namely the quantum discord, to address the
question of the classicality of cosmological perturbations. In
Sec.~\ref{sec:wigner}, we come back again to the question of the
classical limit using Bell ideas. In Sec.~\ref{subsec:critere}, we
explain why the non-positivity of the Wigner function can be taken as
a criterion for the existence of genuine quantum effects. In
Sec.~\ref{subsec:wkb}, we discuss this idea in the context of the
Wentzel-Kramer-Brillouin (WKB) approximation. In
Sec.~\ref{subsec:bellwigner}, we review in detail the paper by John
Bell, mentioned earlier, and show that it is especially relevant for
our purposes. In Sec.~\ref{subsec:wrongbell}, we present the
criticisms that have been made on Bell's letter and, in
Sec.~\ref{subsec:criticsbellwrong}, we comment on these criticisms. In
Sec.~\ref{subsec:correct}, in the light of the previous
considerations, we conclude about the status of Bell's
letter. Finally, in Sec.~\ref{subsec:revzen}, we explain how the whole
situation has been clarified by the publication of a theorem due to
Revzen. In Sec.~\ref{sec:biqv}, based on the previous considerations,
we study whether the Bell inequality can be constructed for CMB
observables and we briefly discuss our results in
Sec.~\ref{sec:discussion}. Finally, in Sec.~\ref{sec:conclusion}, we
briefly present our conclusions.

\section{Inflationary Cosmological Perturbations}
\label{sec:quantfluctu}

\subsection{Classical Perturbations}
\label{subsec:classicalpert}

On large scales, the Universe is homogeneous and isotropic (the
so-called cosmological principle) and is well-described by the
Friedman-Lema\^{\i}tre-Robertson-Walker (FLRW) metric,
${\rm d}s^2=a^2(\eta)(-{\rm d}\eta^2+\delta_{ij}{\rm d}x^i{\rm
  d}x^j)$,
where $\eta$ is the conformal time and $a(\eta)$ the scale factor. The
scale factor describes how the Universe expands. We now have an
accurate picture of the behavior of $a(\eta)$ from epochs possibly
characterized by an energy scale as high as $\sim 10^{15}$ GeV to
present times. This cosmic history constitutes the standard model of
Cosmology, known as the $\Lambda$CDM model~\cite{PeterUzan2009}. This
model is a six parameter model and correctly accounts for all known
cosmological observations. The earliest epoch of this $\Lambda$CDM
model, namely the one which describes the very early Universe, is
known as inflation. It is a phase of accelerated expansion and it is
believed that it was driven by a scalar field, the ``inflaton'', the
physical nature of which is still unknown.

However, in order to understand the large scale structures in our
Universe, such as clusters of galaxies or CMB anisotropies, it is
clearly necessary to go beyond the previous description, namely beyond
the cosmological principle. A crucial remark is that, in the early
Universe, the deviations from homogeneity and isotropy are small
(recall that $\delta T/T\sim 10^{-5}$ on the last scattering
surface). As a consequence, one can use perturbative methods: as a
matter of fact, linear perturbations theory will be
sufficient. Therefore, we perturb the FLRW metric tensor introduced
before and write~\cite{Mukhanov:1990me}
$g_{\mu \nu}=g_{\mu \nu}^{_{\rm FLRW}}(\eta)+\delta g_{\mu
  \nu}(\eta,{\bm x})+\cdots $,
where $g_{\mu \nu}^{_{\rm FLRW}}(\eta)$ is the metric tensor
introduced before which only depends on time since it describes a
homogeneous and isotropic Universe. The perturbed part is
$\delta g_{\mu \nu}(\eta,{\bm x})$, which is supposed to be small
compared to $g_{\mu \nu}^{_{\rm FLRW}}(\eta)$, and which is time, but
also space dependent. It represents small ripples on top of an
expanding Universe, the expansion itself being described by the scale
factor $a(\eta)$. Then, exactly in the same way as a vector can be
decomposed into a curl-free and a divergence-free component (the
Helmhotz theorem that can be found in any textbook on
electromagnetism), a two rank tensor can be decomposed into a scalar,
vector and tensor part, a result known as the Stewart
lemma~\cite{Mukhanov:1990me}. If one restricts ourselves to scalar
perturbations (tensor modes, or primordial gravity waves, can be
treated in a similar fashion and vector modes are absent during
inflation), then the perturbed metric can be written as
\begin{align} 
\label{eq:metric}
\dd s^2=a^2\left(\eta\right)\Bigl\lbrace-\left(1-2\phi\right)
\dd\eta^2+2\left(\partial_iB\right)\dd x^i\dd \eta
+\left[\left(1-2\psi\right)\delta_{ij}+2\partial_i
\partial_jE\right]\dd x^i\dd x^j\Bigr\rbrace .
\end{align}
The above perturbed metric depends on four functions because we need
four functions to write the components of the perturbed metric in
terms of scalar functions only (for instance, as can be seen in the
above equation, the time space component of the metric has been
written in terms of the scalar $B$ since $\delta g_{0i}=\partial_iB$).
Obviously, these four functions are time and space dependent. However,
this description is redundant because of gauge
freedom~\cite{Mukhanov:1990me}. This means that there are
infinitesimal changes of coordinates that can mimic perturbative
solutions. These fictitious solutions must be removed and this is
accomplished in the gauge invariant formalism. It consists in working
with quantities that are invariant under infinitesimal changes of
coordinates. For instance, the gravitational sector can be described
by a single quantity, the so-called Bardeen potential defined by
$\Phi_{_{\rm B}}\left(\eta,\bm{x}\right)
=\phi+\left[a\left(B-E^\prime\right)\right]^\prime/a$,
a prime denoting derivative with respect to conformal time. The
changes in the functions $\phi$, $B$ and $E$ caused by a small
diffeomorphim exactly compensate if the above combination of $\phi$,
$B$ and $E$ is considered, which is the essence of what a
gauge-invariant quantity is. In the same way, the perturbations of
matter can be described by a single quantity. For instance, if one
studies the perturbations during inflation, then this single quantity
is the gauge invariant fluctuation of the inflaton scalar field
$\delta
\varphi^{\left(\mathrm{gi}\right)}\left(\eta,\bm{x}\right)=\delta\varphi
+\varphi^\prime\left(B-E^\prime\right)$,
where the superscript ``gi'' stands for gauge-invariant. Moreover,
since the two above mentioned quantities are related through perturbed
Einstein equations, it is in fact the whole scalar sector that can be
reduced to the study of a single quantity that can be chosen to be
curvature perturbations, usually denoted $\zeta(\eta,{\bm x})$, and
defined by
\begin{align}
\zeta(\eta,{\bm x})=\frac{1}{\Mp\sqrt{2\epsilon_1}}
\left(\delta \varphi^{\left(\mathrm{gi}\right)}
+\frac{\varphi'}{{\cal H}}\Phi_{_{\rm B}}\right),
\end{align}
where ${\cal H}=a'/a$ and $\epsilon_1$ is the first Hubble-flow
function given by $\epsilon_1=1-{\cal H}'/{\cal H}^2$.
$\zeta(\eta,{\bm x})$ is directly related to the perturbed
three-dimensional curvature scalar, hence its name. 

\begin{figure}
\begin{center}
\includegraphics[width=10cm]{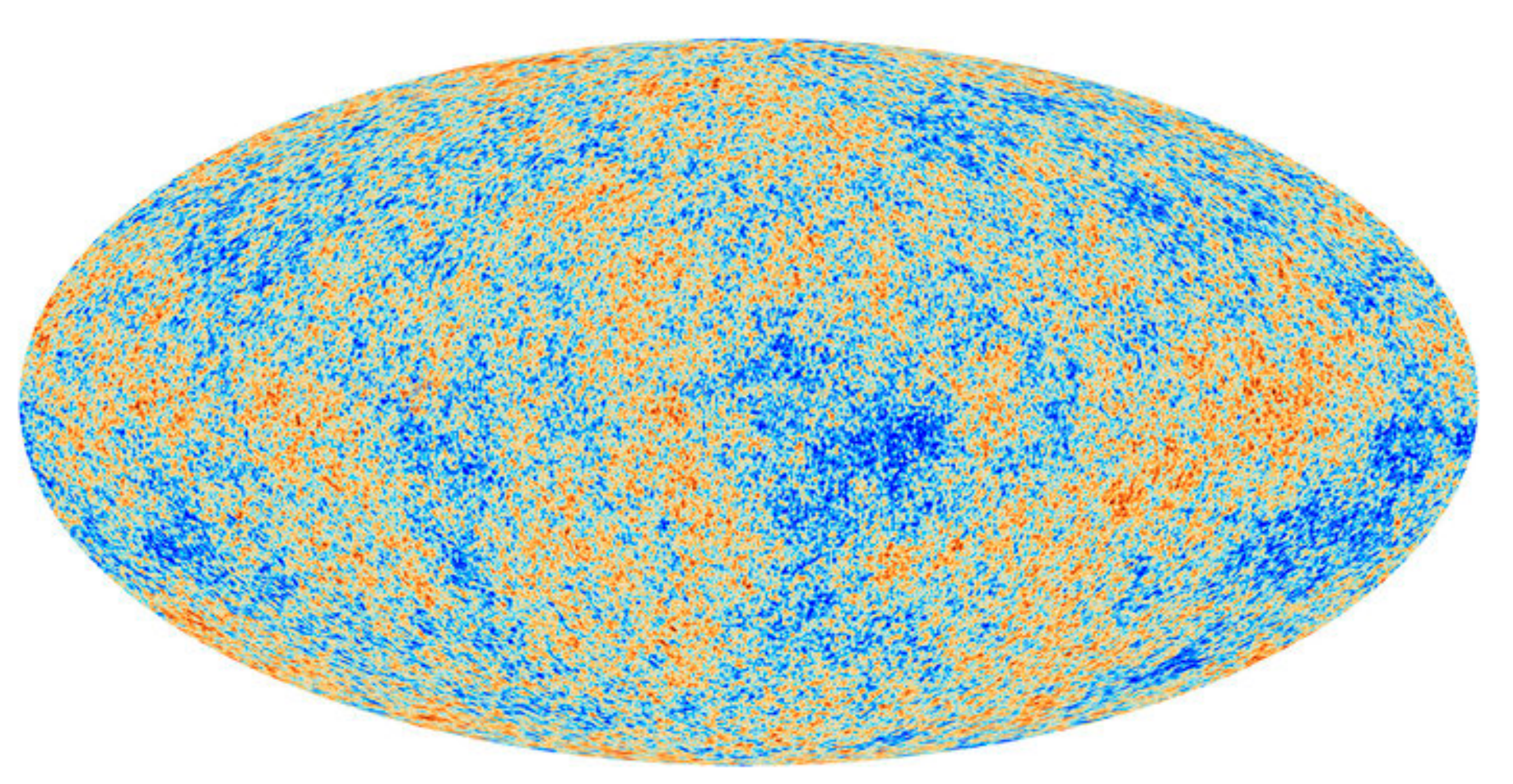}
\end{center}
\caption{Map of the CMB anisotropies obtained by the European Space
  Agency (ESA) Planck
  satellite~\cite{Ade:2015xua,Ade:2015lrj,Ade:2015ava,Aghanim:2018eyx,Akrami:2018odb}. It
  represents the most exquisite measurement of the CMB ever done.}
\label{fig:planck}
\end{figure}

Physically, $\zeta(\eta,{\bm x})$ is a very relevant quantity because,
in the real world, it can be measured (and has been measured). Indeed,
the temperature anisotropy,
\begin{align}
\frac{\delta T}{T}(\theta , \phi)
=\sum _{\ell =2}^{+\infty}\sum _{m=-\ell}^{m=\ell}a_{\ell m}Y_{\ell m}
(\theta,\phi),
\end{align}
where $\theta$ and $\phi$ defines a direction in the sky and
$Y_{\ell m}$ are spherical harmonics, is an observable and has now
been measured by many different experiments. The first one was the
COBE satellite in $1992$~\cite{Bennett:2003bz}. The most recent and
most accurate observation is by The European Space Agency (ESA) Planck
satellite~\cite{Ade:2015xua,Ade:2015lrj,Ade:2015ava,Aghanim:2018eyx,Akrami:2018odb},
see Fig.~\ref{fig:planck}. The so-called Sachs-Wolfe
effect~\cite{Sachs:1967er} relates the presence of small
inhomogeneities, living in three-dimensional space and described by
curvature perturbations $\zeta(\eta,{\bm x})$ to the temperature
anisotropy of Fig.~\ref{fig:planck}, namely
\begin{align}
\label{eq:sw}
\frac{\delta T}{T}({\bm e})
=& \int \frac{{\rm d}{\bm k}}{(2\pi)^{3/2}}
\left[F({\bm k})+i{\bm k}\cdot {\bm e}\, G({\bm k})\right]
\zeta_{\bm k}(\eta_{\rm end})
e^{-i {\bm k}\cdot 
{\bm e}(\eta_{\rm lss}-\eta_0)
+i{\bm k}\cdot {\bm x}_{0}}\, ,
\end{align}
where ${\bm e}$ is a unit vector in the direction of the observation
on the celestial sphere defined by the angles $\theta$ and $\phi$ and
$\zeta_{\bm k}(\eta)$ is the Fourier transform of
$\zeta(\eta,{\bm x})$, namely
$\zeta(\eta,{\bm x})=(2\pi)^{-3/2} \int {\rm d}{\bm k}\zeta_{\bm
  k}(\eta)e^{i{\bm k}\cdot {\bm x}}$.
Notice that, because $\zeta(\eta,{\bm x})$ is real, one has
$\zeta_{\bm k}^*=\zeta_{-{\bm k}}$. The quantities $\eta_{\rm lss}$
and $\eta_0$ are the last scattering surface (lss) and present day
($0$) conformal times, respectively, while ${\bm x}_0$ represents
Earth's location. The last scattering surface is the surface at which
the CMB was emitted. It corresponds to the time at which our Universe
became transparent and is located at a redshift of
$z_{\rm lss}\simeq 1100$.  The functions $F({\bm k})$ and $G({\bm k})$
are the so-called form factors and describe the evolution of the
perturbation in the post-inflationary universe. They are not easy to
calculate as they dependent on the evolution of many different fluids
in interaction. This evolution is usually tracked
numerically~\cite{Seljak:1996is}. But it only depends on known physics
and, therefore, can be subtracted away. Therefore, we see that
temperature fluctuations are in fact directly given by
$\zeta_{\bm k}(\eta_{\rm end})$ evaluated at the end of inflation.

The behavior of curvature perturbations is controlled by the perturbed
Einstein equations, $\delta G_{\mu \nu}=\Mp^{-2}\delta T_{\mu \nu}$.
By definition, these equations are linear partial differential
equations. But they can be transformed into an infinite number of
linear ordinary differential equations by going to Fourier space. One
can then show~\cite{Mukhanov:1990me} that curvature perturbation obeys
\begin{align}
\label{eq:eomzeta}
\left(z\zeta_{\bm k}\right)''
+\left(k^2-\frac{z''}{z}\right)\left(z\zeta_{\bm k}\right)=0,
\end{align}
where we have defined $z\equiv a\Mp\sqrt{2\epsilon_1}$. One recognizes
the equation of motion of an oscillator whose fundamental frequency,
$\omega^2=k^2-z''/z$, is time-dependent. In other words, we deal with
a parametric oscillator: a classical analogy would a pendulum the
length of which can change in time. Here the time dependence is fixed
by $z$, which is only determined by the dynamics of the expansion
since $z$ depends on the scale factor and its derivatives. The
solution to the above equations are easily analyzed. In an
inflationary Universe, the Hubble radius $H^{-1}$ is constant while
the wavelength of a given Fourier mode, which grows proportional to
the scale factor, is stretched beyond the Hubble radius. Therefore,
initially, $k^2\gg z''/z$ (small scales limit) and the quantity
$z\zeta_{\bm k}$ oscillates
\begin{align}
\label{eq:smallscalezeta}
z\zeta_{\bm k}=A_{\bm k}e^{ik\eta}+B_{\bm k}e^{-ik\eta},
\end{align}
where $A_{\bm k}$ and $B_{\bm k}$ are two arbitrary integration
constants. The reason for this behavior is that the wavelength of the
mode is so small that it does not feel the curvature of spacetime and
behaves as if it lived in flat spacetime. In principle, the two
constants $A_{\bm k}$ and $B_{\bm k}$ are fixed by the initial
conditions. At the classical level, it is just unclear what these ones
should be. Then, as time goes on, the mode exits the Hubble radius and
the regime $k^2\ll z''/z$ (large scales limit) becomes relevant. In
that case, the solution can be written
\begin{align}
\label{eq:largescaleszeta}
z\zeta_{\bm k}=C_{\bm k}z+D_{\bm k}z\int ^{\eta}\frac{{\rm d}\tau}{z^2(\tau)},
\end{align}
where $C_{\bm k}$ and $D_{\bm k}$ are two constants of
integrations. The first branch, proportional to $C_{\bm k}$, is the
growing mode and the second one the decaying mode. This can easily be
verified if, for instance, one considers scale factors of the form
$a(\eta)\propto (-\eta)^{1+\beta}$, recalling that inflation
corresponds to $\beta\simeq -2$ and that the conformal time during
inflation is negative and tends to zero (by negative values) as
inflation proceeds. The above solution shows that the growing mode is,
on large scales, constant, namely $\zeta_{\bm k}\simeq C_{\bm k}$,
which means that the curvature perturbation has the advantage (among
others) to be conserved on large scales.

Usually, the properties of CMB anisotropies are characterized by the
correlation functions of $\delta T({\bm e})/T$ which are, thanks to
Eq.~(\ref{eq:sw}), directly related to the correlation functions of
curvature perturbation at the end of inflation. The two-point
correlation function of $\zeta(\eta,{\bm x})$ reads
\begin{align}
\label{eq:classical2pt}
\left \langle \zeta(\eta,{\bm x})\zeta(\eta,{\bm x}+{\bm r})
\right \rangle &=
\int_0^{+\infty} \frac{{\rm d}k}{k} \frac{\sin(kr)}{kr}
\frac{k^3}{2\pi^2}\left \vert 
\zeta_{\bm k}(\eta_{\rm end})\right \vert ^2 ,
\end{align}
where the brackets are supposed to represent an average over some
classical distribution such that
$\langle \zeta_{\bm k}\zeta_{{\bm p}}^*\rangle =\vert \zeta_{\bm
  k}\vert^2\delta({\bm k}-{\bm p})$.
At the end of inflation,
$\zeta_{\bm k}(\eta_{\rm end})\simeq C_{\bm k}$ since, as explained
before, the decaying mode can be neglected. But one needs to specify
the scale dependence of $C_{\bm k}$. This can be done by matching the
large scale regime to the small scale regime which, in practice,
amounts to express $C_{\bm k}$ in terms of $A_{\bm k}$ and
$B_{\bm k}$. The problem is thus moved to determining the scale
dependence of the coefficients $A_{\bm k}$ and $B_{\bm k}$. At the
classical level, as mentioned before, there is just no clear approach
of how this can be done in a well-justified and well-motivated way.

\subsection{Quantum Perturbations}
\label{subsec:quantumpert}

The above considerations, therefore, leave one important question
unanswered: what is the origin of these perturbations? The beauty of
inflation is that it can also provide an answer to this important
question: inflation says that the primordial perturbations originate
from the vacuum quantum fluctuations of the inflaton and gravitational
fields at the beginning of inflation. This means that all structures
in our Universe are nothing but quantum fluctuations stretched over
cosmological distances by the expansion of the Universe and amplified
by gravitational instability.

At the technical level, this means that, now, the perturbed metric
$\delta g_{\mu \nu}$ is viewed as a quantum operator,
$\delta \hat{g}_{\mu \nu}$, satisfying the quantum perturbed Einstein
equations, viewed as equations for quantum operators
$\delta \hat{G}_{\mu \nu}=\Mp^{-2}\delta \hat{T}_{\mu \nu}$. Even more
concretely, in this formulation, curvature perturbation can be viewed
as a (test) quantum scalar field living in the expanding spacetime and
can be written as
\begin{align}
  z(\eta)\hat{\zeta}(\eta,{\bm x})=\frac{1}{(2\pi)^{3/2}}
  \int \frac{{\rm d}{\bm k}}{\sqrt{2k}}\left[\hat{c}_{\bm k}(\eta)
  e^{i{\bm k}\cdot {\bm x}}+\hat{c}_{\bm k}^{\dagger}(\eta)
  e^{-i{\bm k}\cdot {\bm x}}\right],
\end{align}
where $\hat{c}_{\bm k}(\eta)$ and $\hat{c}_{\bm k}^{\dagger}(\eta)$
are the annihilation and creation operators satisfying the usual equal
time commutation relations,
$[\hat{c}_{\bm k}(\eta),\hat{c}_{\bm p}^{\dagger}(\eta)]=\delta({\bm
  k}-{\bm p})$.
Curvature perturbations are then related to the creation and
annihilation operators through
\begin{align}
\label{eq:zetacrea}
z(\eta)\hat{\zeta}_{\bm k}=\frac{1}{\sqrt{2k}}\left(\hat{c}_{\bm k}
+\hat{c}_{-{\bm k}}^{\dagger}\right), \quad 
z(\eta)\hat{\zeta}_{\bm k}'=-i\sqrt{\frac{k}{2}}\left(\hat{c}_{\bm k}
-\hat{c}_{-{\bm k}}^{\dagger}\right).
\end{align}
We notice that $\hat{\zeta}_{\bm k}$ and $\hat{\zeta}_{\bm k}'$ mix creation 
and annihilation operators of momentum ${\bm k}$ and $-{\bm k}$.

The evolution of $\hat{\zeta}(\eta,{\bm x})$ is controlled by the
following Hamiltonian
\begin{align}
\label{eq:hamil}
  \hat{H}=\int _{\setR^3}{\rm d}^3{\bm k}
  \left[\frac{k}{2}\left(\hat{c}_{\bm k}\hat{c}_{\bm k}^{\dagger}
  +\hat{c}_{-{\bm k}}^{\dagger}\hat{c}_{-{\bm k}}\right)
  -\frac{i}{2}\frac{z'}{z}\left(\hat{c}_{\bm k}\hat{c}_{-{\bm k}}
  -\hat{c}_{-{\bm k}}^{\dagger}c_{\bm k}^{\dagger}\right)\right].
\end{align}
This Hamiltonian comes from a second order expansion in $\zeta$ of the
action of GR plus a scalar field (since inflation is driven by this
type of field). This action (and, therefore, the corresponding
Hamiltonian) remains quadratic in $\zeta$ and higher order terms are
ignored because the perturbations are small. As already mentioned,
this is well established at the time of recombination where the
deviations are measured to be of the order $10^{-5}$. Since the
fluctuations grow by gravitational instability, there were certainly
even smaller during inflation.

The Hamiltonian~(\ref{eq:hamil}) is made of two pieces. The first part
describes the free Hamiltonian of a collection of harmonic oscillators
with fundamental frequency $\omega=k$ (as appropriate for massless
excitations). The second piece describes the interaction of the
quantum field $\hat{\zeta}(\eta,{\bm x})$ with the classical
background characterized by the scale factor $a(\eta)$. If space-time
is static (namely Minkowski space-time) then $a'=0$ and the
``time-dependent'' coupling constant $z'/z$ vanishes. This term is
responsible for particle creation. Moreover, the momentum structure of
this second piece in $\hat{c}_{\bm k}\hat{c}_{-{\bm k}}$ or
$\hat{c}_{-{\bm k}}^{\dagger}c_{\bm k}^{\dagger}$ indicates that
particles are created by pairs with opposite momenta (in accordance
with momentum conservation). At this point, one should clarify the
following. In quantum field theory, quadratic action (or Hamiltonian)
usually describes free fields while interactions are described as
higher order terms. There is, however, one exception, namely the case
where a quantum field interacts with a classical source. It is still
described by a quadratic action, the presence of the interaction
manifesting itself only by giving a time dependence to the effective
frequency of the field oscillators. In other words, a free field is
equivalent to a collection of harmonic oscillators while a field in
interaction with a classical source is equivalent to a collection of
parametric oscillators. The classic example is the Schwinger effect,
where a fermionic field interacts with a classical electric
field~\cite{Schwinger:1951nm,Martin:2007bw}. The case of a scalar
field in a cosmological background is another example.

The Heisenberg equation,
$i{\rm d}\hat{c}_{\bm k}/{\rm d}\eta=[\hat{c}_{\bm k},\hat{H}]$,
allows us to calculate the equation of motion of the operator
$\hat{\zeta}_{\bm k}(\eta)$. This leads to
\begin{align}
\label{eq:ceta}
\frac{{\rm d}\hat{c}_{\bm k}}{{\rm d}\eta}=-ik\hat{c}_{\bm k}
+\frac{z'}{z}\hat{c}_{-{\bm k}}^{\dagger},
\end{align}
from which one deduces that
\begin{align}
\left(z\hat{\zeta}_{\bm k}\right)''
+\left(k^2-\frac{z''}{z}\right)\left(z\hat{\zeta}_{\bm k}\right)=0,
\end{align}
that is to say Eq.~(\ref{eq:eomzeta}) but written at the operator
level.

A fundamental assumption of inflation is that the system starts in the
vacuum state $\vert 0\rangle$ defined by the condition
$\hat{c}_{\bm k}(\eta_\uini)\vert 0\rangle =0$. In order to see what
this implies for the field $\hat{\zeta}_{\bm k}(\eta)$, let us first
solve the time dependence of the creation and annihilation operators,
see Eq.~(\ref{eq:ceta}). This can be done by means of a Bogoliubov
transformation, namely
\begin{align}
\hat{c}_{\bm k}(\eta)=u_k(\eta)\hat{c}_{\bm k}(\eta_\uini)
+v_k(\eta)\hat{c}_{-{\bm k}}^{\dagger}(\eta_\uini),
\end{align}
where the functions $u_k(\eta)$ and $v_k(\eta)$ obey
\begin{align}
\label{eqs:uandv}
i\frac{{\rm d}u_k}{{\rm d}\eta} &= ku_k+i\frac{z'}{z}v_k^*, 
\quad
i\frac{{\rm d}v_k}{{\rm d}\eta} = kv_k+i\frac{z'}{z}u_k^*,
\end{align}
and, by definition, have initial conditions $u_k(\eta_\uini)=1$ and
$v_k(\eta_\uini)=0$. Let us notice that $u_k$ and $v_k$ depend on the
modulus of the wavenumber ${\bm k}$ only. The Bogoliubov
transformation allows us to re-express the field expansion as
\begin{align}
\label{eq:bogoexpansion}
  z(\eta)\hat{\zeta}(\eta,{\bm x})=\frac{1}{(2\pi)^{3/2}}
  \int \frac{{\rm d}{\bm k}}{\sqrt{2k}}
\left[\left(u_k+v_k^*\right)\hat{c}_{\bm k}(\eta_\uini)
  e^{i{\bm k}\cdot {\bm x}}+\left(u_k^*+v_k\right)\hat{c}_{\bm k}^{\dagger}(\eta_\uini)
  e^{-i{\bm k}\cdot {\bm x}}\right].
\end{align}
It is easy to verify from Eqs.~(\ref{eqs:uandv}) that the function
$u_k+v_k^*$ obeys the equation
$(u_k+v_k^*)''+(k^2-z''/z)(u_k+v_k^*)=0$, namely the same equation as
$z\zeta_{\bm k}$. Recalling the initial conditions for $u_k$ and $v_k$,
this implies that the mode function in~(\ref{eq:bogoexpansion}),
$(u_k+v_k^*)/\sqrt{2k}$, behaves as
\begin{align}
\lim _{k\eta\rightarrow -\infty}\frac{u_k+v_k^*}{\sqrt{2k}}
=\frac{1}{\sqrt{2k}}e^{ik(\eta-\eta_\uini)},  
\end{align}
in the small scales limit. In other words, the assumption that the
fluctuations are quantum and start in the vacuum state has completely
fixed the initial conditions. As a consequence, in
Eq.~(\ref{eq:smallscalezeta}), one should choose
\begin{align}
A_{\bm k}=\frac{1}{\sqrt{2k}}e^{-ik\eta_\uini}, \quad 
B_{\bm k}=0.
\end{align}
Then, the initial conditions being now known, the calculation of the
power spectrum can be performed explicitly. Indeed, we no longer face
the issues discussed after Eq.~(\ref{eq:classical2pt}): $C_{\bm k}$
can be related to $A_{\bm k}$ and $B_{\bm k}$ but, as just explained,
these ones are now fully determined. The calculation leads to an
almost scale invariant power spectrum,
${\cal P}_{\zeta}(k)=k^3\vert \zeta_{\bm k}(\eta_{\rm
  end})\vert^2/(2\pi^2)\simeq A_{_\mathrm{S}}k^{n_{_\mathrm{S}}-1}$,
where $n_{_{\rm S}}\simeq 1$ plus small corrections that depend on the
model of inflation considered. Scale invariance means that, if
$n_{_{\rm S}}=1$, then ${\cal P}_{\zeta}(k)$ no longer depends on
$k$. Since, according to inflation, $n_{_{\rm S}}\simeq 1$ but
$n_{_{\rm S}}\neq 1$, we have in fact almost scale
invariance. Moreover, measuring the small deviations from scale
invariance allows us to constrain inflation since the corrections, as
just mentioned above, depend on the scenario of inflation. The fact
that $n_{_{\rm S}}\simeq 1$ crucially rests on the choice
$A_{\bm k}\propto 1/\sqrt{2k}$. Had we have another scale dependence
initially for $A_{\bm k}$ and $B_{\bm k}$, the power spectrum would
have been completely different and, generically, far from scale
invariance. Remarkably, according to the most recent data obtained by
the Planck satellite~\cite{Aghanim:2018eyx,Akrami:2018odb}, everything
is precisely consistent with a power spectrum of the form
${\cal P}_{\zeta}(k)=A_{_\mathrm{S}}k^{n_{_\mathrm{S}}-1}$, with
$\ln (10^{10}A_{_\mathrm{S}})=3.044\pm 0.014$ and
$n_{_\mathrm{S}}=0.9649\pm 0.0042$. The value of the spectral index
$n_{_\mathrm{S}}$ is not predicted by inflation (more precisely, if a
model of inflation is given, then it is predicted. But the problem is
that the correct model of inflation is not known). It was known for a
long time that $n_{_\mathrm{S}}$ should be around one but it is only
very recently that Planck demonstrated that $n_{_\mathrm{S}}\neq 1$ at
a significant statistical level (namely more that $5\sigma$).

The Planck results are, therefore, one of the main reasons to trust
inflation and its mechanism of structures formation according to which
structures in the Universe are nothing but quantum fluctuations. This
fascinating conclusion is now well supported by astrophysical data.

\subsection{The Quantum State of Inflationary Perturbations}
\label{subsec:quantumstate}

Before discussing the properties of the state of cosmological
perturbations, let us make the following remark. We saw in
Eqs.~(\ref{eq:zetacrea}) that the definition of $\hat{\zeta}_{\bm k}$
mixes creation and annihilation operators of mode ${\bm k}$ and
$-{\bm k}$. This is because, as already mentioned, particles are
created by pair with opposite momenta. But this is uncommon from a
quantum information point of view where one likes to view the total
system as a collection of subsystems associated to the different
modes. In other words, if ${\cal E}$ is the Hilbert space of the full
system and ${\cal E}_{\bm k}$ the Hilbert space associated to the mode
${\bm k}$, we would like to have
${\cal E}=\otimes_{\bm k}{\cal E}_{\bm k}$. For this reason, we now
introduce, for a fixed mode ${\bm k}$, the ``position''
$\hat{q}_{\bm k}$ and the ``momentum'' $\hat{\pi}_{\bm k}$ defined by
\begin{align}
\label{eq:defqpi}
\hat{q}_{\bm k}=\frac{1}{\sqrt{2k}}\left(\hat{c}_{\bm k}
+\hat{c}_{\bm k}^{\dagger}\right), \quad 
\hat{\pi}_{\bm k}=-i\sqrt{\frac{k}{2}}\left(\hat{c}_{\bm k}
-\hat{c}_{\bm k}^{\dagger}\right).
\end{align}
These two operators are Hermitian. The relation between
$\hat{\zeta}_{\bm k}$ and the position and momentum operators can be
easily derived and reads
\begin{align}
\label{eq:zetaq}
z(\eta)\hat{\zeta}_{\bm k}=\frac12\left[\hat{q}_{\bm k}
+\hat{q}_{-{\bm k}}+\frac{i}{k}\left(\hat{\pi}_{\bm k}
-\hat{\pi}_{-{\bm k}}\right)\right], \quad 
z(\eta)\hat{\zeta}_{\bm k}'=\frac{1}{2i}\left[k\left(\hat{q}_{\bm k}
-\hat{q}_{-{\bm k}}\right)+i\left(\hat{\pi}_{\bm k}
+\hat{\pi}_{-{\bm k}}\right)\right].
\end{align}
In the following, we consider a description of the system based on
these operators since we want to make use of the formalism of quantum
information. In fact, the above definitions allow us to describe
cosmological perturbations as a continuous variable system. A
continuous variable system is a system that is described by Hermitian
operators satisfying canonical commutation relation,
$[\hat{q}_{\bm k},\hat{\pi}_{{\bm k}'}]=i\delta({\bm k}'-{\bm k})$.
The number of degrees of freedom is infinite and labeled by the
wavenumber ${\bm k}$. 

Another idea from Quantum Information Theory that will be playing an
important role in the following is that of bipartite system. Indeed,
for higher than one dimensional system, the set of degrees of freedom
can be split into two subsets. This defines a partition and allows us
to see the whole system as a bipartite system. Then, one can study the
nature of the correlations between the two subsystems which is a way
to assess the ``quantumness'' of the whole system. An important point
is that one can define several partitions for the same system. As an
introductory example, let us consider a four dimensional system with
degrees of freedom $\hat{q}_i$ and vector
$\hat{r}=(\hat{q}_i,\hat{\pi}_i)^{_{\mathrm T}}$, $i=1, \cdots, 4$. We
split the system into two subsystems, $A$ and $B$, and defines a
partition such that ${\cal E}={\cal E}_A\otimes {\cal E}_B$. For
instance, we can choose the subsystem $A$ to contain $\hat{q}_1$ and
$\hat{q}_2$ and, therefore, to be described by
$\hat{r}_A=(\hat{q}_1,\hat{q}_2,\hat{\pi}_1,\hat{\pi}_2)^{_{\rm T}}$
while the system $B$ contains $\hat{q}_3$ and $\hat{q}_4$ and is
described by
$\hat{r}_B=(\hat{q}_3,\hat{q}_4,\hat{\pi}_3,\hat{\pi}_4)^{_{\rm T}}$.
Then, we define the vector $\hat{R}$ by
$\hat{R}=(\hat{r}_A,\hat{r}_B)^{_{\mathrm T}}$. This definition of
$\hat{R}$, namely the way we list its components, is, implicitly, a
definition of a partition. In Cosmology, because of pair creations,
the partition
${\cal E}=\otimes _{{\bm k}\in \setR^{3+}}{\cal E}_{\bm k}\otimes
{\cal E}_{-{\bm k}}$
is, at least at first sight, very natural and, in the following, we
work with it.

After these preliminary remarks, let us come back to the quantum state
of the perturbations. As mentioned earlier, in Cosmology, one starts
from the vacuum state
$\otimes_{\bm k}\vert 0_{\bm k},0_{-{\bm k}}\rangle$ at time
$\eta=\eta_\uini$, that is to say
$\hat{c}_{\bm k}(\eta_\uini)\vert 0\rangle=0$ and
$\hat{c}_{-{\bm k}}(\eta_\uini)\vert 0\rangle =0$. Then, by solving
the Schr\"odinger equation, with the Hamiltonian given by
Eq.~(\ref{eq:hamil}), one can show that this state evolves into a
two-mode squeezed state given by
\begin{align}
\label{eq:quantumstate}
\vert \Psi_{2 {\rm sq}}\rangle=\bigotimes_{\bm k} \frac{1}{\cosh r_k}
\sum_{n=0}^{\infty}e^{-2in \varphi_k}
\tanh ^nr_k \vert n_{\bm k},n_{-{\bm k}}\rangle,
\end{align}
where $\vert n_{\bm k}\rangle $ is an eigenvector of the particle
number operator in the mode ${\bm k}$. $r_k$ and $\varphi_k$ are the
squeezing parameter and squeezing angle, respectively. They are time
dependent functions controlled by the following equations
\begin{align}
\frac{{\rm d}r_k}{{\rm d}\eta}=\frac{z'}{z}\cos(2\varphi_k), \quad
\frac{{\rm d}\varphi_k}{{\rm d}\eta}=-k-\frac{z'}{z}
\coth(2r_k)\sin(2\varphi_k).
\end{align}
The fact that we encounter a squeezed state should not come as a
surprise. It is indeed well known that, while the quantization of an
harmonic oscillator naturally leads to coherent states, the
quantization of a parametric oscillator leads to squeezed
states. Since we saw before that, because of its interaction with the
dynamical scale factor, the field $\hat{\zeta}_{\bm k}$ acquires a
time-dependent fundamental frequency, and, therefore, can be viewed as
parametric oscillator, the whole picture appears to be consistent.

\subsection{Physical Interpretation}
\label{subsec:interpretation}

Let us now come back to the quantum state~(\ref{eq:quantumstate}) and
try to gain physical intuition about it. The corresponding
wavefunction can be expressed as
\begin{align}
\label{eq:squeezewf}
\Psi_{2\rm{sq}}\left(q_{\bm k},q_{-{\bm k}}\right)
=\left\langle q_{\bm k},q_{-{\bm k}}\left \vert 
\frac{1}{\cosh r_k}
\sum_{n=0}^{\infty}e^{-2in \varphi_k}
\tanh ^nr_k \right \vert n_{\bm k},n_{-{\bm k}}\right \rangle
=\frac{e^{A(r_k,\varphi_k)(q_{\bm k}^2+q_{-{\bm k}}^2)-B(r_k,\varphi_k)q_{\bm k}q_{-{\bm k}}}}
{\cosh r_k\sqrt{\pi}\sqrt{1-e^{-4i\varphi_k}\tanh ^2r_k}},
\end{align}
where the functions $A(r_k,\varphi_k)$ and $B(r_k,\varphi_k)$ are
given by
\begin{align}
A(r_k,\varphi_k)=\frac{e^{-4i\varphi_k}\tanh^2 r_k+1}{2(e^{-4i\varphi_k}
\tanh^2 r_k-1)}, \quad B(r_k,\varphi_k)=\frac{2e^{-2i\varphi_k}\tanh r_k}
{e^{-4i\varphi_k}\tanh ^2r_k-1}.
\end{align}

This explicit form for the wave function allows us to understand what,
physically, a two-mode squeezed state
means~\cite{Lvovsky:2014sxa}. Before treating the case of a two-mode
squeezed state however, let us recall some well-known facts. Let us
consider one mode ${\bm k}$. The corresponding vacuum state is a
coherent state, that is to say, a state with wavefunction in position
space given by
\begin{equation}
\Psi_0\left(q_{\bm k}\right)=\frac{1}{\pi^{1/4}}e^{-(q_{\bm k})^2/2},
\end{equation}
which is nothing but the wavefunction for the ground state of the
harmonic oscillator in non-relativistic quantum mechanics. The same
state, written in momentum basis, can be expressed as
$\tilde{\Psi}_0\left(\pi_{\bm
    k}\right)=\frac{1}{\pi^{1/4}}e^{-(\pi_{\bm k})^2/2}$.
From the knowledge of the wave function, one can calculate the
dispersion in field amplitude and momentum. This gives
$\langle \Delta \hat{q}_{\bm k}^2\rangle =\langle \Delta
\hat{\pi}_{\bm k}^2\rangle=1/2$,
which saturates the Heisenberg inequality, namely
$\langle \Delta \hat{q}_{\bm k}^2\rangle \langle \Delta \hat{\pi}_{\bm
  k}^2\rangle=1/4$.

Let us now consider a one-mode squeezed state. Its wave function, in
field amplitude and conjugate momentum basis, can be written as
\begin{equation}
\Psi_R\left(q_{\bm k}\right)=\frac{\sqrt{R}}{\pi^{1/4}}e^{-R^2(q_{\bm k})^2/2},
\quad
\tilde{\Psi}_R\left(\pi_{\bm k}\right)=\frac{1}{\pi^{1/4}\sqrt{R}}
e^{-(\pi_{\bm k})^2/(2R^2)},
\end{equation}
where we have introduced a new parameter, $R$. The physical
interpretation of this parameter can be found by calculating again the
dispersion in position and momentum. One finds
$ \langle \Delta \hat{q}_{\bm k}^2\rangle =1/(2R^2)$ and
$\langle \Delta \hat{\pi}_{\bm k}^2\rangle=R^2/2$. Although they still
saturate the Heisenberg inequality, the two dispersion's are no longer
equal. If $R>1$, then the dispersion in field amplitude is smaller
than that the dispersion in conjugate momentum (and, interestingly
enough, smaller than that of the vacuum state). We say that the state
is squeezed in position or field amplitude, hence its name. Of course,
since one has to satisfy the Heisenberg inequality, the dispersion in
momentum is larger. If $R<1$, we have the opposite situation and the
state is squeezed in momentum.

After these preliminary comments, we now come to the two-mode squeezed
state. As the name of the state indicates, we must now consider two
modes and, of course, we choose ${\bm k}$ and $-{\bm k}$. Following
the tradition in quantum information theory, we can also call mode
${\bm k}$ ``Alice'' and mode $-{\bm k}$ ``Bob''. In field amplitude
basis, the vacuum state of this bipartite system can be written as
\begin{equation}
\Psi_{0}\left(q_{\bm k},q_{-{\bm k}}\right)=\frac{1}{\sqrt{\pi}}
e^{-(q_{\bm k})^2/2-(q_{-{\bm k}})^2/2}
=\frac{1}{\sqrt{\pi}}e^{-(q_{\bm k}-q_{-{\bm k}})^2/4}e^{-(q_{\bm k}+q_{-{\bm k}})^2/4}.
\end{equation}
We see that the position of Alice and Bob are uncorrelated. Then, in a
way which is exactly similar to what has already been done above, we
introduce the following state
\begin{equation}
\label{eq:twomode}
\Psi_{R}\left(q_{\bm k},q_{-{\bm k}}\right)=
\frac{1}{\sqrt{\pi}}e^{-R^2(q_{\bm k}-q_{-{\bm k}})^2/4}
e^{-(q_{\bm k}+q_{-{\bm k}})^2/(4R^2)}
=\frac{1}{\sqrt{\pi}}e^{\overline{A}(R)(q_{\bm k}^2+q_{-{\bm k}}^2)
-\overline{B}(R)q_{\bm k}q_{-{\bm k}}},
\end{equation}
where the squeezing factor $R$ appears again and where the two
functions $\overline{A}(R)$ and $\overline{B}(R)$ are defined by
\begin{align}
\overline{A}(R)\equiv -\frac14\left(R^2+\frac{1}{R^2}\right), \quad 
\overline{B}(R)\equiv -\frac12\left(R^2-\frac{1}{R^2}\right).
\end{align}
The state~(\ref{eq:twomode}) is, by definition, a two-mode squeezed
state. Clearly, since Eqs.~(\ref{eq:twomode}) and~(\ref{eq:squeezewf})
are similar, this means that the state~(\ref{eq:squeezewf}) is also a
two-mode squeezed state. In Eq.~(\ref{eq:twomode}), we have ignored
the squeezed angle and, therefore, we should identify the function
$\overline{A}(R)$ with $A(r_k,0)=-(e^{2r_k}+e^{-2r_k})/4$ which
immediately leads to $r_k=\ln R$. One checks that this is consistent
since $B(r_k,0)=-(e^{2r_k}-e^{-2r_k})/2$ is indeed equals to
$\overline{B}(R)$ and the normalization factor
$\cosh r_k\sqrt{1-e^{-4i\varphi_k}\tanh^2 r_k}$ goes to one when the
squeezing angle is zero. We notice that the position of Alice and Bob
are now correlated and that these correlations are genuinely quantum
since the state~(\ref{eq:twomode}) is an entangled state, namely
$\Psi_{R}\left(q_{\bm k},q_{-{\bm k}}\right)\neq \Psi_{R}\left(q_{\bm
    k}\right) \Psi_{R}\left(q_{-{\bm k}}\right)$.
This means that the state~(\ref{eq:squeezewf}) implies the existence
of genuine quantum correlations between the field amplitudes
$q_{\bm k}$ and $q_{-{\bm k}}$. It is also interesting to remark that
the two-mode squeezed state does not lead to squeezing for Alice or
Bob. Indeed, it is easy to verify that
$\langle \Delta \hat{q}_{\bm k}^2\rangle =\langle \Delta
\hat{q}_{-{\bm k}}^2\rangle=(1+R^4)/(4R^2)$.
These dispersions are always larger than the dispersions obtained for
the vacuum state. This is related to the fact that, if one traces out,
say, Alice's degree of freedom, one is left with a state for Bob that
is not a one-mode squeezed state but a thermal state.

The two-mode squeezed state that is present in Cosmology is quite
peculiar: it is probably the strongest squeezed state ever produced in
Nature. The squeezing factor is often expressed in decibel with the
help of the following definition, see also Eqs.~(14) and~(15) of
Ref.~\cite{Schnabel:2016gdi}
\begin{align}
-10\log_{10}\left(e^{-2r}\right)\, {\rm dB}=\frac{20r}{\ln 10}\, {\rm dB}.
\end{align}
In Cosmology, one can achieve $r\simeq 50$ which means a squeezing of
$\simeq 43\, {\rm dB}$ to be compared with $\sim 15\, {\rm dB}$ which
is the world record in the lab, see Refs.~\cite{2016PhRvL.117k0801V}
and~\cite{PhysRevLett.117.110801}.

\subsection{Gaussian States}
\label{subsec:gaussian}

Another interesting property of a two mode squeezed state is that it
belongs to a wider class of quantum states called ``Gaussian
states''. We indeed check that the wavefunction~(\ref{eq:squeezewf})
is Gaussian. Gaussian states play a fundamental role in quantum
mechanics. They arise in many different branches of Physics such as
Laser Physics, Quantum Field Theory (in curved spacetime or not),
Solid State Physics or Cosmology and they are ubiquitous in Quantum
Information Theory. Gaussian states naturally occur as ground
(coherent or squeezed) or thermal equilibrium states of any physical
quantum system described by a quadratic Hamiltonian. Moreover, with
existing technologies, they are easily manipulable in the lab.

At the technical level, a Gaussian state is a state the characteristic
function of which is a Gaussian. The characteristic function
$\chi(\xi)$ is defined by
\begin{align}
\chi(\xi)={\rm Tr}\left[\hat{\rho}\hat{\cal W}(\xi)\right],
\end{align}
where $\hat{\rho}$ is the density matrix of the quantum state and
where $\hat{\cal W}(\xi)$ is the Weyl operator which can be expressed
as
\begin{align}
\hat{\cal W}(\xi)=e^{i\xi^{_\mathrm{T}}\hat{R}},
\end{align}
where the vector $\hat{R}$ has already been introduced before and is
defined by
$\hat{R}=(k^{1/2}\hat{q}_{\bm k},k^{-1/2}\hat{\pi}_{\bm
  k},k^{1/2}q_{-{\bm k}}, k^{-1/2}\hat{\pi}_{-{\bm k}})^{_\mathrm{T}}$
(here, we have slightly modified the definition by introducing a
$k^{1/2}$ in front of positions and a $k^{-1/2}$ in front of momentum
in order to work with dimensionless quantities). For a two-mode
squeezed state, the characteristic function is indeed a Gaussian
(justifying the fact that it belongs to the class of Gaussian states)
\begin{align}
\label{eq:charagauss}
\chi(\xi_1,\xi_2,\xi_3,\xi_4)=e^{-\xi^{_\mathrm{T}}\gamma \xi/4},
\end{align}
where $\gamma $ is the covariance matrix is given
\begin{align}
\label{eq:covmatrix}
\gamma = 
\begin{pmatrix}
\cosh(2r_k) & 0 & \sinh(2r_k)\cos(2\varphi_k) & \sinh(2r_k)\sin(2\varphi_k) \\
0 & \cosh(2r_k) & \sinh(2r_k)\sin(2\varphi_k) & -\sinh(2r_k)\cos(2\varphi_k)\\
\sinh(2r_k)\cos(2\varphi_k) & \sinh(2r_k)\sin(2\varphi_k) & \cosh(2r_k)
& 0 \\
\sinh(2r_k)\sin(2\varphi_k) & -\sinh(2r_k)\cos(2\varphi_k) & 0 &
\cosh(2r_k)
\end{pmatrix}
.
\end{align}
The covariance matrix is related to the two point correlation function of 
the position and/or momentum operators since $\langle \hat{R}_i\hat{R}_j
\rangle= \gamma_{ij}/2+iJ_{ij}/2$ where the matrix $J$ is defined by
\begin{align}
\label{eq:defJ}
J=
\begin{pmatrix}
0 & 1& 0 &0 \\
-1 & 0 & 0 &0 \\
0 & 0 & 0 & 1 \\
0 & 0 &-1 & 0
\end{pmatrix}
.
\end{align}

Another, equivalent, way to define a Gaussian state is the following:
a Gaussian state is a state which has a Gaussian Wigner function. For
a state with density matrix $\hat{\rho}$, the Wigner function is
defined by
\begin{align}
W(R)=\frac{1}{(2\pi)^2}\int {\rm d}x\, {\rm d}y\, 
e^{-i\pi_{\bm k}x-i\pi_{-{\bm k}}y}
\left\langle q_{\bm k}+\frac{x}{2}
,q_{-{\bm k}}+\frac{y}{2}
\biggl\vert \, \hat{\rho}
\, \biggr\vert q_{\bm k}-\frac{x}{2},q_{-{\bm k}}-\frac{y}{2}\right\rangle.
\end{align}
Physically, the Wigner function is the quantum generalization of the
classical distribution in phase space. It is related to the
characteristic function introduced before by the formula
\begin{align}
W(\xi)=\frac{1}{(2\pi)^4}\int {\rm d}^4\eta \, e^{-i\xi^{_\mathrm{T}}\eta}
\chi(\eta).
\end{align}
Using this result and the characteristic function of a Gaussian state,
see Eq.~(\ref{eq:charagauss}), it is easy to demonstrate that
\begin{align}
W(\xi)=\frac{1}{\pi^2\sqrt{\det \gamma}}e^{-\xi^{_\mathrm{T}}\gamma^{-1}\xi}.
\end{align}
This shows that the Wigner function is also a Gaussian. If one uses
the expression~(\ref{eq:covmatrix}) of the covariance matrix, then the
explicit expression of the Wigner function of a two-mode squeezed
state reads
\begin{align}
W&=\frac{1}{\pi^2}
\exp\biggl[-\left(kq_{\bm k}^2+kq_{-{\bm k}}^2+\frac{\pi_{\bm k}^2}{k}
+\frac{\pi_{-{\bm k}}^2}{k}\right)\cosh(2r_k)
+2\left(q_{\bm k}\pi_{\bm k}+q_{-{\bm k}}\pi_{-{\bm k}}\right)
\sin(2\varphi_k)\sinh(2r_k)
\nonumber \\ &
+2\left(kq_{\bm k}q_{-{\bm k}}-\frac{\pi_{\bm k}\pi_{-{\bm k}}}{k}\right)
\cos(2\varphi_k)\sinh(2r_k)\biggr].
\end{align}

\section{The Quantum-to-Classical Transition of the Cosmological
  Perturbations}
\label{sec:transition}

From the above considerations, why curvature perturbations are viewed
as genuinely quantum should now be clear. However, when CMB
anisotropies are analyzed by astronomers, curvature perturbations are
treated classically without any reference to their quantum origin. Is
this just wrong or can we justify this approach by claiming that some
sort of quantum-to-classical transition took place in the early
Universe?  This question is reminiscent of the question of classical
limit in Quantum Mechanics. It is known that this problem is subtle
and we will argue that, in the context of Cosmology, it is even more
subtle than in ordinary situations.

\subsection{Stochastic Description?}
\label{subsec:stocha}

The fact that a two-mode squeezed state is Gaussian implies, as
discussed before, that its Wigner function is positive definite. In
fact, one can show that the only states for which this is the case are
precisely the Gaussian states~\cite{HUDSON1974249}. This leads to the
idea that the Wigner function could be used as a classical
distribution over phase space. If so, this would mean that there
exists a classical, stochastic, description of the properties of the
system. This would indicate that a quantum-to-classical transition has
taken place. At the technical level, the previous argument can be
formulated as follows. Let us consider a function $O$ of position and
momentum, namely
$O(q_{\bm k},\pi_{\bm k},q_{-{\bm k}},\pi_{-{\bm k}})$.  According to
the above considerations, one would define the classical average of
$O$ as
\begin{align}
\label{eq:meanstocha}
\langle O\rangle_{\rm stocha}=\int O(R)W(R){\rm d}^4R.
\end{align}
Using the correspondence principle, one can define
the corresponding operator $\hat{O}$ as
$\hat{O}=O(\hat{q}_{\bm k},\hat{\pi}_{\bm k},\hat{q}_{-{\bm
    k}},\hat{\pi}_{-{\bm k}})$.
If the system is classical, then one must have
\begin{align}
\langle \hat{O}\rangle =\left \langle 
O\right \rangle_{\rm stocha}.
\end{align}
We now establish under which conditions the above equation holds. Let
us first define the Weyl transform of the operator $\hat{O}$ by
\begin{align}
\label{eq:defweyltrans}
\tilde{O}(q_{\bm k},\pi_{\bm k},q_{-{\bm k}},\pi_{-{\bm k}})
=
\int {\rm d}x\, {\rm d}y\,
e^{-i\pi_{\bm k}x-i\pi_{-{\bm k}}y}
\left\langle q_{\bm k}+\frac{x}{2}
,q_{-{\bm k}}+\frac{y}{2}
\biggl\vert \, \hat{O}
\, \biggr\vert q_{\bm k}-\frac{x}{2},q_{-{\bm k}}-\frac{y}{2}\right\rangle.
\end{align}
It is of course reminiscent of the Wigner function. In fact, up to a
factor $(2\pi)^{-2}$ the Wigner function is the Weyl transform of the
density matrix, namely $\tilde{\rho}=(2\pi)^2W$. The Weyl transform
associates a function in phase space to any operator in Hilbert
space. The fundamental property of the Weyl transform is that, for two
operators $\hat{A}$ and $\hat{B}$, one has
\begin{align}
\label{eq:trab}
{\rm Tr}\left(\hat{A}\hat{B}\right)=\frac{1}{(2\pi)^2}
\int \tilde{A}(R)\tilde{B}(R)
{\rm d}^4R,
\end{align}
where the seemingly awkward factor $(2\pi)^2$ comes from the fact that
the subspace we consider here is four-dimensional. It follows that
\begin{align}
\label{eq:meano}
\langle \hat{O}\rangle ={\rm Tr}(\hat{\rho}\hat{O})=\frac{1}{(2\pi)^2}
\int \tilde{\rho}(R)
\tilde{O}(R) {\rm d}^4R=\int 
\tilde{O}(R) W(R){\rm d}^4R.
\end{align}
Comparing the above formula to Eq.~(\ref{eq:meanstocha}), we see that quantum 
and stochastic averages coincide if 
\begin{align}
\label{eq:proper}
O(R)=\tilde{O}(R).
\end{align}
Therefore, whether or not a quantum-to-classical transition takes
place can be summarized to the above equation and to whether it holds
in general. For instance, it is easy to show that it is always valid
for any power of the operator $\hat{q}_{\bm k}^m$, namely
$\widetilde{q_{\bm k}^m}=q_{\bm k}^m$. In the same fashion, one also
has $\widetilde{\pi_{\bm k}^m}=\pi_{\bm k}^m$. However,
$\widetilde{q_{\bm k}\pi_{\bm k}}=q_{\bm k}\pi_{\bm k}+i/2$. For quadratic 
combinations, one can summarize the previous results as 
\begin{align}
\label{eq:weylquadra}
\widetilde{R_jR_k}=R_jR_k+\frac12\left[\hat{R}_j,\hat{R}_k\right]
=R_jR_k+\frac{i}{2}J_{jk},
\end{align}
where the matrix $J$ has been defined in
Eq.~(\ref{eq:defJ}). Moreover, any combination of operators of mode
${\bm k}$ and mode $-{\bm k}$ has a trivial Weyl transform, for
instance
$\widetilde{q_{\bm k}^{m_1}\pi_{-{\bm k}}^{m_2}}=q_{\bm
  k}^{m_1}\pi_{-{\bm k}}^{m_2}$.
Using Eqs.~(\ref{eq:zetaq}), one can now calculate
$\widetilde{\zeta_{\bm k}^2}$ in order to see whether a stochastic
calculation of the curvature perturbations power spectrum is
possible. The previous results imply that
\begin{align}
z^2(\eta)\widetilde{\zeta_{\bm k}^2}&=\frac{1}{4}
\biggl[\widetilde{q_{\bm k}^2}
+\widetilde{q_{-{\bm k}}^2}
+\widetilde{q_{\bm k}q_{-{\bm k}}}
+\widetilde{q_{-{\bm k}}q_{{\bm k}}}
+\frac{i}{k}\left(\widetilde{q_{\bm k}\pi_{\bm k}}
+\widetilde{\pi_{\bm k}q_{\bm k}}\right)
-\frac{i}{k}\left(\widetilde{q_{-{\bm k}}\pi_{-{\bm k}}}
+\widetilde{\pi_{-{\bm k}}q_{-{\bm k}}}\right)
\nonumber \\ &
+\frac{i}{k}\left(\widetilde{q_{-{\bm k}}\pi_{\bm k}}
-\widetilde{q_{\bm k}\pi_{-{\bm k}}}
+\widetilde{\pi_{\bm k}q_{-{\bm k}}}
-\widetilde{\pi_{-{\bm k}}q_{\bm k}}\right)
-\frac{1}{k^2}\left(
\widetilde{\pi_{\bm k}^2}
+\widetilde{\pi_{-{\bm k}}^2}
-\widetilde{\pi_{\bm k}\pi_{-{\bm k}}}
-\widetilde{\pi_{-{\bm k}}\pi_{{\bm k}}}\right)
\biggr].
\end{align}
Among all the terms in the above expression, only the last four ones
on the first line have a non trivial Weyl transform. However, if
$\widetilde{q_{\bm k}\pi_{\bm k}}$ and
$\widetilde{\pi_{\bm k}q_{\bm k}}$ have, separately, a non-trivial
Weyl transform, the sum of these two terms actually has a trivial Weyl
transform because the additional factor $i/2$ originating from
Eq.~(\ref{eq:weylquadra}) cancel out. Therefore, we conclude that
$\widetilde{\zeta_{\bm k}^2}=\zeta_{\bm k}^2$. As a consequence, the
quantum two-point correlation function of curvature perturbations,
$\langle \hat{\zeta}(\eta,{\bm x})\hat{\zeta}(\eta,{\bm y})\rangle$
can be exactly reproduced in a classical, stochastic, approach. One
can also show that this is the case for
$\langle \hat{\zeta}(\eta,{\bm x}) \hat{\zeta}'(\eta,{\bm
  y})+\hat{\zeta}'(\eta,{\bm x})\hat{\zeta}(\eta,{\bm y})\rangle$
or
$\langle \hat{\zeta}'(\eta,{\bm x})\hat{\zeta}'(\eta,{\bm y})\rangle$.
Notice that this is true whatever the state of the system. Of course,
in order for Eq.~(\ref{eq:meanstocha}) to be not only mathematically
correct but also physically meaningful, the distribution $W$ has to
be positive definite and this is the case only for Gaussian
states. The identification of the quantum and stochastic correlators
being valid for any states, it is obviously valid for any Gaussian
states, in particular for any values of the squeezing parameter and
angle. It is sometimes claimed that this identification is possible
only in the limit $r_k\rightarrow +\infty$ and we see that this is not
just the case.

However, for higher order correlators, the story is different. Rather
than a long demonstration, let us take an example. One can show that
$\widetilde{q_{\bm k}^2\pi_{\bm k}^2}=q_{\bm k}^2\pi_{\bm
  k}^2+2iq_{\bm k}\pi_{\bm k}-1/2$
and
$\widetilde{\pi_{\bm k}^2q_{\bm k}^2}=\pi_{\bm k}^2q_{\bm
  k}^2-2i\pi_{\bm k}q_{\bm k}-1/2$.
This implies that
$\widetilde{q_{\bm k}^2\pi_{\bm k}^2+\pi_{\bm k}^2q_{\bm k}^2}=q_{\bm
  k}^2\pi_{\bm k}^2+\pi_{\bm k}^2q_{\bm k}^2-1$.
We see that this particular correlator has a non trivial Weyl
transform. However, if one uses Eqs.~(\ref{eq:zetaq}) to calculate the
higher order correlation function of curvature perturbations, one
finds
\begin{align}
\widetilde{\zeta_{\bm k}^3}=\zeta_{\bm k}^3, \quad \widetilde{\zeta_{\bm k}^4}
=\zeta_{\bm k}^4.
\end{align}
Some combinations participating to $\widetilde{\zeta_{\bm k}^3}$ and
$\widetilde{\zeta_{\bm k}^4}$ have a non trivial Weyl transform but
these extra contributions all cancel out to produce the above
result. Again, this result is obtained without the help of the large
squeezing limit. In fact, one can even show that it is valid for any
power of $\zeta_{\bm k}$, namely
$\widetilde{\zeta_{\bm k}^n}=\zeta_{\bm k}^n$, see Eq.~(13.5) in
Ref.~\cite{hall2013quantum}.

\subsection{Quantum Discord and Inflation}
\label{subsec:discord}

The previous results seem to indicate that a system described by a
two-mode squeezed state is classical since the correlation functions
of curvature perturbations can also be obtained by mean of a
stochastic process. Even quantities such as
$q_{\bm k}^2\pi_{\bm k}^2+\pi_{\bm k}^2q_{\bm k}^2$ become identical
to their Weyl transform in the large squeezing limit. So there is
indeed a quantum-to-classical transition and this would explain why
the astronomers can treat the perturbations classically. However, we
have already noticed that a two-mode squeezed state is also an
entangled state which, on the contrary, is usually viewed as the
prototype of a non-classical state. Moreover, in the limit
$r\rightarrow +\infty$, one has
$\Psi_R(q_{\bm k},q_{-{\bm k}})\propto \delta(q_{\bm k}-q_{-{\bm
    k}})$,
which is an Einstein Podolsky Rosen (EPR) quantum
state~\cite{Einstein:1935rr}, also considered as ``the'' state that
can be used to illustrate the non-intuitive features of Quantum
Mechanics.

In addition, more recently, in the context of Quantum Information
Theory, a quantitative measure of the ``quantumness'' of a system has
been introduced. This measure is called the ``quantum discord'', see
Refs.~\cite{Ollivier:2001fdq} and~\cite{2001JPhA...34.6899H}. Very
briefly, the main idea is the following. In order to decide whether a
system is quantum or classical, one can divide it into two parts (a
``bipartite'' system as already discussed before) and look whether the
correlations between the two subsystems can be understood classically
or not. In the cosmological context, as already discussed as well, the
two subsystems can be taken to be the mode ${\bm k}$ and the mode
$-{\bm k}$. Let us first discuss the idea classically. For this
purpose, let us consider two random variables $a$ and $b$ having a
joint probability distribution $p(a_i,b_j)$ (the indices $i$ and $j$
label the possible realizations; of course, it can be a continuous
index if we deal with continuous random variables). Each random
variable has a distribution that can be obtained from the joint
distribution by marginalization, namely $p(a_i)=\sum_jp(a_i,b_j)$ and
$p(b_j)=\sum _ip(a_i,b_j)$. Then, the mutual information is given by
\begin{align}
\label{eq:mutualinf}
{\cal I}(a,b)\equiv S[p(a_i)]+S[p(b_j)]-S[p(a_i,b_j)],
\end{align}
where $S[p(a_i)]\equiv -\sum_ip(a_i)\ln [p(a_i)]$ is the Shannon
entropy. As is well-known, the entropy quantifies the {\it
  uncertainty} of the possible outcomes $a_i$ endowed with probability
distribution $p(a_i)$. For instance, if there are only two possible
outcomes, $a_1$ and $a_2$, with probability $p$ and $1-p$, then
$S[p(a_i)]=-p\ln p-(1-p)\ln(1-p)$. If $p=0$ or $p=1$, then
$S[p(a_i)]=0$. The first case corresponds to a situation where the
probability of having $a_1$ vanishes and the probability of having
$a_2$ is one. The second case corresponds to the opposite
situation. Clearly, in both cases, the outcomes are certain and,
therefore, the uncertainty is zero. The uncertainty is maximal if
$p=1/2$ which is also very intuitive.

Coming back to Eq.~(\ref{eq:mutualinf}), this quantity is a measure of
the correlations between the two subsystems since, when they are
independent, the joint distribution factorizes,
$p(a_i,b_j)=p(a_i)p(b_j)$ and ${\cal I}(a,b)=0$. We can also view it
as the ``distance'' between two distributions also known as the
Kullback and Leibler divergence~\cite{kullback1951}. If they are
one-dimensional, then the distance between two distributions $p(a_i)$
and $p(b_j)$ is defined by
$S[p(a_i)\vert \vert \, p(b_j)]\equiv \sum_ip(a_i)\ln[p(a_i)/p(b_i)]$;
if we deal with two-dimensional distributions, $p(a_i,b_j)$ and
$p(c_k,d_\ell)$, then it is
$S[p(a_i,b_j)\vert \vert \, p(c_k,d_\ell)]\equiv \sum
_{ij}p(a_i,b_j)\ln[p(a_i,b_j)/p(c_i,d_j)]$
and so on. Let us remark, however, that this is not a real distance
since it is not symmetric. It is easy to show that the mutual
information discussed above is nothing but the distance between the
joint distribution and the product of the two marginalized
distributions, namely\footnote{The calculation proceeds as follows:
\begin{align}
\sum
_{ij}p(a_i,b_j)\ln\left[\frac{p(a_i,b_j)}{p(a_i)p(b_j)}\right]
&=\sum _{ij}p(a_i,b_j)\ln[p(a_i,b_j)]
-\sum _{ij} p(a_i,b_j)\ln[p(a_i)]
-\sum _{ij} p(a_i,b_i)\ln [p(b_j)]
\\
&=\sum _{ij}p(a_i,b_j)\ln[p(a_i,b_j)]
-\sum _i \ln [p(a_i)]\sum _j p(a_i,b_j)
-\sum _j \ln [p(b_j)]\sum _i p(a_i,b_j)
\\
&=\sum _{ij}p(a_i,b_j)\ln[p(a_i,b_j)]
-\sum _i \ln [p(a_i)]p(a_i)
-\sum _j \ln [p(b_j)]p(b_j)
={\cal I}(a,b)
\end{align}}
 \begin{align}
{\cal I}(a,b)=S[p(a_i,b_j)\vert \vert \, p(a_k)p(b_\ell)].
\end{align}
When the two random variables are independent the distance between
$p(a_i,b_j)$ and $p(a_i)p(b_j)$ vanishes.

One can also discuss the above concepts in another way. Let
$p(b_j\vert a_i)$ be the probability to observe $b_j$ given that
$a_i$ has been observed. Then the uncertainty associated with the
outcomes $b_j$ is defined by
$-\sum_j p(b_j\vert a_i)\ln [p(b_j\vert a_i)]$. Of course, this
quantity depends on the measured quantity $a_i$. In order to have the 
total uncertainty one can average the previous quantity using the 
distribution $p(a_i)$. This leads us to define the conditional entropy by
\begin{align}
  S(b\vert a)\equiv -\sum_i p(a_i)\sum_j p(b_j\vert a_i)\ln [p(b_j\vert a_i)].
\end{align}
Then, if one uses the Bayes theorem,
$p(a_i,b_j)=p(b_j\vert a_i)p(a_i)$, one can show that
\begin{align}
\label{eq:mutualinf2}
{\cal I}(a,b)=S[p(b_j)]-S(b\vert a)=S[p(a_i)]-S(a\vert b).
\end{align}

Let us now discuss the previous considerations again but in Quantum
Mechanics. First of all, let us recall that the entropy $S$ of a
system characterized by the state $\hat{\rho}$ is defined by
$S=-{\rm Tr}(\hat{\rho}\log_2\hat{\rho})$. The interpretation of this
quantity is the same as before. Then we can define the
quantum-mechanical mutual information by the following expression
\begin{align}
{\cal I}({\bm k},-{\bm k})=
S[\hat{\rho}({\bm k})]
+S[\hat{\rho}(-{\bm k})]
-S[\hat{\rho}({\bm k},-{\bm k})],
\end{align}
where the density matrices $\hat{\rho}({\bm k})$ and
$\hat{\rho}(-{\bm k})$ are obtained from
$\hat{\rho}({\bm k},-{\bm k})$ by tracing out the degrees of freedom
associated with $-{\bm k}$ and ${\bm k}$, respectively. The non
trivial part comes from the quantum-mechanical generalization of the
expression of ${\cal I}$ expressed in terms of the conditional
entropy, see Eq.~(\ref{eq:mutualinf2}). This expression is based on
conditional probabilities which deal with the concept of observing an
outcome given that another outcome has been observed or measured. It
is well known that the concept of measurement is subtle in Quantum
Mechanics and, in some sense, highly ``non-classical''. So let us
suppose that we perform a measurement on the system $-{\bm
  k}$.
Measurements in Quantum Mechanics are represented by projectors and we
note $\hat{\Pi}_j$ the projector associated to the measurement of the
system $-{\bm k}$ (it is therefore an operator living in the Hilbert
space associated to the subsystem $-{\bm k}$). After the measurement,
the state of the system is
$\hat{\rho}({\bm k},-{\bm k})\hat{\Pi}_j/p_j$ with probability
$p_j={\rm Tr}[\hat{\rho}({\bm k},-{\bm k})\hat{\Pi}_j]$. If we only
have access to the system ${\bm k}$, we trace out degrees of freedom
associated with the system $-{\bm k}$ and we arrive at
$\hat{\rho}({\bm k},\hat{\Pi}_j)={\rm Tr}_{-{\bm k}}[\hat{\rho}({\bm
  k},-{\bm k})\hat{\Pi}_j/p_j]$.
This allows us to calculate probabilities for outcomes associated with
the system ${\bm k}$ given that a measurement has been performed on
the sub system $-{\bm k}$. In this sense, this is the equivalent of
the classical $p(b_j\vert a_i)$. Based on Eq.~(\ref{eq:mutualinf2}),
we define by analogy
\begin{align}
{\cal J}({\bm k},-{\bm k})=S[\hat{\rho}({\bm k})]-\sum_j p_j 
S[\hat{\rho}({\bm k},\hat{\Pi}_j)].
\end{align}
In Quantum Mechanics, contrary to the case where classical probability
calculus applies, ${\cal I}({\bm k},-{\bm k})$ and
${\cal J}({\bm k},-{\bm k})$ need not to coincide. In fact, we use
this difference as a signature of the fact that the system is not
classical. This leads us to define the ``quantum discord'' by
\begin{align}
\delta ({\bm k},-{\bm k})=\min_{\hat{\Pi}_j}
\left[{\cal I}({\bm k},-{\bm k})-{\cal J}({\bm k},-{\bm k})\right],
\end{align}
where a minimization over the set of all possible projectors is
performed in order for the discord to be independent of the choice of
a particular $\hat{\Pi}_j$.

\begin{figure}
\begin{center}
\includegraphics[width=10cm]{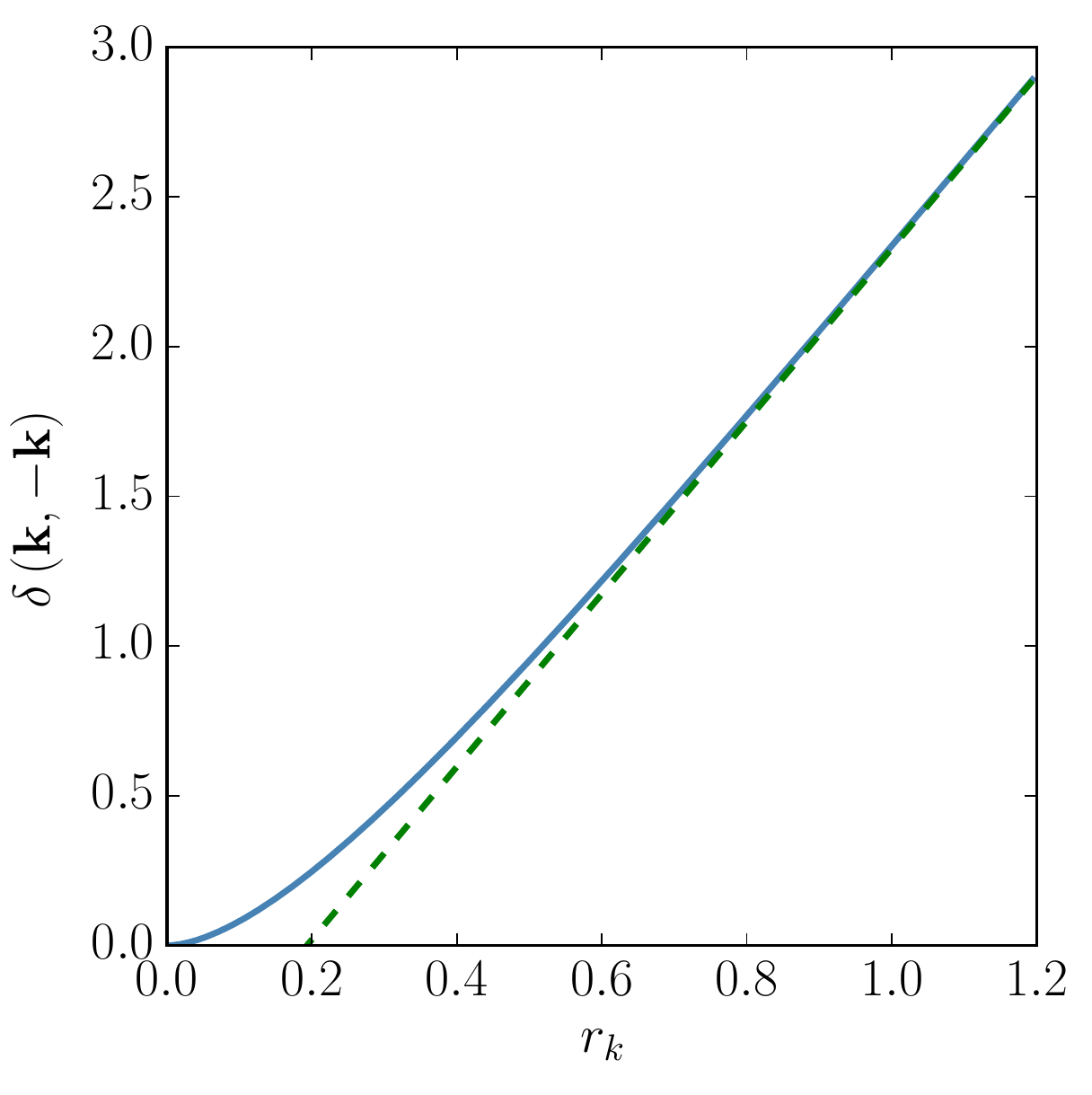}
\end{center}
\caption{Quantum discord of a two-mode squeezed state as a function of
  the squeezing parameter $r_k$ (solid blue line). The dashed green
  line represents the large $r_k$ expansion of
  $\delta({\bm k},-{\bm k})$. Figure taken from
  Ref.~\cite{Martin:2015qta}.}
\label{fig:discord}
\end{figure}

Having defined what the quantum discord is, one can now calculate it
for a two-mode squeezed state. Straightforward manipulations lead
to~\cite{Martin:2015qta}
\begin{align}
\delta({\bm k},-{\bm k})=\cosh^2 r_k\log_2\left(\cosh ^2r_k\right)
-\sinh^2 r_k\log_2\left(\sinh ^2r_k\right).
\end{align}
Let us notice that the discord does not depend on the squeezing angle.
The discord is represented in Fig.~\ref{fig:discord}. For a vanishing
squeezing parameter, the discord is zero and then grows with
$r_k$. For large values of $r_k$, it simply grows linearly since
$\delta({\bm k},-{\bm k})=2r_k/\ln 2-2-1/\ln 2+{\cal O}(e^{-2r_k})$.
Therefore, one concludes that a two-mode squeezed state is not a
classical state at all, at least if one accepts the quantum discord as
a meaningful criterion for classicality.

\section{EPR, EPW and Cosmology}
\label{sec:wigner}

\subsection{Negativity of the Wigner Distribution 
as a Criterion of Non-Classicality}
\label{subsec:critere}

We have reached a seemingly paradoxical stage. On one hand, the
considerations in Sec.~\ref{subsec:stocha} seem to indicate that the
system is classical. Everything can indeed be described by means of a
stochastic distribution, namely the Wigner function. This one is
positive definite because we deal with a Gaussian state and the Weyl
transform of any power of curvature perturbation is ``trivial'', which
indicates that any quantum correlation function can be obtained as a
stochastic correlation function using the Wigner distribution. On the
other, a two-mode squeezed state tends to an EPR state and, moreover,
modern criterion designed in the context of Quantum Information
Theory, such as the quantum discord calculated in the last section,
Sec.~\ref{subsec:discord}, unambiguously shows that the system is
quantum. How can we understand something which, at first sight, looks
like a contradiction? It should be added that, although interesting in
general, this question is especially relevant in Cosmology which, as
argued before, is the situation in Physics where the strongest
squeezing and, therefore, the largest discord, are obtained.

In fact this question has a fascinating history although it was
realized only very recently that Cosmology is probably the most
interesting context to discuss it. To our knowledge, it started in
$1986$ with the letter ``{\it EPR correlations and EPW
  distributions}''~\cite{1986NYASA.480..263B} that J.~Bell presented
in a conference organized by the New York Academy of Sciences and
which is also reproduced in his famous book ``{\it Speakable and
  unspeakable in quantum mechanics}'' (this is chapter
$21$)~\cite{1987suqm.book.....B}. The letters ``EPW'' in the title
stand for ``Eugene Paul Wigner'' since Bell's letter is dedicated to
Professor E.~P.~Wigner. Amusingly enough, the inflationary mechanism
for structure formation was invented only a few years before and, soon
after Bell's letter, in $1990$, Grishchuk and
Sidorov~\cite{Grishchuk:1990bj} realized for the first time that a
two-mode squeezed state was involved in the inflationary
scenario. Remarkably, the Grishchuk and Sidorov paper contains a
calculation of the Wigner function of cosmological perturbations.

\begin{figure}
\begin{center}
\includegraphics[width=10cm]{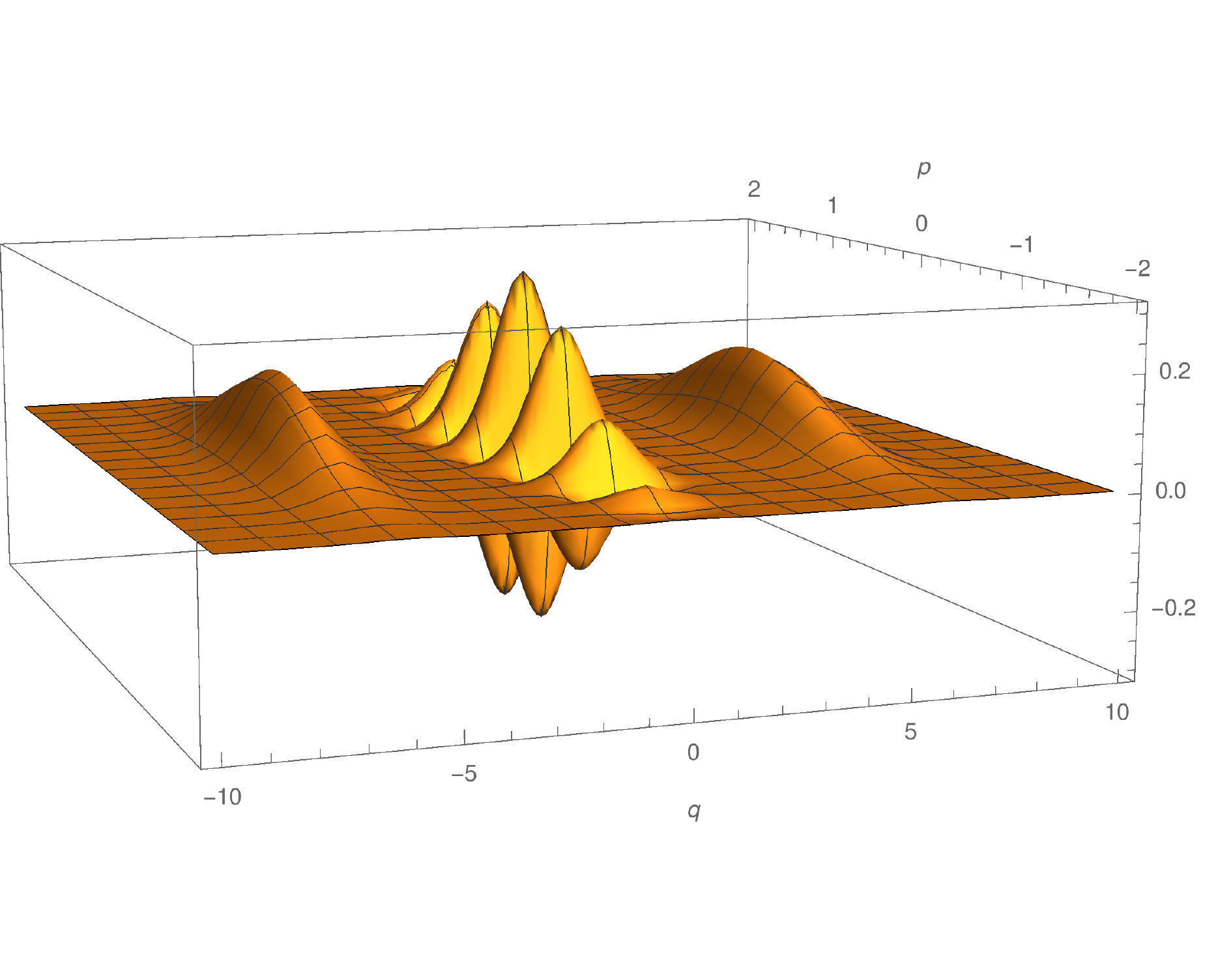}
\end{center}
\caption{Wigner function~(\ref{eq:wignercat}) for the Schr\"odinger
  cat state~(\ref{eq:catwf}). The two bumps corresponds to
  $W_{\pm}(q,p)$ in Eq.~(\ref{eq:wignercat}) and would be the Wigner
  function if the state was given by $\Phi_{\pm }(q)$ only. The
  oscillations in between correspond to $W_{\rm int}(q,p)$ in
  Eq.~(\ref{eq:wignercat}) and are present because of the
  superposition of the two wavepackets. In this region, it is obvious
  from the figure that the Wigner function can be negative. Thus the
  superposition of two wavepackets, a genuine quantum effect, is
  associated to region in phase space where the Wigner function can
  take negative values. We have used $q_0=6$, $p_0=0$ and
  $m=\omega=1$.}
\label{fig:wignercat}
\end{figure}

The main idea of Bell's paper is to relate the presence of
non-classical, quantum, correlations to the non-positivity of the
Wigner function. The idea that a negative Wigner function signals
non-classicality is intuitive since a classical probability function
must be positive. It is best illustrated in the case of a
Schr\"odinger cat state (we consider for simplicity but without loss
of generality, a one-dimensional system)
\begin{align}
\label{eq:catwf}
\Psi_{_{\rm CAT}}(q)=\frac{N_{_{\rm CAT}}}{\sqrt{2}}
\left[\Phi_+(q)+\Phi_-(q)\right],
\end{align}
with 
\begin{align}
\Phi_{\pm}(q)=\left(\frac{m\omega}{\pi }\right)^{1/4}
\exp\left[-\frac{m\omega}{2}(q\pm q_0)^2+ip_0(q\pm q_0)\right],
\end{align}
and $N_{_{\rm CAT}}=[1+e^{-m\omega q_0^2}\cos(2q_0p_0)]^{-1/2}$ in
order for the wave function~(\ref{eq:catwf}) to be properly
normalized. Inserting this expression into the definition of the
Wigner function\footnote{For a one-dimensional system, it reads
\begin{align}
W(q,p)=\frac{1}{2\pi}\int {\rm d}x \left\langle q-\frac{x}{2}
\biggl \vert \Psi\right \rangle \left \langle \Psi 
\biggr \vert q+\frac{x}{2}\right \rangle e^{ipx}.
\end{align}}
one arrives at the following expression, see Ref.~\cite{2004JOptB...6..396K}
\begin{align}
\label{eq:wignercat}
W_{_{\rm CAT}}(q,p)=W_+(q,p)+W_-(q,p)+W_{\rm int}(q,p),
\end{align}
where $W_\pm(q,p)$ represents the Wigner function of a single wave packet, 
namely
\begin{align}
W_\pm(q,p)=\frac{N^2_{_{\rm CAT}}}{2\pi}e^{-m\omega (q\pm q_0)^2
-\frac{1}{m\omega}(p-p_0)^2},
\end{align}
and $W_{\rm int}(q,p)$ is an interaction term
\begin{align}
\label{eq:intW}
W_{\rm int}(q,p)=\frac{N^2_{_{\rm CAT}}}{\pi}\cos (2pq_0)
e^{-m\omega q^2-\frac{1}{m\omega}(p-p_0)^2}.
\end{align}
This Wigner function is represented in Fig.~\ref{fig:wignercat}. We
see two peaks where the Wigner function is positive and, in between, a
series of oscillations due to the cosine term in Eq.~(\ref{eq:intW})
where the Wigner function can be negative. The oscillations in the
Wigner function are clearly due to interferences between the two wave
packets. Therefore, interferences, which are a typical quantum
phenomenon, are responsible for the non-positivity of the Wigner
function, hence the idea to view the non-positivity of the Wigner
function as a criterion for classicality.

\subsection{WKB}
\label{subsec:wkb}

Given the considerations presented in the previous section, it is
interesting to calculate the Wigner function of a WKB state since the
WKB approximation is often viewed as a way to study the classical
limit in Quantum Mechanics. Surprisingly, the calculation of the WKB
Wigner function is not as straightforward as one might think at first
sight. In Cosmology, it has even a controversial and rich history. The
WKB Wigner function has indeed been applied to various questions in
Cosmology such as the interpretation of the wavefunction of the
Universe (Quantum Cosmology) and the quantum-to-classical transition
of inflationary perturbations, this last topic being obviously
especially relevant for the present article.

Although the original calculation of the semi-classical Wigner
function was performed by M.~Berry in $1976$~\cite{Berry:1977zz}, it
started to be applied in Cosmology only at the end of the $80$'s, in
Ref.~\cite{Halliwell:1987eu}. The question of how to interpret the
wavefunction of the Universe in Quantum Cosmology was the issue
tackled in this article. Usually, the solution of the Wheeler-de Witt
equation makes sense only in the WKB approximation because this is the
regime where positive probabilities can be extracted from this
formalism (recall that the Wheeler-de Witt equation is similar to a
Klein-Gordon equation and, hence, does not always lead to positive
probabilities). Then, the idea was to look for correlations in the WKB
Wigner function. The calculation of Ref.~\cite{Halliwell:1987eu}
proceeds as follows. Inserting the WKB wavefunction,
\begin{align}
\label{eq:wfwkb}
\Psi(q,t)=C(q,t)\exp\left[\frac{i}{\hbar}S(q,t)\right],
\end{align}
where we have provisionally re-established $\hbar$ and where $S(q,t)$
is the classical action of the system, into the definition of the
Wigner function, one arrives at
\begin{align}
W_{_{\rm WKB}}(q,p)=\frac{1}{2\pi\hbar}
\int {\rm d}x \, C^*\left(q-\frac{1}{2}\hbar x\right)
C\left(q+\frac{1}{2}\hbar x\right)
\exp\left[-\frac{i}{\hbar}px
-\frac{i}{\hbar}S\left(q-\frac{1}{2}\hbar x\right)
+\frac{i}{\hbar}S\left(q+\frac{1}{2}\hbar x\right)\right].
\end{align}
Expanding the amplitude and the phase in $\hbar$, one obtains
\begin{align}
W_{_{\rm WKB}}(q,p)\simeq \frac{1}{2\pi \hbar}\int {\rm d}x
\left[\vert C(q)\vert ^2+{\cal O}(\hbar^2)\right]
\exp\left[-\frac{i}{\hbar}px +\frac{i}{\hbar}
\frac{\partial S}{\partial q}x+{\cal O}(\hbar^2)\right],
\end{align}
that is to say, by performing the integration over $x$
\begin{align}
\label{eq:wkbwignerwrong}
W_{_{\rm WKB}}(q,p)=\vert C(q)\vert^2\delta 
\left(p-\frac{\partial S}{\partial q}\right).
\end{align}
Therefore, the conclusion of Ref.~\cite{Halliwell:1987eu} was that the
WKB approximation is really a classical limit in the sense that the
Wigner function is positive definite and peaked over the classical
trajectories $p=\partial S/\partial q$ with a weigh given by the
squared amplitude $\vert C(q)\vert^2$.

However, after the publication of Ref.~\cite{Halliwell:1987eu}, it was
pointed out in Refs.~\cite{Habib:1990hz} and~\cite{Anderson:1990vc}
that the calculation of the WKB Wigner
function~(\ref{eq:wkbwignerwrong}) is in fact unjustified and that,
moreover, the correct formula was derived, as already mentioned, by
M.~Berry in Ref.~\cite{Berry:1977zz}. The trouble with
Eq.~(\ref{eq:wkbwignerwrong}) is that one cannot truncate the
expansion of the phase and, then, perform the integration over $x$. It
is true that the higher order terms are proportional to powers of
$\hbar$ (which goes to zero) but also to powers of $x$ (the range of
which is the entire real axis) so that it is unclear whether the
contributions of higher order terms are really negligible. The correct
method consists in fact in using the saddle point approximation. This 
leads to 
\begin{align}
\label{eq:wkbwigner}
W_{_{\rm WKB}}(q,p)=2\sqrt{2}\frac{\left[\frac32 A(q,p)\right]^{1/6}}
{\left[\frac{\partial E}{\partial q}\biggl\vert_2
\frac{\partial E}{\partial p}\biggl \vert_1
-\frac{\partial E}{\partial p}\biggl \vert_2\frac{\partial E}{\partial q}
\biggl\vert_1
\right]^{1/2}}
{\rm Ai}\left\{-\left[\frac32 A(q,p)\right]^{2/3}\right\},
\end{align}
where ${\rm Ai}(z)$ is a Airy function~\cite{AbraSteg72}. In the above
expression, $E$ is the energy shell, namely the quantity such that the
Hamiltonian of the system satisfies $H(p,q)=E$. The points $1$ and $2$
are the points of coordinates $q\pm x_0(q,p)$, $p(q\pm x_0)$
satisfying the stationary phase condition
$\left[p\left(q-x_0/2,E\right)+p\left(q+x_0/2,E \right)\right]=2p$.
They lie on the classical trajectory and their position is determined
such that the arithmetic mean of their momentum is $p$. Finally,
$A(q,p)$ is the area comprised between the chord $1-2$ and the
classical torus $H=E$. One can show that, when the Wigner function is
known, the above formula~(\ref{eq:wkbwigner}) matches very well the
exact result in the regime where the WKB condition is valid. But the
most important property of Eq.~(\ref{eq:wkbwigner}) is that it is not
positive definite and usually displays oscillations in phase
space. This shows that the semi-classical limit cannot be viewed as
being a truly classical regime.

The WKB approximation has also been applied to the
quantum-to-classical transition of cosmological perturbations in
Ref.~\cite{Albrecht:1992kf}. In that paper, it is claimed that this
transition is achieved because the quantum state of the perturbations
precisely satisfies the WKB approximation on super Hubble
scales. Based on what we have just seen about the WKB Wigner function,
this last statement should be taken with great care. In fact, as we
are going to see, the behavior of cosmological perturbations on large
scales is especially relevant when it comes to semi-classical methods
in phase space.

Based on the fact that the quantum state of the perturbations is a
squeezed state, Ref.~\cite{Albrecht:1992kf} considers a simplified
model consisting in an inverted harmonic oscillator whose Lagrangian
is given by
\begin{align}
L=\frac12 m\dot{q}^2+\frac12 kq^2,
\end{align}
where the potential term has the ``wrong'' sign, $V(x)=-kq^2/2$. As we
are going to see, the state of the oscillator evolves into a strongly
squeezed state which justifies to consider this simple system. The
corresponding Hamiltonian reads $H=p^2/(2m)-kq^2/2$, with
$p=\partial L/\partial \dot{q}=m\dot{q}$, and is not bounded from
below. Then, creation and annihilation operators can be introduced in
the standard fashion
\begin{align}
  \hat{q}=\sqrt{\frac{1}{2m\omega}}\left(\hat{c}
  +\hat{c}^{\dagger}\right), \quad 
  \hat{p}=-i\sqrt{\frac{m\omega }{2}}\left(\hat{c}-\hat{c}^{\dagger}\right),
\end{align}
where $\omega^2=k/m$. This allows us to express the Hamiltonian as
\begin{align}
\hat{H}=-\frac{\omega}{2}\left(\hat{c}^2+\hat{c}^{\dagger }{}^2\right).
\end{align}
Of course, the most striking property of this Hamiltonian is the
presence of an overall minus sign which is just the consequence of the
inverted nature of the oscillator. Then, the equations of motion are
given by
\begin{align}
\frac{{\rm d}\hat{c}}{{\rm d}t}=-i\left[\hat{c},\hat{H}\right]
=i\omega \hat{c}^{\dagger}, \quad 
\frac{{\rm d}\hat{c}^{\dagger}}{{\rm d}t}
=-i\left[\hat{c}^{\dagger},\hat{H}\right]=-i\omega \hat{c}.
\end{align}
As usual, they can be solved by mean of a Bogoliubov transformation,
namely
\begin{align}
\hat{c}(t)=u(t) \hat{c}(t_\uini)+v(t)\hat{c}^{\dagger}(t_\uini),
\quad
\hat{c}^{\dagger}(t)=u^*(t)\hat{c}^{\dagger}(t_\uini)
+v^*(t)\hat{c}(t_\uini),
\end{align}
where the functions $u(t)$ and $v(t)$ satisfy the following equations
\begin{align}
\frac{{\rm d}u}{{\rm d}t}=i\omega v^*, \quad 
\frac{{\rm d}v}{{\rm d}t}=i\omega u^*,
\end{align}
with initial conditions $u(0)=1$ and $v(0)=0$ (here, for simplicity,
we have taken $t_\uini=0$). Combining these two first order
differential equations, one can obtain one second order equation for
$u$ (and/or $v$) which reads
\begin{align}
\frac{{\rm d}^2u}{{\rm d}t^2}=\omega ^2 u.
\end{align}
which gives $u(t)=\cosh(\omega t)$ and $v(t)=i\sinh(\omega t)$. As a 
consequence, the operator $\hat{c}(t)$ can be rewritten as
\begin{align}
\hat{c}(t)=\cosh(\omega t)\hat{c}_\uini+e^{i\pi/2}\sinh(\omega t)
\hat{c}_{\uini}^{\dagger}
=\hat{S}^{\dagger}\hat{c}_\uini \hat{S},
\end{align}
where we have written $\hat{c}(t_\uini)=\hat{c}_\uini$ and where the 
operator $\hat{S}$ is defined by
\begin{align}
\hat{S}=\exp\left[\frac{r}{2}\left(\hat{c}^2_\uini e^{-2i\phi}
-\hat{c}^{\dagger}_\uini{}^2e^{2i\phi}\right)\right].
\end{align}
with squeezing parameter and angle given by $r=\omega t$ and
$\phi=-\pi/4$. $\hat{S}$ is the squeezing operator and is responsible,
as announced above, for the appearance of squeezed states in the
problem.

In order to mimic the behavior of cosmological perturbations, we
assume that the initial state of the system is the vacuum
$\vert 0\rangle$. Then, the state at a subsequent time $t$ can be
found using techniques based on the operator ordering
theorem~\cite{barnett2002methods} which allows us to rewrite the
operator $\hat{S}$ as
\begin{align}
\hat{S}=\exp\left[e^{-2i\phi}\tanh \vert r\vert 
\frac{\hat{c}_\uini^{\dagger}{}^2}{2}
\right]
\exp\left[-\frac12 (\hat{c}_\uini \hat{c}_\uini^{\dagger}
+\hat{c}_\uini^{\dagger }\hat{c}_\uini)
\ln(\cosh \vert r\vert)\right]
\exp\left[-e^{2i\phi}\tanh \vert r\vert \frac{\hat{c}_\uini^2}{2}
\right],
\end{align}
from which it follows that
\begin{align}
\vert \Psi\rangle =\hat{S}\vert 0\rangle 
=\frac{1}{\sqrt{\cosh r}}\sum_{p=0}^{\infty}
\frac{e^{-2ip\phi}}{2^p p!}
\sqrt{2p!}
\tanh ^p r \vert 2p\rangle.
\end{align}
This state is a one-mode squeezed state. It slightly differs from the
two-mode squeezed state considered before in
Eq.~(\ref{eq:quantumstate}). In particular, we see that the sum is
only on states with an even number of particles. Since $r=\omega t$,
the squeezing goes to infinity in the large time limit.

Our next move is to calculate the wavefunction. It reads
\begin{align}
\Psi(q)=\left \langle q\vert \Psi\right \rangle
=\frac{e^{-q^2/2}}{\pi^{1/4}\sqrt{\cosh r}}
\sum_{p=0}^{\infty}
\frac{e^{-2ip\phi}}{2^{2p} p!}
\tanh ^p r H_{2p}(q),
\end{align}
where $H_{2p}(z)$ is a Hermite polynomial of order
$2p$~\cite{AbraSteg72} and appears in the $q$-representation of the
state $\vert 2p\rangle$. Then, using
$\sum_{p=0}^{\infty}t^nH_{2n}(x)/n!=e^{4tx^2/(1+4t)}/\sqrt{1+4t}$
(here, one has $4t=e^{-2i\phi}\tanh r$) and recalling that
$\phi=-\pi/4$ one arrives at
\begin{align}
\label{eq:wkbonemode}
\Psi(q)=\left[\pi \cosh (2r)\right]^{-1/4}
\exp\left[-\frac{q^2}{2\cosh(2r)}+\frac{i}{2}q^2\tanh (2r)
-\frac{i}{2}\arctan \left(\tanh r\right)\right],
\end{align}
where one verifies that this wavefunction is correctly normalized. 

As noticed in Ref.~\cite{Albrecht:1992kf}, it can be written in a WKB
form~(\ref{eq:wfwkb}) with
\begin{align}
\label{eq:Conemode}
C=\left[\pi \cosh (2r)\right]^{-1/4}
\exp\left[-\frac{q^2}{2\cosh(2r)}\right], \quad 
S=\frac{1}{2}q^2\tanh (2r)
-\frac{1}{2}\arctan \left(\tanh r\right),
\end{align}
and, in the large time limit or, equivalently, strong squeezing limit,
the semi-classical approximation is extremely well-verified since the
WKB condition is satisfied
$\vert C \partial_q S/\partial _qC\vert =\sinh (2r)\gg 1$. Hence the
claim that cosmological perturbations on super Hubble scales, which is
equivalent to strong squeezing, are ``semi-classical''. It is then
tempting to go from ``semi-classical'' to ``classical'' and consider
that the quantum-to-classical transition has been achieved. However,
as seen above, given that the WKB Wigner function is not positive
definite, one should a priori resist to this temptation.

However, as it is often the case, Cosmology introduces a new twist in
this question. Using Eq.~(\ref{eq:wkbonemode}), one can calculate the
Wigner function. The result reads
\begin{align}
\label{eq:onemodewigner}
W(q,p)=\frac{1}{\pi}e^{-q^2/\cosh(2r)}
e^{-\cosh (2r)\left[p-q\tanh(2r)\right]^2}.
\end{align}
One obtains a Gaussian, which is consistent with the fact that the
wavefunction~(\ref{eq:wkbonemode}) is a Gaussian. This means that this
Wigner function, contrary to the WKB Wigner
function~(\ref{eq:wkbwigner}), is positive definite which, at first
sight, seems surprising since $\Psi(q)$ in Eq.~(\ref{eq:wkbonemode})
satisfies the WKB approximation. Moreover, writing
$\epsilon=1/[4\cosh(2r)]$ which, in the strong squeezing limit, goes
to zero, representing the Dirac function by
$ \delta _{\epsilon}(x)=e^{-x^2/(4\epsilon)}/[2\sqrt{\pi \epsilon}] $
and, finally, noticing that $p=\partial S/\partial q=q\tanh (2r)$, the
Wigner function~(\ref{eq:onemodewigner}) can be re-written as
\begin{align}
  W(q,p)=\vert C\vert ^2 \delta_{\epsilon}\left[p-q\tanh(2r)\right]
  =\vert C\vert ^2\delta_{\epsilon}\left(p-\frac{\partial S}{\partial q}
  \right),
\end{align}
where $C$ is given in Eq.~(\ref{eq:Conemode}). This equation is
nothing but Eq.~(\ref{eq:wkbwignerwrong}) which was, as discussed in
Refs.~\cite{Habib:1990hz} and~\cite{Anderson:1990vc}, supposed to be
incorrect! What happened is in fact very simple. It was pointed out
before that the expansion of the phase performed in
Ref.~\cite{Halliwell:1987eu} is not justified because the order of the
various terms of this expansion is in fact indeterminate. There is of
course one exception to this claim which is when the calculation of
Ref.~\cite{Halliwell:1987eu} is not an expansion but is exact. This is
exactly what happens here since the phase is quadratic in $q$. So
ignoring higher order terms, which is usually unjustified, is, in the
present case, totally valid simply because these higher order terms
are just not present. This is consistent with the fact that the Wigner
function of Gaussian is a Gaussian and, therefore, is positive
definite. This shows how peculiar and subtle is the
quantum-to-classical transition of cosmological perturbations is.

\subsection{Bell's Paper on the Wigner Function}
\label{subsec:bellwigner}

After these preliminary considerations, let us now come back to the
letter written by J.~Bell in $1986$~\cite{1986NYASA.480..263B}. Based
on the original EPR article~\cite{Einstein:1935rr}, Bell imagines a
situation where there are two free particles traveling in space along
a given axis (the particles can propagate in both directions). Then,
Bell assumes that one can measure position (of the two particles)
only. As he notices himself, this slightly differs from the standard
EPR argument where it is also assumed that momenta can be
measured. The article makes use of the ``two-time'' Wigner function
defined by
\begin{align} 
W(q_1,q_2,p_1,p_2,t_1,t_2)&=\frac{1}{(2\pi)^2}
\int {\rm d}y_1\, {\rm d}y_2\, e^{-ip_1y_1-ip_2y_2}
\nonumber \\ & \times 
\Psi\left(q_1+\frac{y_1}{2},q_2+\frac{y_2}{2},t_1,t_2\right)
\Psi^*\left(q_1-\frac{y_1}{2},q_2-\frac{y_2}{2},t_1,t_2\right),
\end{align}
where $\Psi(q_1,q_2,t_1,t_2)$ is the wavefunction of the system. If
one considers two freely-moving particles, then $\Psi$ satisfies the
Schr\"odinger equations,
$i\partial \Psi/\partial t_{1,2}=\hat{p}_{1,2}^2/(2m_{1,2})\Psi$ and,
as a consequence, the Wigner function obeys
$(\partial/\partial t_{1,2}+p_{1,2}\partial/\partial q_{1,2})W=0$.
This means that
$W(q_1,q_2,p_1,p_2,t_1,t_2)=W(q_1-p_1t_1,q_2-p_2t_2,p_1,p_2,0,0)$ and
this allows us to calculate the Wigner function at any times from the
sole knowledge of the initial wavefunction. Of course, the calculation
of $W(q_1-p_1t_1,q_2-p_2t_2,p_1,p_2,0,0)$ still requires the knowledge
of the wavefunction and, in the following, several
possibilities are considered.

Then, Bell proceeds and shows how his famous inequality can be
implemented in the situation described before. More precisely, he does
so in the so-called Clauser, Horne, Shimony and Holt (CHSH)
formulation\footnote{Amusingly, throughout his paper, Bell mixes the
  acronyms and refers to this inequality as the CHHS inequality
  although he cites the correct reference (\ie the correct journal,
  issue and page), namely Ref.~\cite{PhysRevLett.24.549}. A closer look at
  the list of references shows that he has also permuted the two
  ``H'', that is to say he puts Holt before Horne while, in the
  original paper, this is the opposite. In fact, Bell has simply taken
  the liberty to re-establish the alphabetical order!} which supposes
to deal with quantities that can only take the values $\pm 1$. This is
why, usually, the CHSH inequality is experimentally tested with spin
variables. However, in the case considered by Bell, the particles are
spinless and, as already mentioned, we only have access to position
measurements. Although he does not present it exactly in this way,
what Bell does to circumvent this problem is to introduce the two
following operators
\begin{align}
\hat{S}_1(t_1)={\rm sgn}\left(\hat{q}_1+\frac{q_0}{2}, t_1\right), \quad 
\hat{S}_2(t_2)={\rm sgn}\left(\hat{q}_2-\frac{q_0}{2}, t_2\right),
\end{align}
which represent the sign of $q_1+q_0/2$ and $q_2-q_0/2$ at times $t_1$
and $t_2$, respectively, $q_0$ being an arbitrary position. Clearly,
the spectra of $\hat{S}_1$ and $\hat{S}_2$ only consists of two
values, namely $\pm 1$, as required. Interestingly enough, this is
exactly what is done in
Refs.~\cite{PhysRevLett.82.2009,PhysRevLett.88.040406,2004PhLA..324..415G}
and, then in the context of Cosmology, in Refs.~\cite{Martin:2016tbd}
and~\cite{Martin:2017zxs}. Therefore, remarkably, Bell's paper already
contains the idea of fictitious spin operators that, as we will see
later on, can be used in order to design a cosmic Bell experiment.

Then, once we have discrete variables, one can just mimic the usual
approach which, as reminded above, is formulated in terms of
spins. The first step consists in defining the two-point correlators
\begin{align}
\label{eq:defcorrelE}
E(t_1,t_2)\equiv \left \langle \Psi \left \vert 
\hat{S}_1(t_1)\hat{S}_2(t_2)\right \vert \Psi \right \rangle,
\end{align}
where $\vert \Psi \rangle$ is the quantum state in which the system is
placed. Let us also remark that the two times $t_1$ and $t_2$ play the
role of the polarizers settings in the standard CHSH
formulation. Following the usual considerations, one can then prove
that
\begin{align}
\label{eq:chsh}
B(t_1,t_2,t_1',t_2')=E(t_1,t_2)+E(t_1,t_2')+E(t_1',t_2)-E(t_1',t_2')\le 2,
\end{align}
if the correlators are interpreted as stochastic averages and if
locality holds~\cite{Maudlin:2014tia}. On the contrary, in Quantum
Mechanics, one just has $B(t_1,t_2,t_1',t_2')<2\sqrt{2}$, hence the
idea to look for experimental configurations for which
$2<B(t_1,t_2,t_1',t_2')<2\sqrt{2}$. In the following, Bell inequality
violation (or CHSH inequality violation, we use the two expressions
indifferently) will refer to as a situation where
$B(t_1,t_2,t_1',t_2')>2$.

As discussed at the beginning of this section, Bell wants to relate
the non-positivity of the Wigner function to a violation of the CHSH
inequality. Technically, the link between the inequality and the
Wigner function is expressed as follows. It is easy to check that the
quantities $S_1$ and $S_2$ are such that $\widetilde{S_1}=S_1$ and
$\widetilde{S}_2=S_2$, see the discussion above in
Sec.~\ref{subsec:stocha} and, as a consequence, thanks to
Eqs.~(\ref{eq:trab}), (\ref{eq:meano}), the expression of the
correlator $E(t_1,t_2)$ can be rewritten as
\begin{align}
E(t_1,t_2)&=\int _{-\infty}^{\infty}{\rm d}q_1
\int _{-\infty}^{\infty}{\rm d}q_2\int _{-\infty}^{\infty}{\rm d}p_1
\int _{-\infty}^{\infty}{\rm d}p_2\,
W(q_1,q_2,p_1,p_2,t_1,t_2) \widetilde{S}_1(t_1)\widetilde{S}_2(t_2)
\\ &
=
\int _{-\infty}^{\infty}{\rm d}q_1\int_{-\infty}^{\infty}{\rm d}q_2
\int_{-\infty}^{\infty}{\rm d}p_1\int _{-\infty}^{\infty}{\rm d}p_2\,
W(q_1-p_1t_1,q_2-p_2t_2,p_1,p_2,0,0) 
\widetilde{S}_1(t_1)\widetilde{S}_2(t_2)
\\ &
\label{eq:reprho}
=
\int _{-\infty}^{\infty}{\rm d}q_1
\int _{-\infty}^{\infty}{\rm d}q_2\,
\widetilde{S}_1(t_1)\widetilde{S}_2(t_2)
\rho(q_1,q_2,t_1,t_2),
\end{align}
where the two-time distribution probability $\rho(q_1,q_2,t_1,t_2)$ is
defined by
$\rho(q_1,q_2,t_1,t_2)=\int _{-\infty}^{\infty}{\rm
  d}p_1\int_{-\infty}^{\infty} {\rm
  d}p_2W(q_1-p_1t_1,q_2-p_2t_2,p_1,p_2,0,0)$.
The above equations exactly represent the relation needed as it allows
us to estimate the left hand side of Eq.~(\ref{eq:chsh}) in terms of
the Wigner function. Using the definition of $\widetilde{S}_1$ and
$\widetilde{S}_2$, one can also show that
\begin{align}
\label{eq:E}
E(t_1,t_2)&=1-2\biggl[\int_0^{+\infty}{\rm d}\bar{q}_1
\int _{-\infty}^0{\rm d}\bar{q}_2 \, \rho\left(\bar{q}_1-\frac{q_0}{2},
\bar{q}_2+\frac{q_0}{2},t_1,t_2\right)
\nonumber \\ &
+\int_{-\infty}^0{\rm d}\bar{q}_1
\int _0^{+\infty}{\rm d}\bar{q}_2 \, \rho\left(\bar{q}_1-\frac{q_0}{2},
\bar{q}_2+\frac{q_0}{2},t_1,t_2\right)\biggr],
\end{align}
where we used the fact that the Wigner function is normalized to one.

To go further and concretely calculate the correlators, and, hence,
verify whether the CHSH inequality is violated or not, one needs to
specify the state in which the system is placed. The first example
considered by Bell is simply the original EPR
wavefunction~\cite{Einstein:1935rr} (supposed to hold at initial
times),
$\Psi_{_{\rm EPR}}(q_1,q_2)=N_{_{\rm EPR}}\delta(q_1-q_2+q_0)$, where
$N_{_{\rm EPR}}$ is a normalization constant. Then, he calculates the
corresponding Wigner function and finds
\begin{align}
\label{eq:eprwigner}
  W_{_{\rm EPR}}(q_1,q_2,p_1,p_2,0,0)=\frac{N_{_{\rm EPR}}^2}{4\pi}
\delta (q_1-q_2+q_0)\delta(p_1+p_2).
\end{align}
Bell remarks that this Wigner function is positive everywhere and that
``{\it the EPR correlations are precisely those between two classical
  particles in independent free classical motion. With the wave
  function (8) {\rm [namely the EPR wavefunction]}, then, there is no
  non-locality problem when the incompleteness of the wave function
  description is admitted}''. Therefore, Bell explicitly relates
classicality (namely no violation of the CHSH inequality) and
positivity of the Wigner function. 

However, it is interesting to notice that he does not explicitly
verify the violation of the CHSH inequality in that case [namely he
does not explicitly calculate the two-point correlators for the EPR
wavefunction nor the combination~(\ref{eq:chsh})], despite the fact
that he clearly suggests it is not violated. Although he does not
explain why, one can maybe guess the reason. If one takes the EPR
Wigner function~(\ref{eq:eprwigner}) and tries to calculate
$\rho(q_1,q_2,t_1,t_2)$ one finds an infinite result. Indeed, the
first integration, say on $p_1$, kills the Dirac delta function
$\delta(p_1+p_2)$. Then it remains an integral over $p_2$ of something
which does not depend on $p_2$, which gives an infinite result. This
remark is interesting because it turns out to be at the core of all
the literature that is devoted to this question and to the Bell's
paper: this will be discussed at length in the following sections. The
reason for this problem is in fact related to the normalization
$N_{_{\rm EPR}}$. Indeed, the wavefunction
$\Psi_{_{\rm EPR}}(q_1,q_2)$ is not properly normalized. A correct way
to normalize it is to modify the wavefunction such that it now reads
\begin{align}
\label{eq:eprwf}
\Psi_{_{\rm EPR}}(q_1,q_2)=\left(\frac{\varepsilon}{b}2\sqrt{2}\right)^{1/2}
e^{-(q_1+q_2)^2/(2b^2)}\frac{1}{\varepsilon \sqrt{\pi}}
e^{-(q_1-q_2+q_0)^2/\varepsilon^2}.
\end{align}
This is suggested by Bell himself in the continuation of his
article. Strictly speaking, in Bell's paper, this trick is not applied
to the EPR wavefunction but to another wavefunction considered later
on in his article [and, in the present paper as well, see
Eq.~(\ref{eq:psiWnegative})]. Here, we have just anticipated his guess
and have applied it to the EPR wavefunction. We will comment more
about this point in the following. The ``new''
wavefunction~(\ref{eq:eprwf}) is made of three pieces: a normalization
factor, a new factor depending on a new parameter $b$,
$e^{-(q_1+q_2)^2/(2b^2)}$ and the last factor depending on another new
parameter $\varepsilon$,
$e^{-(q_1-q_2+q_0)^2/\varepsilon^2}/(\varepsilon\sqrt{\pi})$. This
last factor is just a finite representation of the Dirac function
$\delta(q_1-q_2+q_0)$ which is recovered in the limit
$\varepsilon \rightarrow 0$. The second factor is necessarily present
to make the wavefunction normalizable. Even if one uses the finite
representation of the Dirac function, it is not possible to make
$\int \vert \Psi_{_{\rm EPR}}\vert^2 {\rm d}q_1{\rm d}q_2$ finite
without the factor $e^{-(q_1+q_2)^2/(2b^2)}$ as it is easy to see by
introducing new variables $q_1\pm q_2$ in the previous
integral. Finally, the first factor is the normalization coefficient
implied by the two other factors.

In his article, Bell claims\footnote{Again, this in fact applies to
  another wavefunction in the paper. However, since the problem is
  exactly the same for the EPR wavefunction, we take the liberty to
  propagate this statement to the situation studied here.} that the
second factor can be ignored by taking the limit
$b\rightarrow \infty$. However, we see that this limit, as well as the
limit $\varepsilon \rightarrow 0$, are very problematic due to the
normalization factor. For this reason, in the following, we work with
Eq.~(\ref{eq:eprwf}) without taking any limit. Since the wavefunction
is Gaussian, the calculations are tractable. Indeed, using the
wavefunction~(\ref{eq:eprwf}), it is easy to calculate the Wigner
function which reads
\begin{align}
\label{eq:wignerepr}
W_{_{\rm EPR}}(q_1,q_2,p_1,p_2,0,0)=\frac{1}{\pi^2}
e^{-b^2(p_1+p_2)^2/4-(q_1+q_2)^2/b^2}
e^{-\varepsilon^2(p_1-p_2)^2/8-2(q_1-q_2+q_0)^2/\varepsilon^2}.
\end{align}
Of course, since the state~(\ref{eq:eprwf}) is Gaussian, we find that
the Wigner function is also a Gaussian, see
Sec.~\ref{subsec:gaussian}. One checks that the Wigner function is
properly normalized,
$\int {\rm d}q_1{\rm d}q_2{\rm d}p_1{\rm d}p_2 W_{_{\rm EPR}}=1$,
which is a consistency check of Eq.~(\ref{eq:wignerepr}). The next
move consists in calculating the distribution $\rho$. Recall that this
requires the calculation of
$W_{_{\rm EPR}}(q_1-p_1t_1,q_2-p_2t_2,p_1,p_2,0,0)$. Using
Eq.~(\ref{eq:wignerepr}), this can be expressed as
\begin{align}
W_{_{\rm EPR}}(q_1-p_1t_1,q_2-p_2t_2,p_1,p_2,0,0)
=\frac{1}{\pi^2}e^{-(q_1+q_2)^2/b^2-2(q_1-q_2+q_0)^2/\varepsilon^2}
e^{-\frac12 P^{\rm T}AP-P^{\rm T}J},
\end{align}
where one has defined
\begin{align}
\displaystyle
A &=
\begin{pmatrix}
\displaystyle
\frac{b^2}{2}+\frac{\varepsilon^2}{4}+\left(\frac{2}{b^2}
+\frac{4}{\varepsilon^2}\right)t_1^2 & 
\displaystyle
\frac{b^2}{2}
-\frac{\varepsilon^2}{4}+\left(\frac{2}{b^2}
-\frac{4}{\varepsilon^2}\right)t_1t_2 \\
\\
\displaystyle
\frac{b^2}{2}
-\frac{\varepsilon^2}{4}+\left(\frac{2}{b^2}
-\frac{4}{\varepsilon^2}\right)t_1t_2 &
\displaystyle
\frac{b^2}{2}+\frac{\varepsilon^2}{4}+\left(\frac{2}{b^2}
+\frac{4}{\varepsilon^2}\right)t_2^2
\end{pmatrix}, 
\end{align}
and 
\begin{align}
J(q_1,q_2)&=\left(
\begin{matrix}
\displaystyle
-\frac{2}{b^2}(q_1+q_2)t_1-\frac{4}{\varepsilon^2}(q_1-q_2+q_0)t_1 \\
\\
\displaystyle
-\frac{2}{b^2}(q_1+q_2)t_2+\frac{4}{\varepsilon^2}(q_1-q_2+q_0)t_2
\end{matrix}
\right),
\end{align}
and $P^{\rm T}=(p_1,p_2)$. 
\begin{figure}
\begin{center}
\includegraphics[width=10cm]{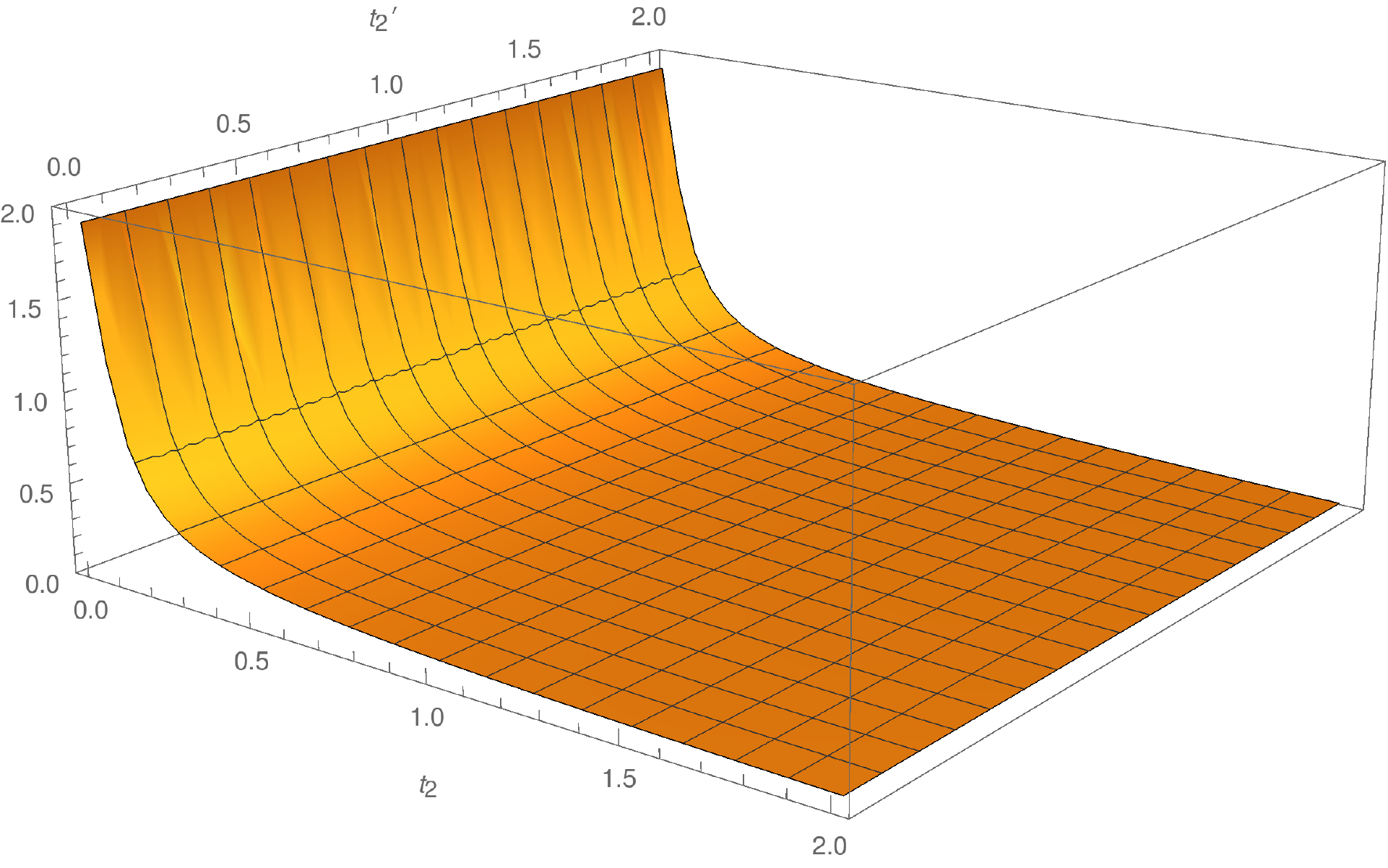}
\end{center}
\caption{Bell quantity $B(0,t_2,0,t_2')$ as a function of the two
  times $t_2$ and $t_2'$ calculated for the EPR state, a state which
  has a positive definite Wigner function. As is apparent from the
  plot, one always has $B(0,t_2,0,t_2')<2$ which means that Bell
  inequality is never violated.}
\label{fig:bellepr}
\end{figure}
Then, the distribution $\rho$ can then be straightforwardly evaluated
by applying well-known formula for Gaussian integrals. This leads to
\begin{align}
\label{eq:rhoepr}
\rho_{_{\rm EPR}}(q_1,q_2,t_1,t_2)=\frac{2}{\pi}
e^{-(q_1+q_2)^2/b^2-2(q_1-q_2+q_0)^2/\varepsilon^2}
\frac{1}{\sqrt{\det A}}e^{\frac12 J^{\rm T}A^{-1}J}.
\end{align}
We are now in a position where the correlator
$E_{_{\rm EPR}}(t_1,t_2)$ can be calculated. Plugging the above
equation~(\ref{eq:rhoepr}) into Eq.~(\ref{eq:E}), one obtains
\begin{align}
E_{_{\rm EPR}}(t_1,t_2)=&1-\frac{4}{\pi \sqrt{\det A}}
\biggl[\int_0^{+\infty}{\rm d}\bar{q}_1
\int _{-\infty}^0{\rm d}\bar{q}_2 
e^{-(\bar{q}_1-q_0/2+\bar{q}_2+q_0/2)^2/b^2-2(\bar{q}_1-q_0/2
-\bar{q}_2-q_0/2+q_0)^2/\varepsilon^2}
\nonumber \\ & \times
e^{\frac12 J^{\rm T}(\bar{q}_1-q_0/2,\bar{q}_2+q_0/2)A^{-1}J(\bar{q}_1-q_0/2,\bar{q}_2+q_0/2)}
\nonumber \\ &
+\int_{-\infty}^0{\rm d}\bar{q}_1
\int _0^{+\infty}{\rm d}\bar{q}_2 
e^{-(\bar{q}_1-q_0/2+\bar{q}_2+q_0/2)^2/b^2-2(\bar{q}_1-q_0/2
-\bar{q}_2-q_0/2+q_0)^2/\varepsilon^2}
\nonumber \\ & \times
e^{\frac12 J^{\rm T}(\bar{q}_1-q_0/2,\bar{q}_2+q_0/2)A^{-1}J(\bar{q}_1-q_0/2,\bar{q}_2+q_0/2)}
\biggr].
\end{align}
We see that, in the arguments of all exponentials, $q_0$ cancel out
and makes the correlation function independent of $q_0$. This makes
sense since this quantity is arbitrary and, therefore, a physical
result cannot depend on its value. This also implies that the
arguments of the exponentials are in fact quadratic in $\bar{q}_1$ and
$\bar{q}_2$. As a consequence, by a simple change of variable, the
second term in the square bracket is in fact equal to the first
one. Finally, one obtains
\begin{align}
\label{eq:intEepr}
E_{_{\rm EPR}}(t_1,t_2)=&1-\frac{8}{\pi \sqrt{\det A}}
\int_0^{+\infty}{\rm d}\bar{q}_1
\int _{-\infty}^0{\rm d}\bar{q}_2 
e^{c_1\bar{q}_1^2+c_2\bar{q}_2^2+c_3\bar{q}_1\bar{q}_2},
\end{align}
where the coefficients $c_1$, $c_2$ and $c_3$ are defined by
\begin{align}
c_1 &=-\frac{2b^2+\varepsilon^2}{2 \det A}
\left(1+\frac{8 t_2^2}{b^2 \varepsilon^2}\right), \quad
c_2 =-\frac{2b^2+\varepsilon^2}{2 \det A}
\left(1+\frac{8 t_1^2}{b^2 \varepsilon^2}\right), \quad
c_3 =\frac{2b^2-\varepsilon^2}{ \det A}
\left(1-\frac{8 t_1t_2}{b^2 \varepsilon^2}\right), \\
\det A &= \frac{1}{2b^2\varepsilon^2}
\left[b^4\varepsilon^4+64 t_1^2t_2^2+(4b^4+\varepsilon^4)(t_1+t_2)^4
+4b^2\varepsilon^2(t_1-t_2)^2\right].
\end{align}
The integral~(\ref{eq:intEepr}) can easily be performed and one obtains
\begin{align}
\label{eq:EEPR}
E_{_{\rm EPR}}(t_1,t_2)=&1-\frac{8}{\pi \sqrt{\det A}}
\frac{\sqrt{\det A}}{8}
\left\{\pi 
-2 \arctan \left[\frac{(2b^2-\varepsilon^2)
\left[1-8t_1t_2/(b^2\varepsilon^2)\right]}{4 \sqrt{\det A}}\right]
\right\}\\
&=\frac{2}{\pi}\arctan \left[\frac{(2b^2-\varepsilon^2)
\left[1-8t_1t_2/(b^2\varepsilon^2)\right]}{4 \sqrt{\det A}}\right].
\end{align}
As required $E_{_{\rm EPR}}(t_1,t_2)$ varies between $-1$ and $1$. It
is interesting to notice that the form of the correlator is really
typical of what one obtains with pseudo spin operators, see for
instance Eq.~(26) of Ref.~\cite{Martin:2017zxs}: the resemblance is 
striking. 

Having obtained the correlators, it is now easy to verify whether the
CHSH inequality~(\ref{eq:chsh}) is violated or not. In
Fig.~\ref{fig:bellepr}, we have represented the quantity
$B(0,t_2,0,t_2')$ defined in Eq.~(\ref{eq:chsh}) and evaluated with
the correlator obtained above. We see that this quantity is always
smaller than two so that Bell inequality is never violated.

Therefore, we have confirmed Bell's suggestion that the EPR state,
which has a positive Wigner function, does not lead to any Bell
inequality violation. This was done with a method that avoided the
technical issues present in Bell's letter. In short, we conclude that
Bell calculations are problematic but, despite those issues, the
overall result, at least for the example of the EPR state, is correct.

\begin{figure}
\begin{center}
\includegraphics[width=10cm]{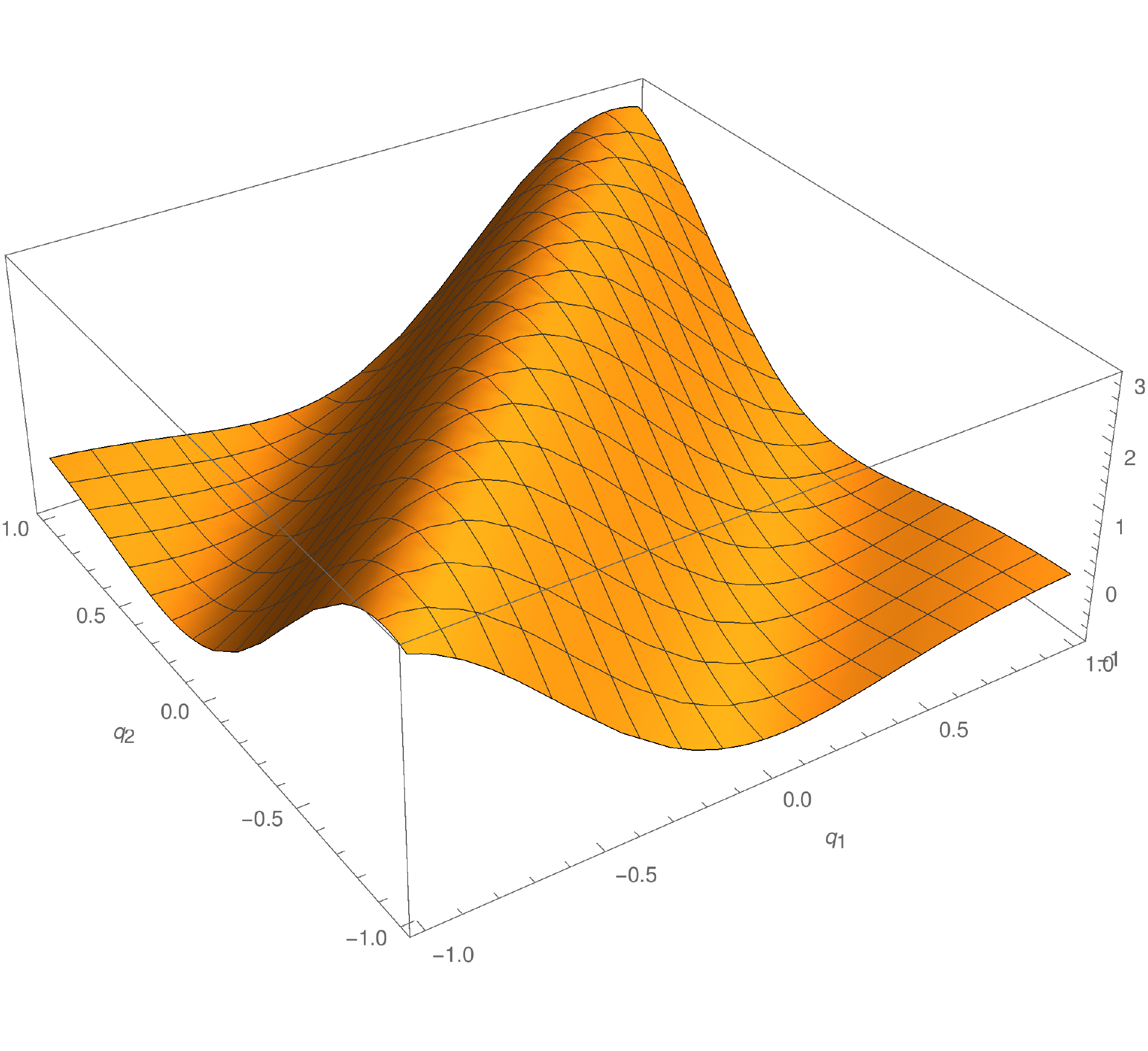}
\end{center}
\caption{Wigner distribution~(\ref{eq:wignerbell}) for the
  state~(\ref{eq:psiWnegative}). The most striking property of this
  $W_{_{\rm BELL}}$ is that it can be negative. We have used
  $p_1=p_2=0$ and $q_0=0.1$ and have plotted
  $\sqrt{\pi}W_{_{\rm BELL}}/N_{_{\rm BELL}}^2$.}
\label{fig:wignerbell}
\end{figure}

After this description of the warm-up example in Bell's paper, we now
come to the core of it. As noticed by Bell and as we have seen in
detail previously, the EPR state corresponds to a positive Wigner
function. However, Bell remarks that this is not always the case and
that, for other wavefunctions, the Wigner distribution can take
negative values [see, for instance, the case of the Schr\"odinger cat
state~(\ref{eq:catwf})]. The next and crucial step of Bell's paper is
then to study the CHSH inequality for states corresponding to Wigner
functions that are not positive definite. In particular, Bell
considers the case of the following initial wavefunction
\begin{align}
\label{eq:psiWnegative}
\Psi_{_{\rm BELL}}(q_1,q_2)\propto
\left[\left(q_1-q_2+q_0\right)^2-2a^2\right]e^{-(q_1-q_2+q_0)^2/(2a^2)},
\end{align}
where $a$ is a free parameter. Bell notices that this wavefunction is
not properly normalized but he suggests that it could easily made so
by including a factor $e^{(q_1+q_2)^2/(2b^2)}$, where $b$ is a new
parameter\footnote{Here, we recover the origin of the trick that we
  have already used previously for the EPR wavefunction.}. Notice
that, while only the difference $q_1-q_2$ appeared in
Eq.~(\ref{eq:psiWnegative}), considering this extra factor introduces
an additional dependence in $q_1+q_2$ in the wavefunction. It can be
checked that the wavefunction
\begin{align}
\label{eq:psireal}
\Psi_{_{\rm BELL}}(q_1,q_2)=\sqrt{\frac{8}{11\pi a^5 b}}
\left[\left(q_1-q_2+q_0\right)^2-2a^2\right]
e^{-(q_1-q_2+q_0)^2/(2a^2)}e^{-(q_1+q_2)^2/(2b^2)},
\end{align}
is indeed correctly normalized. However, Bells argues that the limit
$b\rightarrow \infty$ can be taken from the very beginning so that we
can work with~(\ref{eq:psiWnegative}) and ignore the more complicated
form~(\ref{eq:psireal}). The justification given by Bell is that only
relative probabilities will be calculated in the rest of his
article. So, for all practical purposes, he argues that one can work
with Eq.~(\ref{eq:psiWnegative}), replacing the proportionality symbol
by a $q_1$ and $q_2$ independent ``normalization constant''
$N_{_{\rm BELL}}$. In other words, $N_{_{\rm BELL}}$ which, according
to Eq.~(\ref{eq:psireal}), reads
$N_{_{\rm BELL}}=\sqrt{8/(11\pi a^5b)}e^{-(q_1+q_2)^2/b^2}$, is
treated as a simple constant. Then, it is easy to calculate the
corresponding Wigner function which, if the choice $a=1$ is made
(namely the value considered by Bell in his letter), reads
\begin{align}
\label{eq:wignerbell}
W_{_{\rm BELL}}(q_1,q_2,p_1,p_2,0,0)&=\frac{N_{_{\rm BELL}}^2}{\sqrt{\pi}}
e^{-(q_1-q_2+q_0)^2}e^{-(p_1-p_2)^2/4}
\biggl[\frac{11}{4}-5(q_1-q_2+q_0)^2+\left(\frac{p_1-p_2}{2}\right)^2
\nonumber \\ & 
+2\left(\frac{p_1-p_2}{2}\right)^2(q_1-q_2+q_0)^2
+\left(\frac{p_1-p_2}{2}\right)^4+(q_1-q_2+q_0)^4\biggr]
\delta(p_1+p_2).
\end{align}
This equation is in perfect agreement with Eq.~(13) of Bell's
letter~\cite{1986NYASA.480..263B} and, in addition, allows us to
identify ``$K$'' the ``{\it unimportant constant}'' introduced by Bell
in the above equation. Comparing with Eq.~(13) of
Ref.~\cite{1986NYASA.480..263B}, we have
$K=N_{_{\rm BELL}}^2/\sqrt{\pi}$. The main point of this example is
that the Wigner function~(\ref{eq:wignerbell}) can be negative in some
region as can be seen in Fig.~\ref{fig:wignerbell}.

Then, we follow the same procedure as the one already used and
explained for the EPR wavefunction: the next step consists in
calculating the distribution $\rho$. Using the definition of this
quantity, see below Eq.~(\ref{eq:reprho}), one arrives at
\begin{align}
\rho_{_{\rm BELL}}(q_1,q_2,t_1,t_2)&=N_{_{\rm BELL}}^2(1+\tau^2)^{-5/2}
e^{-(q_1-q_2+q_0)^2/(1+\tau^2)}
\nonumber \\ & \times
\left[(q_1-q_2+q_0)^4+(2\tau^2-4)(q_1-q_2+q_0)^2+\tau^4+5\tau^2+4\right],
\end{align}
where $\tau\equiv t_1+t_2$. This expression coincides with Eq.~(15) of
Ref.~\cite{1986NYASA.480..263B} and, again, we can identify the
constant $K'$ in Ref.~\cite{1986NYASA.480..263B}, namely
$K'=N_{_{\rm BELL}}^2$. The final step consists in using the
distribution $\rho_{_{\rm BELL}}$ in Eq.~(\ref{eq:E}) in order to
calculate the two-point correlators of the pseudo spin operators. This
leads to
\begin{align}
\label{eq:bellE}
E_{_{\rm BELL}}(t_1,t_2)=1-5N_{_{\rm BELL}}^2(1+\tau^2)^{-1/2}\left
(\tau^2+\frac25\right).
\end{align}
This expression is identical to Eq.~(18) of
Ref.~\cite{1986NYASA.480..263B} and, therefore, is in perfect
agreement with Bell calculations. Once more this allows us to identify
the constant called $K''$ by Bell and we have
$K''=5N_{_{\rm BELL}}^2$. Finally, one can compute the quantity
$B(t_1,t_2,t_1',t_2')$ given in Eq.~(\ref{eq:chsh}). One arrives at
\begin{align}
\label{eq:fullbell}
B_{_{\rm BELL}}(t_1,t_2,t_1',t_2')&=
2-5N_{_{\rm BELL}}^2\left[1+\left(t_1+t_2\right)^2\right]^{-1/2}
\left[\left(t_1+t_2\right)^2+\frac25\right]
\nonumber \\ &
-5N_{_{\rm BELL}}^2\left[1+\left(t_1+t_2'\right)^2\right]^{-1/2}
\left[\left(t_1+t_2'\right)^2+\frac25\right]
\nonumber \\ &
-5N_{_{\rm BELL}}^2\left[1+\left(t_1'+t_2\right)^2\right]^{-1/2}
\left[\left(t_1'+t_2\right)^2+\frac25\right]
\nonumber \\ &
+5N_{_{\rm BELL}}^2\left[1+\left(t_1'+t_2'\right)^2\right]^{-1/2}
\left[\left(t_1'+t_2'\right)^2+\frac25\right],
\end{align}
which is our final result. We have an explicit form of the Bell
operator mean value which allows us to test the CHSH inequality.

\begin{figure}
\begin{center}
\includegraphics[width=10cm]{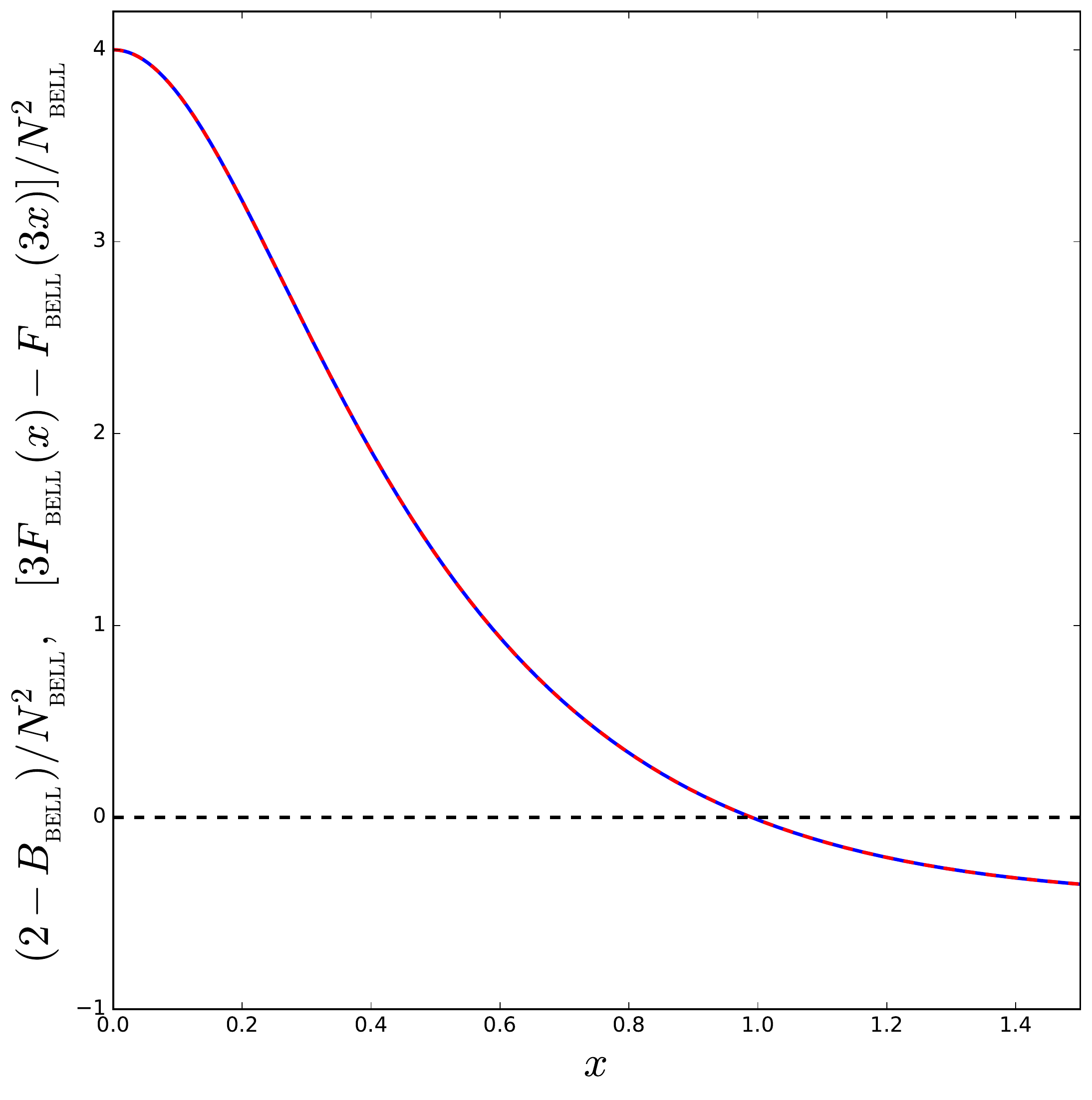}
\end{center}
\caption{The solid blue curve represents the quantity
  $[2-B_{_{\rm BELL}}(t_1,t_2,t_1',t_2')]/N_{_{\rm BELL}}^2$ with
  $t_1=-2x$, $t_2=x$, $t_1'=0$ and $t_2'=3x$ and
  $B_{_{\rm BELL}}(t_1,t_2,t_1',t_2')$ given by
  Eq.~(\ref{eq:fullbell}). The dotted red curve is
  $[3F_{_{\rm BELL}}(x)-F_{_{\rm BELL}}(3x)]/N_{_{\rm BELL}}^2$ where
  the function $F_{_{\rm BELL}}(x)$ is defined in
  Eq.~(\ref{eq:bellE}). Evidently, the two curves match perfectly
  well. Since they are negative for some values of $x$, we have
  violation of the CHSH inequality.}
\label{fig:chsh_B}
\end{figure}

In his paper, Bell calculates $B_{_{\rm BELL}}(t_1,t_2,t_1',t_2')$ for
$t_1=-2x$, $t_2=x$, $t_1'=0$ and $t_2'=3x$. If, given
Eq.~(\ref{eq:bellE}), one writes, as Bell does,
$E_{_{\rm BELL}}(t_1,t_2)=1-F_{_{\rm BELL}}(\tau)$, which defines
$F_{_{\rm BELL}}(\tau)=-5N_{_{\rm
    BELL}}^2(1+\tau^2)^{-1/2}(\tau^2+2/5)$,
then one has
$B_{_{\rm BELL}}(-2x,x,0,3x)=2-F_{_{\rm BELL}}(-x)-F_{_{\rm
    BELL}}(x)-F_{_{\rm BELL}}(x)+F_{_{\rm BELL}}(3x)$.
In fact, since $F_{_{\rm BELL}}(\tau)$ is a function of $\tau^2$ only,
the previous equation reduces to
$B_{_{\rm BELL}}(-2x,x,0,3x)=2-[3F_{_{\rm BELL}}(x)-F_{_{\rm
    BELL}}(3x)]$ or, more explicitly,
\begin{align}
\label{eq:BellB}
B_{_{\rm BELL}}(-2x,x,0,3x)=2-5N_{_{\rm BELL}}^2
\left[3(1+x^2)^{-1/2}\left(x^2+\frac25\right)
-(1+9x^2)^{-1/2}\left(9x^2+\frac25\right)\right].
\end{align}
In Fig.~\ref{fig:chsh_B}, we have have plotted the quantity
$[2-B_{_{\rm BELL}}(-2x,x,0,3x)]/N_{_{\rm BELL}}^2$ (solid blue
line). When this quantity is negative, the CHSH inequality is
violated. We see on the plot that this is indeed the case provided
$x>0.989761$, a conclusion also reached by Bell. Notice that one can
obtain this result regardless of the value of $N_{_{\rm BELL}}$. In
Fig.~\ref{fig:chsh_B}, we have also represented the quantity
$[3F_{_{\rm BELL}}(x)-F_{_{\rm BELL}}(3x)]/N_{_{\rm BELL}}^2$ (dashed
red line) where $F_{_{\rm BELL}}$ has been defined above. Evidently,
according to the previous considerations, it should exactly coincide
with $[2-B_{_{\rm BELL}}(-2x,x,0,3x)]/N_{_{\rm BELL}}^2$ and one
checks that this is indeed the case. The condition
$3F_{_{\rm BELL}}(x)-F_{_{\rm BELL}}(3x)\ge 0$ for a non violation of
the CHSH inequality is Eq.~(25) of Bell's paper.

Based on the previous considerations, Bell concludes that the
non-positivity of the Wigner function~(\ref{eq:wignerbell}) implies a
violation of the CHSH inequality. He also adds that ``{\it I do not
  know that the failure of $W$ to be non-negative is a sufficient
  condition in general for a locality paradox}''. Although it is fair
to say that the main message is not explicitly expressed in this way,
it is however clear that Bell's letter suggests a (one-to-one?)
correspondence between a violation of the CHSH inequality and the
non-positivity of the Wigner distribution. This argument seems to be
supported by the EPR example (positive Wigner function and no
violation of CHSH) and by the example we have just studied (no
positive definite Wigner function and violation of CHSH). After all,
the non-positivity of the Wigner function certainly signals that some
genuine quantum effects are at play and, when this is the case, it is
natural to think that Bell inequality could be violated. Therefore, at
first sight, this conclusion appears to be meaningful and correct. It
has very important consequences for Cosmology. Indeed, as we have
already seen, cosmological perturbations are placed in a two-mode
squeezed state, which is a Gaussian state, and, therefore, has a
positive definite Wigner function. As a consequence, Bell's result, if
true, would imply that no violation of his inequality can be observed
in the CMB.

\subsection{Is Bell's Paper Wrong?}
\label{subsec:wrongbell}

In $1997$, the article~\cite{Johansen:1997wz} was published by
L.~Johansen in {\it Physics Letters A}. In brief, this article claims
that Bell's paper is wrong. The main argument is that working with
Eq.~(\ref{eq:psiWnegative}), namely with a wave function
$\Psi_{_{\rm BELL}}(q_1,q_2)=N_{_{\rm BELL}}
\left[\left(q_1-q_2+q_0\right)^2-2a^2\right]e^{-(q_1-q_2+q_0)^2/(2a^2)}$,
where $N_{_{\rm BELL}}$ is just viewed as a constant, is incorrect
because, as we have already noticed in the previous section, this
wavefunction is not correctly normalized. Ref.~\cite{Johansen:1997wz}
quotes the book by A.~Peres, ``{\it Quantum Theory: Concepts and
  Methods}'' where, on p.~80, one is warned that not normalizing
properly the wavefunction can lead to negative probabilities or to
probabilities larger than one.

Then, Ref.~\cite{Johansen:1997wz} makes his argument more precise and
states that Bell's mistake is in fact to treat the normalization
factor $N_{_{\rm BELL}}$ as time independent. However, what
Ref.~\cite{Johansen:1997wz} does in practice is in fact much more
interesting for the question discussed here. The idea is to consider a
Wigner function which is positive definite, then use Bell's
mathematical trick described above and, finally, show that this
implies of violation of the CHSH inequality. Since, according to
Ref.~\cite{Johansen:1997wz}, one cannot have a CHSH inequality
violation if the Wigner function is positive definite, it follows that
Bell's mathematical trick and, therefore, Bell's result must be
incorrect. This is a {\it reductio ad absurdum} proof. What is
especially interesting is the fact that the correspondence positivity
of the Wigner distribution versus impossibility to violate Bell
inequality is taken for granted or is considered as obvious by the
author.

Let us now study in detail the results of
Ref.~\cite{Johansen:1997wz}. The starting point is the following Wigner 
function 
\begin{align}
\label{eq:wignerjohan}
W_{_{\rm J}}(q_1,q_2,p_1,p_2)=\frac{1}{\pi^2}
e^{-[(q_1-q_2)/\sqrt{2}-q_0]^2-[(p_1-p_2)/\sqrt{2}-p_0]^2}
e^{-s^2(q_1+q_2)^2/2-(p_1+p_2)^2/(2s^2)},
\end{align}
which is obviously positive definite. In fact, this is the product of
the Wigner function of a coherent state with the Wigner function of a
squeezed state. One can also check that this Wigner function is
correctly normalized. Then, Ref.~\cite{Johansen:1997wz} proceeds and
applies Bell's trick consisting in working with unnormalized states to
the Wigner function~(\ref{eq:wignerjohan}). In order to see what it
means in the present context, the best is to calculate the Wigner
function associated with the wavefunction~(\ref{eq:psireal}). Recall
that the wavefunction~(\ref{eq:psireal}) is the correctly normalized
version of the unnormalized wavefunction~(\ref{eq:psiWnegative})
considered before and used by Bell to show that a non-positive Wigner
function may cause a CHSH inequality violation. Assuming the
wavefunction~(\ref{eq:psireal}), the corresponding Wigner function
reads
\begin{align}
\label{eq:realwignerbell}
W_{_{\rm BELL}}(q_1,q_2,p_1,p_2)&=\frac{4}{11\pi^3a^5 b}
b\sqrt{\pi}e^{-b^2(p_1+p_2)^2/4-(q_1+q_2)^2/b^2}
\sqrt{\pi}a^5e^{-a^2(p_1-p_2)^2/4-(q_1-q_2+q_0)^2/a^2}
\nonumber \\ &
\left\{\frac{11}{4}+\left[\frac{(q_1-q_2+q_0)^2}{a^2}
+\frac{a^2}{4}(p_1-p_2)^2\right]^2+\frac{a^2}{4}(p_1-p_2)^2
-5\frac{(q_1-q_2+q_0)^2}{a^2}\right\}.
\end{align}
As discussed before, Bell claims that one can take the limit
$b\rightarrow \infty$ from the very beginning, which consists in
killing the term proportional to $(q_1+q_2)^2$ and boosting the term
proportional to $(p_1+p_2)^2$ in the argument of the first
exponential. In the limit $b\rightarrow \infty$, one has
$b\sqrt{\pi}e^{-b^2(p_1+p_2)^2/4}\rightarrow 2\pi \delta(p_1+p_2)$ and
recalling that
$N_{_{\rm BELL}}^2=8/(11 \pi a^5b)e^{-(q_1+q_2)^2/b^2}$, we see that
we exactly arrive at the Wigner function considered by Bell in his
article, namely Eq.~(\ref{eq:wignerbell}) (for $a=1$, which is the
choice made in Bell's article). The conclusion is that Bell's limit or
trick is equivalent to killing the term proportional to $(q_1+q_2)^2$
and boosting the term $\propto (p_1+p_2)^2$ in the Wigner
function. 

Therefore, coming back to Eq.~(\ref{eq:wignerjohan}) and to the
article~\cite{Johansen:1997wz}, this corresponds to taking the limit
$s\rightarrow 0$, leading to
\begin{align}
\label{eq:WignerJ}
W_{_{\rm J}}(q_1,q_2,p_1,p_2)=\frac{K}{\pi}
e^{-[(q_1-q_2)/\sqrt{2}-q_0]^2-[(p_1-p_2)/\sqrt{2}-p_0]^2}
\delta(p_1+p_2),
\end{align}
where $K=\sqrt{2/\pi}se^{-s^2(q_1+q_2)^2/2}$. We see that this Wigner
function also contains a Dirac function of $p_1+p_2$ as in
Eq.~(\ref{eq:wignerbell}), which confirms that Bell's limit has indeed
been correctly implemented in Eq.~(\ref{eq:wignerjohan}). As already
noticed before, Ref.~\cite{Johansen:1997wz} remarks that $K$ is what
Bell calls an ``{\it an unimportant constant}''. Here, we have
calculated $K$ in term of $s$, which is not done in
Ref.~\cite{Johansen:1997wz}.

Then, we repeat once more the standard procedure. We first calculate
$\rho_{_{\rm J}}(q_1,q_2,t_1,t_2)$ by integrating the Wigner function
over $p_1$ and $p_2$. We find
\begin{align}
\label{eq:rhojoha}
\rho_{_{\rm J}}(q_1,q_2,t_1,t_2)=\frac{K}{\sqrt{2\pi}}\frac{1}{\sqrt{1+\tau^2}}
e^{-[(q_1-q_2)/\sqrt{2}-q_0(\tau)]^2/(1+\tau^2)},
\end{align}
where $\tau\equiv (t_1+t_2)/2$ (recall that Bell defines $\tau$ as
$t_1+t_2$) and $q_0(\tau)\equiv q_0+p_0\tau$. This expression exactly
coincides with Eq.~(10) of Ref.~\cite{Johansen:1997wz}. The only
difference is that the constant $K$ is divided by $\sqrt{\pi}$ instead
of $\sqrt{2\pi}$ in the above expression. This factor $\sqrt{2}$ is
just due to the fact that Ref.~\cite{Johansen:1997wz} has a slightly
different definition of $K$: according to its Eq.~(3), it is indeed
the overall constant for the Wigner function~(\ref{eq:WignerJ}) if the
Dirac function appearing is written as $\delta[(p_1+p_2)/\sqrt{2}]$,
while in our case the Dirac function is simply written
$\delta(p_1+p_2)$. This difference accounts for the $\sqrt{2}$ between
the two expressions.

Although Ref.~\cite{Johansen:1997wz} is supposed to mimic Bell's paper
exactly, there are other differences between the two articles. One,
which is only a detail, is that Ref.~\cite{Johansen:1997wz} defines
the sign operators, or pseudo spin operators, with $q_0=0$, namely
$\hat{S}_1(t_1)={\rm sgn}\left(\hat{q}_1, t_1\right)$ and
$\hat{S}_2(t_2)={\rm sgn}\left(\hat{q}_2, t_2\right)$. However, this
does not affect the discussion since it was shown before that Bell's
result does not depend on $q_0$. This also means that Eq.~(\ref{eq:E})
now reads
$E_{_{\rm J}}(t_1,t_2)=1-2\left[\int_0^{+\infty}{\rm d}q_1 \int
  _{-\infty}^0{\rm d}q_2 \rho\left(q_1, q_2,t_1,t_2\right)
  +\int_{-\infty}^0{\rm d}q_1 \int _0^{+\infty}{\rm d}q_2
  \rho\left(q_1, q_2,t_1,t_2\right)\right]$.
Inserting Eq.~(\ref{eq:rhojoha}) in this last expression, one finds
that
\begin{align}
\label{eq:Ejoha}
E_{_{\rm J}}(t_1,t_2)=1-2\sqrt{2}K\left\{\frac{\sqrt{1+\tau^2}}{\sqrt{\pi}}
e^{-q_0^2(\tau)/(1+\tau^2)}+q_0(\tau){\rm erf}
\left[\frac{q_0(\tau)}{\sqrt{1+\tau^2}}\right]
\right\},
\end{align}
which coincides with Eq.~(14) of Ref.~\cite{Johansen:1997wz} (up to
the factor $\sqrt{2}$ already mentioned above). Following Bell,
Ref.~\cite{Johansen:1997wz} simply defines $F_{_{\rm J}}(\tau)$ by
$E_{_{\rm J}}(t_1,t_2)=1-F_{_{\rm J}}(\tau)$, which means 
that 
\begin{align}
F_{_{\rm J}}(\tau)=
2\sqrt{2}K\left\{\frac{\sqrt{1+\tau^2}}{\sqrt{\pi}}
e^{-q_0^2(\tau)/(1+\tau^2)}+q_0(\tau){\rm erf}
\left[\frac{q_0(\tau)}{\sqrt{1+\tau^2}}\right]
\right\},
\end{align}
in agreement with Eq.~(14) of this paper.

\begin{figure}
\begin{center}
\includegraphics[width=10cm]{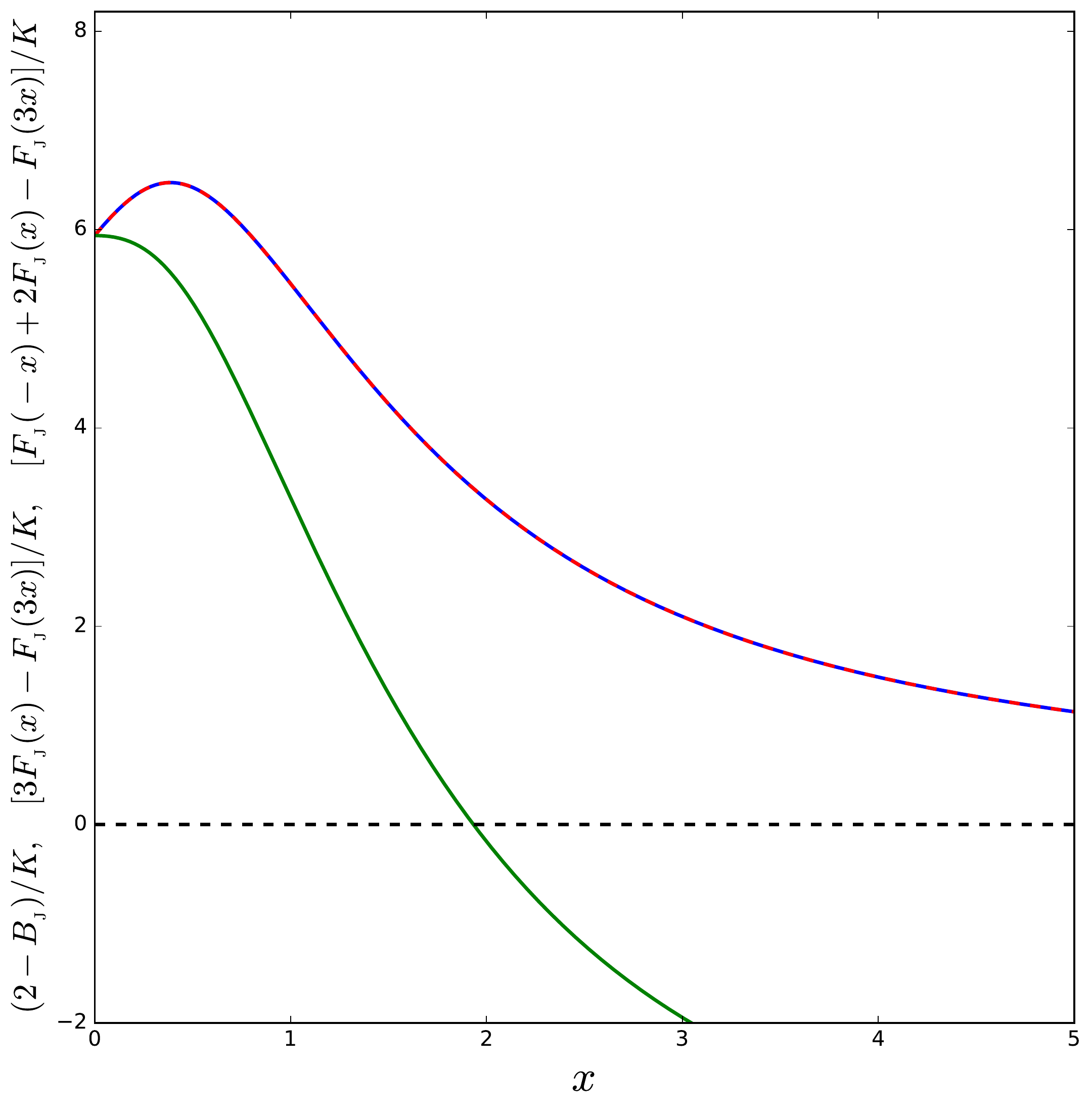}
\end{center}
\caption{The blue solid line is
  $[2-B_{_{\rm J}}(t_1,t_2,t_1',t_2')]/K$ with $t_1=-2x$, $t_2=x$,
  $t_1'=0$ and $t_2'=3x$, and $q_0=1$, $p_0=-1$, the function
  $E_{_{\rm J}}(t_1,t_2)$ appearing in the definition of
  $B_{_{\rm J}}$ being given in~(\ref{eq:Ejoha}). The red dashed line
  represents $[F_{_{\rm J}}(-x)+2F_{_{\rm J}}(x)-F_{_{\rm J}}(3x)]/K$.
  The two functions are identical and remain positive. Finally, the
  solid green line is $[3F_{_{\rm J}}(x)-F_{_{\rm J}}(3x)]/K$, which
  is incorrectly interpreted as
  $[2-B_{_{\rm J}}(t_1,t_2,t_1',t_2')]/K$ by
  Ref.~\cite{Johansen:1997wz}. This function, contrary to
  $F_{_{\rm J}}(-x)+2F_{_{\rm J}}(x)-F_{_{\rm J}}(3x)$, can be
  negative suggesting (incorrectly) that the CHSH inequality could be
  violated.}
\label{fig:chsh_J}
\end{figure}

Finally, Ref.~\cite{Johansen:1997wz} computes the mean value of the
Bell operator given in Eq.~(\ref{eq:BellB}), namely for $t_1=-2x$,
$t_2=x$, $t_1'=0$, $t_2'=3x$. Following Bell,
Ref.~\cite{Johansen:1997wz} studies the function
$3F_{_{\rm J}}(x)-F_{_{\rm J}}(3x)$ which, if it takes negative
values, signals a violation of the CHSH inequality, see the discussion
around Eq.~(\ref{eq:BellB}). Ref.~\cite{Johansen:1997wz} notices that,
if, for instance, one chooses $q_0=1$ and $p_0=-1$, this is precisely
the case in the limit $x \gtrsim 1$, see Fig.~1 of this article. In
Fig.~\ref{fig:chsh_J}, we have checked that, indeed, the function
$3F_{_{\rm J}}(x)-F_{_{\rm J}}(3x)$ can be negative, see the green
solid line. Notice that the scales in Fig.~\ref{fig:chsh_J} and in
Fig.~1 of Ref.~\cite{Johansen:1997wz} do not coincide because of the
slight difference in the definition of $\tau$ already signaled before.

Therefore, by using Bell's trick, Ref.~\cite{Johansen:1997wz} arrives
at a violation of the CHSH inequality starting with a Wigner function
which is positive definite.  According to this paper, this is
impossible because a positive Wigner function necessarily implies that
the CHSH inequality cannot be violated. As a consequence,
Ref.~\cite{Johansen:1997wz} concludes that the only way out is that
Bell's trick, and therefore his entire paper, is incorrect. The deep
reason for this mistake is that ``{\it one is not allowed to assume
  that $K$ is time independent}''.

\subsection{Are Criticisms Against Bell's Paper Wrong?}
\label{subsec:criticsbellwrong}

Let us now examine in more details the considerations of
Ref.~\cite{Johansen:1997wz} presented in the previous section.

A first remark is that Ref.~\cite{Johansen:1997wz} has $p_0\neq 0$
while there is no $p_0$ in Bell's article. This difference turns out
to be crucial because, if Bell's function $F_{_{\rm BELL}}$ defined in
Eq.~(\ref{eq:bellE}) only depends on $\tau^2$, $F_{_{\rm J}}$ in
Ref.~\cite{Johansen:1997wz} defined in Eq.~(\ref{eq:Ejoha}) depends on
$\tau$ precisely because of the presence of $p_0$ [since
$q_0(\tau)=q_0+p_0\tau$]. Only when $p_0=0$ this becomes a function of
$\tau^2$ only. This turns out to have drastic consequences. Indeed,
since $F_{_{\rm J}}(-x)\neq F_{_{\rm J}}(x)$, one cannot say that
$F_{_{\rm J}}(-x)+2F_{_{\rm J}}(x)-F_{_{\rm J}}(3x)=3F_{_{\rm
    J}}(x)-F_{_{\rm J}}(3x)$
and a signature of a CHSH inequality violation is
$F_{_{\rm J}}(-x)+2F_{_{\rm J}}(x)-F_{_{\rm J}}(3x)\le 0$ and no
longer $3F_{_{\rm J}}(x)-F_{_{\rm J}}(3x)\le 0$; and it turns out that
if $3F_{_{\rm J}}(x)-F_{_{\rm J}}(3x)$ does become negative (see the
solid green line in Fig.~\ref{fig:chsh_J}, this is not the case for
$F_{_{\rm J}}(-x)+2F_{_{\rm J}}(x)-F_{_{\rm J}}(3x)$ as shown in
Fig.~\ref{fig:chsh_J} (see the dashed red line). This clearly
invalidates the whole reasoning of Ref.~\cite{Johansen:1997wz}: Bell's
mathematical trick has not led to a fake CHSH violation for the Wigner
function~(\ref{eq:wignerjohan}).

Second, Ref.~\cite{Johansen:1997wz} claims that Bell's mistake is to
have ignored the time dependence of the constant $K$. But the constant
$K$ was calculated before and reads
$K=\sqrt{2/\pi}se^{-s^2(q_1+q_2)^2/2}$. It does not contain any time
dependence so this argument is incorrect as well.

Third, let us notice that the Wigner function considered in
Ref.~\cite{Johansen:1997wz}, namely Eq.~(\ref{eq:wignerjohan}), is
nothing but a special case of the EPR Wigner
function~(\ref{eq:wignerepr}) considered in
Sec.~\ref{subsec:bellwigner}. Indeed, if one takes $b=2/s$,
$\varepsilon=2$, $q_0\rightarrow \sqrt{2}q_0$, then
Eq.~(\ref{eq:wignerepr}) becomes identical to
Eq.~(\ref{eq:wignerjohan}) with $p_0=0$. But we have established
before that, if the state of the system is the EPR state, then no CHSH
violation can occur. This means that if we had found a violation of
the CHSH inequality starting from the Wigner
function~(\ref{eq:wignerjohan}), this would have indeed indicated a
mathematical inconsistency somewhere, as argued by
Ref.~\cite{Johansen:1997wz}. But this is not the case since the CHSH
inequality is never violated as can be seen in Fig.~\ref{fig:chsh_J}
where $(2-B_{_{\rm J}})/K$ is represented (solid blue line) and is
always positive, a conclusion already obtained before from the plot of
the combination
$[F_{_{\rm J}}(-x)+2F_{_{\rm J}}(x)-F_{_{\rm J}}(3x)]/K$ represented
by the dashed red line. Therefore, this is an additional reason why
the argument of Ref.~\cite{Johansen:1997wz}, which is entirely based
on this belief, would have remained, in any case, problematic. We
therefore conclude that the criticisms put forward in
Ref.~\cite{Johansen:1997wz} against Bell's paper are incorrect.

\subsection{Correct or not Correct?}
\label{subsec:correct}

Before closing this section, let us examine again Bell results. After
all, the fact that the criticisms against them are wrong does not mean
that Bell's paper is correct. As guessed in
Ref.~\cite{Johansen:1997wz}, we believe that Bell's treatment of the
wavefunction normalization is problematic. We have established before
that
\begin{align}
N_{_{\rm BELL}}=\sqrt{\frac{8}{11\pi a^5b}}e^{-(q_1+q_2)^2/b^2}.
\end{align}
Bell's main idea is to ignore the dependence in $q_1+q_2$ by sending
$b$ to infinity in the argument of the exponential. But we see in the
above expression that, in fact, this sends the whole normalization
factor, and therefore the whole wavefunction to zero! The same remark
of course applies to the exact Wigner function calculated in
Eq.~(\ref{eq:realwignerbell}). Therefore, this procedure cannot be
correct. Moreover, $N_{_{\rm BELL}}$ is a function of $q_1$ and $q_2$
and, as a consequence, when one integrates $\rho_{_{\rm BELL}}$ over
$\bar{q}_1$ and $\bar{q}_2$ in Eq.~(\ref{eq:bellE}), it is simply
incorrect to treat $N_{_{\rm BELL}}$ as a constant. Therefore, we also
reach the conclusion that Bell's article is incorrect even if for
completely different reasons from those put forward in
Ref.~\cite{Johansen:1997wz}.

\subsection{Revzen's Theorem}
\label{subsec:revzen}

We are apparently in a complex situation: we have found that Bell's
paper establishing a connection between the non-positivity of the
Wigner function and a CHSH inequality violation is incorrect but we
have also reached the conclusion that the criticisms expressed against
that paper are wrong as well! Moreover, all these authors seem to
agree that, if the Wigner function is positive definite, then no CHSH
inequality violation can occur which, we recall, would have important
conceptual consequences for Cosmology.

In fact, the situation was clarified in $2004$ in
Refs.~\cite{2005PhRvA..71b2103R} and~\cite{Revzen2006-REVTWF}. In
these papers, Revzen establishes that, under certain conditions that
we are going to describe, Bell inequality can be violated even if the
Wigner distribution is positive definite. Notice that this both
invalidates Bell's paper~\cite{1986NYASA.480..263B}, since Revzen's
result shows that a Bell inequality violation is not necessarily
associated to a negative Wigner function, but also
Ref.~\cite{Johansen:1997wz} since its reasoning was entirely based on
the fact that violating Bell's inequality is impossible if the Wigner
function is positive. In his paper, Revzen mentions Bell's
paper~\cite{1986NYASA.480..263B} but says that ``{\it Bell's original
  argument that nonnegativity of Wigner's function suffices to
  preclude Bell inequality violation was shown to be inaccurate}'' in
Ref.~\cite{Johansen:1997wz}. He adds that ``{\it Difficulties in
  handling normalization of the EPR state considered by Bell were
  shown to involve a misleading factor}''. As explained in the last
section, this description is not very accurate since we have just
shown that the criticisms of Ref.~\cite{Johansen:1997wz} are in fact
not valid.

Let us now come to the main result obtained by Revzen. In brief,
Revzen shows that Bell inequality can be violated even if the Wigner
function is positive definite provided the variables considered are
``improper'', namely if the Weyl transform [defined in
Eq.~(\ref{eq:defweyltrans})] of an operator takes different values
than the spectrum of that operator. Let us give an example of a proper
and improper operators. First, let us consider the pseudo spin
operators used by Bell, namely $\hat{S}={\rm sgn}(\hat{q},t)$. Its
Weyl transform is given by
\begin{align}
\widetilde{S}&=\int {\rm d}x \, 
e^{-i\pi x}\left \langle q+\frac{x}{2}\biggl \vert
{\rm sgn}(\hat{q})\biggr\vert q-\frac{x}{2}\right \rangle
=\int {\rm d}x\, e^{-i\pi x}{\rm sgn}\left(q-\frac{x}{2}\right)
\left \langle q+\frac{x}{2}\biggl \vert
 q-\frac{x}{2}\right \rangle
\nonumber \\ &
=\int {\rm d}x\, e^{-i\pi x}{\rm sgn}\left(q-\frac{x}{2}\right)\delta(x)
={\rm sgn}({q}).
\end{align}
Therefore this operator is proper since its Weyl transform takes
values $\pm 1$ which are exactly the values taken by the spectrum of
the operator. This explains why, in the EPR state, Bell inequality was
not violated in Sec.~\ref{subsec:bellwigner}. This was a situation
where the Wigner function was positive and the operator used to
construct the Bell operator was proper.

Let us now give an example of an improper operator. Let us consider
the following operator
\begin{align}
\label{eq:defimproper}
\hat{s}_z=\sum_{n=0}^{\infty}
\left(\vert 2n+1\rangle \langle 2n+1\vert 
-\vert 2n\rangle 
\langle 2n\vert \right).
\end{align}
The reason for the notation $\hat{s}_z$ will be clarified below. Here
$\vert n\rangle$ is an eigenvector of the number operator. It is easy
to show that the spectrum of this operator is $\pm 1$ because
$\hat{s}_z^2=\hat{\mathbb{I}}$. The matrix element of $\hat{s}_z$ is
given by $\langle m\vert \hat{s}_z\vert m'\rangle=\pm \delta_{mm'}$
with a plus sign if $m$ is odd and a minus sign if m is even. This
allows us to rewrite $\hat{s}_z$ as
\begin{align}
\hat{s}_z=-\int _{-\infty}^{\infty}{\rm d}q \vert q\rangle \langle -q\vert.
\end{align}
Indeed, one can show that this leads to the same matrix element, namely
\begin{align}
\langle m\vert \hat{s}_z\vert m'\rangle
=-\int_{-\infty}^{+\infty}{\rm d}q\langle m\vert 
q\rangle \langle -q\vert m'\rangle 
=\frac{-(-1)^{m'}}{\sqrt{\pi 2^{m+m'}m!m'!}}
\int_{-\infty}^{+\infty}{\rm d}qH_m(q)
H_{m'}(q)e^{-q^2} 
=-(-1)^m\delta _{mm'},
\end{align}
where $H_m$ is a Hermite polynomial of order
$n$~\cite{AbraSteg72}. It follows that
\begin{align}
\widetilde{s_z}&=-\int {\rm d}x \, e^{-i\pi x}
\left \langle q+\frac{x}{2}\biggl \vert\int {\rm d}\bar{q}
\biggl \vert \bar{q}\biggr \rangle \biggl \langle 
-\bar{q}\biggr \vert q-\frac{x}{2}
\right \rangle 
=-\int {\rm d}x\int {\rm d}\bar{q}\, 
e^{-i\pi x}\delta\left(q+\frac{x}{2}-\bar{q}\right)
\delta\left(-\bar{q}-q+\frac{x}{2}\right)
\nonumber \\ &
=-\delta (-2q)
\int {\rm d}x\, e^{-i\pi x}
=-\pi\delta (q)\delta(\pi).
\end{align}
Clearly, the Weyl transform of $\hat{s}_z$ does not take values $\pm 1$ 
and, as a consequence, this operator is improper.

The Revzen theorem rests on Eqs.~(\ref{eq:trab})
and~(\ref{eq:meano}). Indeed, according to these equations, the mean
value of an operator is the average of its Weyl transform weighted by
the Wigner function. If the Weyl transform takes the same values as
the spectrum of the operator, it means that any quantum average can be
obtained through the usual, classical, laws of random variables. But,
if this is so, Bell's theorem precisely tells us that its inequality
cannot be violated.

We conclude that the history of the relationship between the
possibility of violating Bell inequality and the positivity of the
Wigner function is a long, chaotic and rich one. For Cosmology, this
question is absolutely crucial since the Wigner function is, in this
case, positive definite. The Revzen theorem establishes the
possibility of a Bell inequality violation in the sky, a fascinating
possibility indeed. In his paper, Revzen precisely discusses his
theorem with the help of a two-mode squeezed state. What was not
realized before is that Cosmology provides a perfect situation to
illustrate this problem. It was not realized by the cosmology
community because the issues related to quantum foundations are,
usually, far from its everyday interests and it was not realized by
people working on Quantum Mechanics because the inflationary mechanism
and the fact that cosmological perturbations are placed in a two-mode
squeezed state was largely ignored by people working in this field. In
fact, given that Cosmology is the part of physics where the largest
squeezing is achieved, one can even argue that it is the most
interesting situation to discuss the issues tackled in this section.

\section{Bell Inequality Violation in the CMB?}
\label{sec:biqv}

We have just seen that, even if the CMB is placed in a quantum state
with positive Wigner function, Revzen's theorem is compatible with
Bell inequality violation in the sky. Based on this result, the next
question is of course to identify improper variables in the CMB that
could be used for that purpose. In fact, it turns out that it is
possible to build improper pseudo spin operators out of a continuous
variable, here of course taken to be the Fourier amplitude of
curvature perturbations. A first example has been considered by
Banaszek and Wodkiewics (BM) in Ref.~\cite{PhysRevLett.82.2009} and
Chen, Pan, Hou and Zhang in Ref.~\cite{PhysRevLett.88.040406}. They
have defined the following operators
\begin{eqnarray}
\label{eq:defSx1}
\hat{s}_x\left({\bm k}\right)&=& \sum_{n=0}^{\infty}
\left(\vert 2n_{\bm k}+1\rangle \langle 2n_{\bm k}\vert +\vert 2n_{\bm k}\rangle 
\langle 2n_{\bm k}+1\vert \right), \\
\label{eq:defSy1}
\hat{s}_y\left({\bm k}\right) &=& i\sum_{n=0}^{\infty}
\left(\vert 2n_{\bm k}\rangle \langle 2n_{\bm k}+1\vert 
-\vert 2n_{\bm k}+1\rangle 
\langle 2n_{\bm k}\vert \right),\\
\label{eq:defSz1}
\hat{s}_z\left({\bm k}\right) &=& \sum_{n=0}^{\infty}
\left(\vert 2n_{\bm k}+1\rangle \langle 2n_{\bm k}+1\vert 
-\vert 2n_{\bm k}\rangle 
\langle 2n_{\bm k}\vert \right),
\end{eqnarray}
where $\vert n_{\bm k}\rangle$ are the eigenvectors of the particle
number operator. These operators are spin operators because they
satisfy $\left[\hat{s}_x,\hat{s}_y\right]=2i\hat{s}_z$,
$\left[\hat{s}_x,\hat{s}_z\right]=-2i\hat{s}_y$ and
$\left[\hat{s}_y,\hat{s}_z\right]=2i\hat{s}_x$ and that their spectrum
is $\pm 1$. Notice that the $z$-component is precisely the state we
discussed in Eq.~(\ref{eq:defimproper}).

Another example of fictitious spin operators are those introduced by
Gour, Khanna, Mann and Revzen (GKMR) in~\Ref{2004PhLA..324..415G},
which are given by
\begin{align}
\label{eq:defSxalternative}
\hat{\cal S}_x &=\int _0^{+\infty} {\rm d}q_{\bm k}
\left(\vert {\cal E}_{\bm k}\rangle \langle {\cal O}_{\bm k}\vert
+\vert {\cal O}_{\bm k}\rangle \langle {\cal E}_{\bm k}\vert \right),\\
\label{eq:defSyalternative}
\hat{\cal S}_y &=i\int _0^{+\infty} {\rm d}q_{\bm k}
\left(\vert {\cal O}_{\bm k}\rangle \langle {\cal E}_{\bm k}\vert
-\vert {\cal E}_{\bm k}\rangle \langle {\cal O}_{\bm k}\vert \right),\\
\label{eq:defSzalternative}
\hat{\cal S}_z &=-\int _0^{+\infty} {\rm d}q_{\bm k}
\left(\vert {\cal E}_{\bm k}\rangle \langle {\cal E}_{\bm k}\vert
-\vert {\cal O}_{\bm k}\rangle \langle {\cal O}_{\bm k}\vert \right),
\end{align}
where $\vert {\cal E}_{\bm k}\rangle $ and
$\vert {\cal O}_{\bm k}\rangle$ are defined by
\begin{align}
\label{eq:defEO}
\vert {\cal E}_{\bm k}\rangle &=\frac{1}{\sqrt{2}}\left(\vert q_{\bm k}\rangle 
+\vert -q_{\bm k}\rangle \right), \quad
\vert {\cal O}_{\bm k}\rangle =\frac{1}{\sqrt{2}}\left(\vert q_{\bm k}\rangle 
-\vert -q_{\bm k}\rangle \right).
\end{align}
Here, $\vert q_{\bm k}\rangle $ is an eigenstate of the position
operator for the mode ${\bm k}$. Let us notice that, in principle, it
is not an eigenstate of curvature perturbations, see also
Eq.~(\ref{eq:linkzetaq}) below.

There exists a third way to define fictitious spin operators as shown
by Larsson in Ref.~\cite{2004PhRvA..70b2102L}. The $z$-component of
the Larsson spin operators can be defined as
\begin{align}
\label{eq:defsz}
\hat{S}_z(\ell)
& =\sum_{n=-\infty}^{\infty}(-1)^n\int _{n\ell}
^{(n+1)\ell}{\rm d}q_{\bm k} \vert q_{\bm k}\rangle 
\langle q_{\bm k}\vert \, ,
\end{align}
where $\ell$ is a free parameter that can be arbitrarily chosen. The 
other components can then be introduced once the step operator
\begin{align}
  \hat{S}_+(\ell)
  =\sum_{n=-\infty}^{\infty}\int_{2n\ell}^{(2n+1)\ell}{\rm
  d}q_{\bm k}\left\vert q_{\bm k}\rangle \langle q_{\bm k}
+\ell\right\vert,
\label{eq:defsplus}
\end{align}
has been defined [and
$\hat{S}_-(\ell)=\hat{S}_+^{\dagger}(\ell)$]. They are given by
$\hat{S}_x(\ell) =\hat{S}_+(\ell)+\hat{S}_-(\ell)$,
$\hat{S}_y(\ell) =-i\left[\hat{S}_+(\ell)-\hat{S}_-(\ell)\right]$.

It can be checked that, among each set of fictitious spin operators,
there are at least two spin components that are improper
operators~\cite{Martin:2016nrr,Martin:2016tbd,Martin:2017zxs}. As
discussed before, this implies that, despite the positivity of the
Wigner function, they can be used to define a quantity that violates
Bell inequality. What should be done is just to follow the pioneering
ideas of Bell's paper and implement Bell inequality with the pseudo
spin operators in its CHSH version.

\begin{figure}[t]
\centering
\includegraphics[width=6cm]{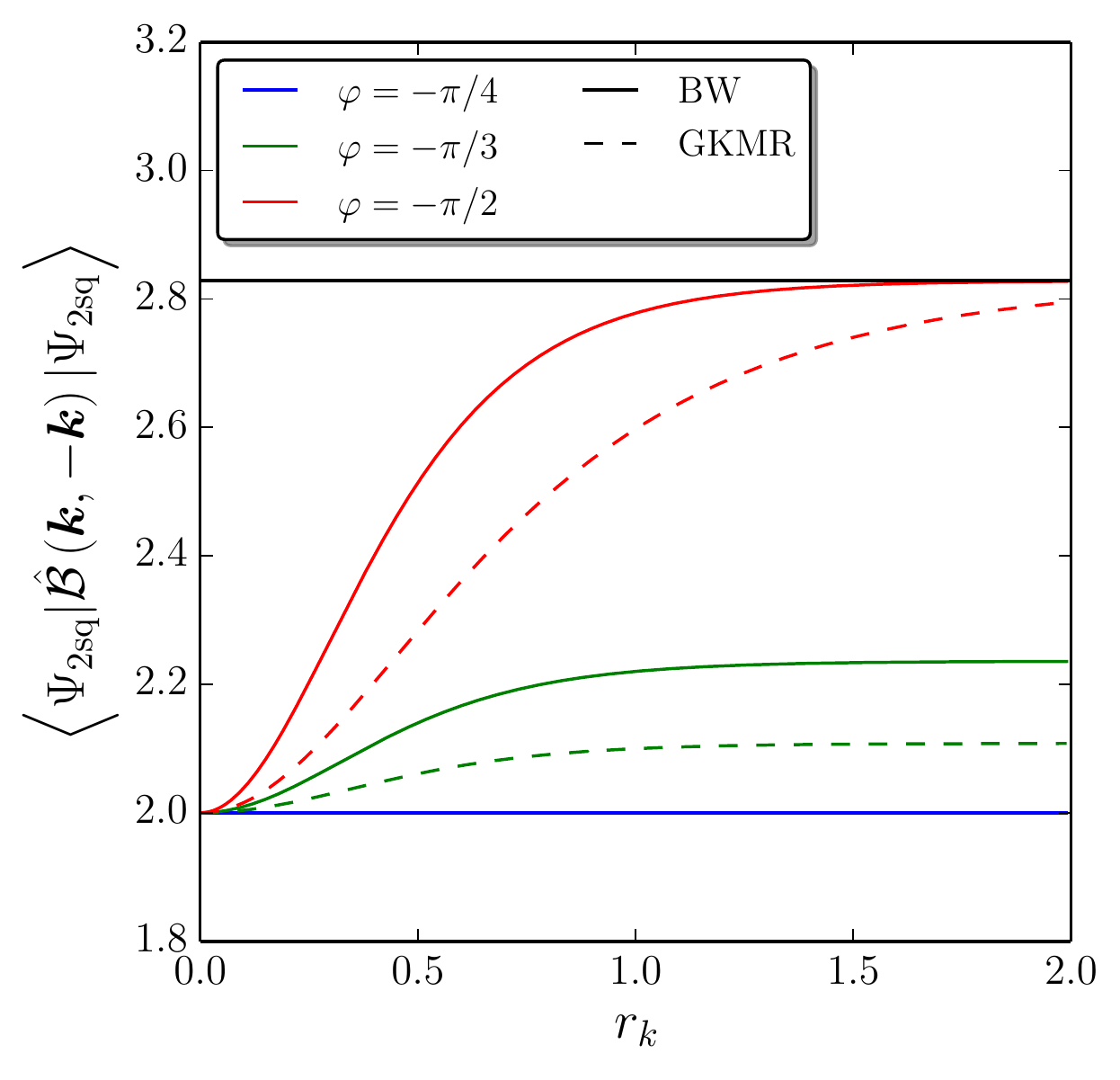}
\includegraphics[width=6.3cm]{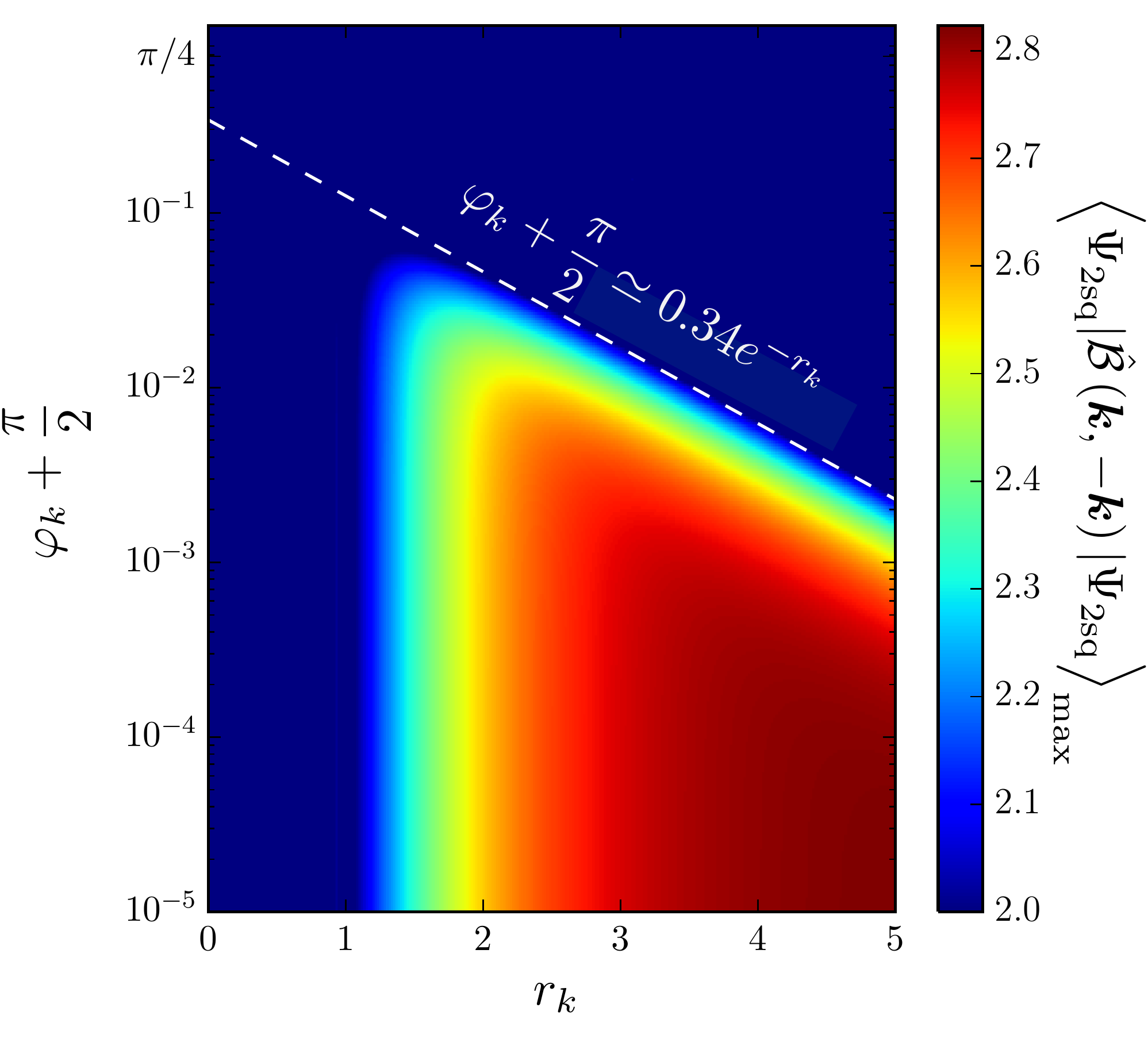}
\caption{Left panel: Mean value of the Bell operators for the BW
  (solid lines) and GKMR (dashed lines) fictitious spin operators
  versus the squeezing parameter $r_k$, for different values of the
  squeezing angle $\varphi_k$. We notice that Bell inequality is
  violated since
  $\vert \langle \Psi_{2\, {\rm sq}} \vert \hat{{\cal B}} \left({\bm
      k},-{\bm k}\right)\vert \Psi_{2\, {\rm sq}}\rangle\vert \ge 2$.
  Right panel: same quantity for the Larsson fictitious spin
  operators. The value of the Bell operator mean is indicated by a
  color code where the maximum over $\ell $ is taken. If less than
  two, the color is blue otherwise the corresponding value is
  indicated by the color bar. Bell inequality is violated in some
  region of the squeezing parameter space corresponding to large
  squeezing and small squeezing angle. Credit:
  Refs.~\cite{Martin:2016nrr,Martin:2016tbd,Martin:2017zxs}.}
\label{fig:bellspinoperator}
\end{figure}

In practice, this is done as follows. First, we view the CMB has a
bipartite system made of two sub-systems corresponding to mode
${\bm k}$ and $-{\bm k}$. Second, we calculate the following quantity
\begin{align}
\langle \Psi_{2\, {\rm sq}} \vert \hat{{\cal B}}
\left({\bm k},-{\bm k}\right)\vert \Psi_{2\, {\rm sq}}\rangle =
E\left(\theta_n,\theta_m\right)
+E\left(\theta_n,\theta_{m'}\right)
+E\left(\theta_{n'},\theta_m\right)
-E\left(\theta_{n'},\theta_{m'}\right),
\end{align}
where the two-point correlation function $E({\bm n},{\bm m})$ is
defined by
$E({\bm n},{\bm m})=\langle \Psi_{2\,{\rm sq}}\vert {\bm n}\cdot
\hat{{\mathfrak S}}\left({\bm k}\right) \otimes {\bm m}\cdot
\hat{{\mathfrak S}}\left(-{\bm k}\right) \vert \Psi_{2\,{\rm
    sq}}\rangle$,
$\hat{{\mathfrak S}}$ denoting a spin operator of one of the three
types introduced before. The vector ${\bm n}$ is a unit vector that
can be written as
${\bm n}=\left(\sin \theta_n\cos\varphi_n, \sin \theta_n \sin
  \varphi_n, \cos \theta_n\right)$
(in the following, we choose vanishing azimuthal angles). We have
calculated the mean of the Bell operator for the three sets of pseudo
spin operators introduced before, see Refs.~\cite{Martin:2016tbd}
and~\cite{Martin:2017zxs}, and the result is displayed in
Fig.~\ref{fig:bellspinoperator}. For each of them, we find that Bell
inequality can be violated (and, as a consistency check, we find that
$\vert \langle \Psi_{2\, {\rm sq}} \vert \hat{{\cal B}} \left({\bm
    k},-{\bm k}\right)\vert \Psi_{2\, {\rm sq}}\rangle\vert
<2\sqrt{2}$, the Cirel'Son bound~\cite{1980LMaPh...4...93C}).

\section{Discussion}
\label{sec:discussion}

Let us recap what has been achieved. According to inflation, curvature
perturbations that source CMB anisotropies are placed in a two-mode
squeezed state. This state has a positive Wigner function but,
nevertheless, is highly non-classical. One way to highlight this
non-classical nature is by studying the Bell inequality. It turns out
that, from the curvature perturbation, which is a continuous variable,
one can extract dichotomic spin operators (following Bell's paper),
which allows us to study the Bell inequality in its CHSH
incarnation. We have checked that, if the system is placed in a two
mode squeezed state, then this inequality is indeed
violated. Observing this violation would be the definitive proof that
CMB anisotropies are of quantum mechanical origin. Is it possible in
practice?

The first question is what it means to ``measure'' the spin
operators. Concretely, we measure the temperature anisotropy. But,
through the so-called Sachs-Wolfe effect, $\delta T/T$ is in fact a
direct measurement of curvature perturbation, as explained in
Eq.~(\ref{eq:sw}). The definition of the spin operators introduced
before, however, involves $\hat{q}_{\bm k}$ and not
$\hat{\zeta}_{\bm k}$. These two quantities are related by
\begin{align}
\label{eq:linkzetaq}
  \hat{q}_{\bm k}=\frac{z}{2}\left(\hat{\zeta}_{\bm k}
  +\hat{\zeta}_{-{\bm k}}\right)
  +\frac{z}{2k}\left(\hat{\zeta}_{\bm k}'-\hat{\zeta}_{-{\bm k}}'\right).
\end{align}
We see that the knowledge of $\hat{\zeta}_{\bm k}$ is not sufficient
to infer $\hat{q}_{\bm k}$. However, the amplitude of
$\hat{\zeta}_{\bm k}'$ is in fact related to the decaying mode as was
established before in Eq.~(\ref{eq:largescaleszeta}). Since the
curvature perturbation is conserved on large scale, this decaying mode
is in fact negligible. If one accepts this reasoning, then a
measurement of $\hat{\zeta}_{\bm k}$ is equivalent to a measurement of
$\hat{q}_{\bm k}$.

The next question is whether this allows us to measure the spin
operators? Let us discuss this question for the GKMR operators defined
in \Eqs{eq:defSxalternative}, (\ref{eq:defSyalternative})
and~(\ref{eq:defSzalternative}) (the same conclusion applies to the
two other sets). It is interesting to notice that
\begin{align}
  \left\langle q_{\bm k}\left\vert \left[\hat{\cal S}_x({\bm k}),
\hat{q}_{\bm k}\right]
  \right\vert q_{\bm k}'\right\rangle =0,
\end{align}
which means that $\hat{\cal S}_x({\bm k})$ can be, in principle,
inferred from a measurement of $\hat{q}_{\bm k}$. But for
$\hat{\cal S}_y$ one has
\begin{align}
 \langle q_{\bm k}\vert \left[\hat{\cal
    S}_y({\bm k}),\hat{q}_{\bm k}\right] \vert q_{\bm k}'\rangle
&=\left(q_{\bm k}-q'_{\bm k}\right)\langle -q_{\bm k} \vert q'_{\bm
  k}\rangle+q'_{\bm k}\langle q_{\bm k} \vert -q'_{\bm k}\rangle
+q_{\bm k}\langle q_{\bm k} \vert q'_{\bm k}\rangle
\nonumber \\ &= q_{\bm k} \left[ \delta\left(q_{\bm k}+q_{\bm k}^\prime\right) 
+ \delta\left(q_{\bm k}-q_{\bm k}^\prime\right)\right]\neq 0\, . 
\end{align}
This means that measuring $\hat{q}_{\bm k}$ is not sufficient to
measure the $y$-component. This implies that we would need to measure
another non commuting operator that can only be
$\hat{\zeta}_{\bm k}'$. There are two problems with this idea. First,
we would need to measure the system again and it is not clear what it
means when what is measured is the Universe itself! Maybe, by applying
some ergodic reasoning, this would mean measuring different horizon
patch on the last scattering surface? But, even if we succeed in doing
so, measuring the derivative of curvature perturbation means measuring
the decaying mode. Although it is a priori possible in principle, it
practices it is clearly impossible. Recall that its amplitude is
typically of the order $e^{-50}$, which seems out of reach for ever.

\section{Conclusions}
\label{sec:conclusion}

Let us now present our conclusions. In this article, we have studied
the quantum mechanical aspects of inflationary
perturbations. According to inflation, CMB anisotropies and large
scale structures are nothing but quantum fluctuations amplified by
gravitational instability and stretched to cosmological scales by the
cosmic expansion. This raises two questions: first, data analysis in
Cosmology is usually done without any reference to the quantum origin
of the perturbations since these ones look classical to us. Then, how
can we understand this quantum-to-classical transition? Second, if the
perturbations are really of quantum-mechanical origin, is there a
signature of this origin left over somewhere in the cosmological data?

It seems fair to acknowledge the frustrating aspect of the results
established above. We have shown that the quantum mechanical origin of
the perturbations is still encoded in the CMB map which, in some
sense, contains many Schr\"odinger cats, a fascinating conclusion
indeed! However, highlighting this signature essentially appears to be
impossible in practice due to the smallness of the signal. We have
therefore to find another method to check the quantum origin of the
galaxies. A suggestion recently made in Ref.~\cite{Martin:2016nrr} is
to use Leggett-Garg inequality rather than Bell inequality since the
former one only requires the measurement of a single spin
component. The price to pay, however, is that it should be done at
different times that is to say at different redshifts. Maybe a future
experiment such as Euclid~\cite{Laureijs:2011gra} could be useful for
that purpose since it plans to perform a ``tomography'' of the power
spectrum. The quest continues!

To address these tricky questions, we have also shown that the ideas
and questions that John Bell discusses in his letter ``{\it EPR
  correlations and EPW distributions}'', despite the technical
problems of his paper, are crucial to investigate these issues. They
allow us to better understand the quantum-to-classical transition of
the fluctuations and they help us to imagine what could be an
unambiguous signature of their origin.

The contribution of John Bell to Cosmology is usually summarized by
his letter ``{\it Quantum mechanics for cosmologists}'' (this is
chapter $15$ of the book ``{\it Speakable and unspeakable in quantum
  mechanics}'') where he emphasizes that the interpretational issues
of Quantum Mechanics are exacerbated in the context of Cosmology, see
the famous quote ``{\it Was the world wavefunction waiting to jump for
  thousands of millions of years until a single-celled living creature
  appeared? Or did it have to wait a little longer for some more
  highly qualified measurer -- with a Ph.~D.?}''. Here, we have argued
that the article ``{\it EPR correlations and EPW distributions}'' is
another, unrecognized, Bell's contribution to Cosmology and to the
theory of inflation which is probably even more important than ``{\it
  Quantum mechanics for cosmologists}''. The intriguing point is that
Bell's contribution was written before the cosmology community, thanks
to Grishchuk and Sidorov, realized that the inflationary perturbations
are placed in a two-mode squeezed state, namely an entangled state
with positive definite Wigner function.

Once more, John Bell has been a precursor!

\vspace{6pt}

\acknowledgments{I would like to thank the organizers, Guiseppe Dito
  and Hermano Velten, for the invitation to the "Estate Quantistica''
  conference held in Scalea (Italy) in June $2018$. It is a pleasure
  to thank Vincent Vennin for a long-term collaboration on the issues
  discussed in this paper and a careful reading of the manuscript. I
  also would like to thank Amaury Micheli for helpful comments. This
  contribution is dedicated to Julio Fabris, Richard Kerner and
  Winfried Zimdahl on the occasion of their birthday. I am especially
  indebted to Richard Kerner and Julio Fabris for everything I have
  learned from them.}

\conflictsofinterest{The author declares no conflict of interest.} 

\reftitle{References}

\end{document}